\documentclass[twocolumn, times]{aastex62}
\usepackage{amsmath,amstext}
\usepackage{comment}
\usepackage{longtable}
\shorttitle{The Optical, X-ray and Radio Properties of EP241021a}
\newcommand{\ergs}{${\rm erg \ cm^{-2} \ s^{-1}}$ }
\newcommand{\erg}{${\rm erg \ s^{-1}}$ }

\def\ltsima{$\; \buildrel < \over \sim \;$}
\def\simlt{\lower.5ex\hbox{\ltsima}}
\def\gtsima{$\; \buildrel > \over \sim \;$}
\def\simgt{\lower.5ex\hbox{\gtsima}}
\newcommand{\msun}{{\rm\,M$_\odot$}}

\newcommand{\srcs}{{\rm\,EP241021a}}
\newcommand{\src}{{\rm\,EP241021a }}

\begin{document}

%\begin{comment}
\title{
%Onset of delayed radio flares from a tidal disruption event in the center of a dwarf galaxy %hosting an active nucleus
%Delayed radio flares from an optical and X-ray bright tidal disruption event in the center of a dwarf galaxy 
%Delayed and fast rising radio flares from an optical and X-ray detected tidal disruption event in the center of a dwarf galaxy 
%Delayed radio flares from an optical and X-ray detected tidal disruption event  by a candidate intermediate-mass black hole
%The Einstein Probe Transient 
EP241021a: A Months-Duration X-ray Transient with Luminous Optical and Radio Emission 
%driven by relativistic jet ejections 
%possibly powered by a jetted tidal disruption event 
}

%\begin{comment}
%\correspondingauthor{X. W.~Shu \& L. Yang} 
%\email{xwshu@ahnu.edu.cn; ahnuyl@ahnu.edu.cn}

%\author{EP collaboration}
%\author{Fabao~Zhang }
%\affil{Department of Physics, Anhui Normal University, Wuhu, Anhui, 241002, China}

\author[0000-0002-7020-4290]{Xinwen~Shu }
\affil{Department of Physics, Anhui Normal University, Wuhu, Anhui, 241002, China}
%\email{xwshu@ahnu.edu.cn}

\author{Lei Yang}
\affil{Department of Physics, Anhui Normal University, Wuhu, Anhui, 241002, China} %xwshu@ahnu.edu.cn}
% ahnuyl@ahnu.edu.cn

\author{Haonan Yang}
\affil{National Astronomical Observatories, Chinese Academy of Sciences, Beijing 100101, China} 
% hnyang@nao.cas.cn

\author{Fan Xu}
\affil{Department of Physics, Anhui Normal University, Wuhu, Anhui, 241002, China} %xwshu@ahnu.edu.cn}
% fanxu@ahnu.edu.cn

\author{Jin-Hong Chen}
\affil{Department of Physics, The University of Hong Kong, Pokfulam Road, Hong Kong, China} 
% chenjh2@hku.hk 

\author{Rob A. J. Eyles-Ferris}
\affil{School of Physics and Astronomy, University of Leicester, University Road, Leicester, LE1 7RH, UK} 
% raje1@leicester.ac.uk 

\author{Lixin Dai}
\affil{Department of Physics, The University of Hong Kong, Pokfulam Road, Hong Kong, China} %xwshu@ahnu.edu.cn}
% lixindai@hku.hk

\author{Yunwei Yu}
\affil{Institute of Astrophysics, Central China Normal University, Wuhan 430079, China} %xwshu@ahnu.edu.cn}
% yuyw@ccnu.edu.cn

\author{Rong-Feng Shen}
\affil{School of Physics and Astronomy, Sun Yat-Sen University, Zhuhai 519082, China} %xwshu@ahnu.edu.cn}
% shenrf3@mail.sysu.edu.cn

\author{Luming Sun}
\affil{Department of Physics, Anhui Normal University, Wuhu, Anhui, 241002, China} %xwshu@ahnu.edu.cn}
% sunluming@ahnu.edu.cn 

\author{Hucheng~Ding }
\affil{Department of Physics, Anhui Normal University, Wuhu, Anhui, 241002, China} %xwshu@ahnu.edu.cn}
% dinghc@ahnu.edu.cn

\author{WeiKang Zheng }
\affil{Department of Astronomy, University of California, Berkeley, CA 94720-3411, USA} 
% weikang@berkeley.edu 

\author{Ning Jiang}
\affil{Department of Astronomy, University of Science and Technology of China, Hefei, Anhui 230026, China}
% jnac@ustc.edu.cn

\author{Wenxiong Li }
\affil{National Astronomical Observatories, Chinese Academy of Sciences, Beijing 100101, China} 
% liwx@bao.ac.cn

\author{Ning-Chen Sun}
\affil{School of Astronomy and Space Science, University of Chinese Academy of Sciences, Beijing 100049, China}
\affil{National Astronomical Observatories, Chinese Academy of Sciences, Beijing 100101, China}
\affil{Institute for Frontier in Astronomy and Astrophysics, Beijing Normal University, Beijing 102206, China}
% sunnc@ucas.ac.cn

\author{Dong Xu }
\affil{National Astronomical Observatories, Chinese Academy of Sciences, Beijing 100101, China}
\affil{Altay Astronomical Observatory, Altay, Xinjiang 836500, China}
% dxu@nao.cas.cn

%\author{WeiKang Zheng }
%\affil{Department of Astronomy, University of California, Berkeley, CA 94720-3411, USA} 
% weikang@berkeley.edu 

\author{Zhumao Zhang}
%\author{Fabao~Zhang }
\affil{Department of Physics, Anhui Normal University, Wuhu, Anhui, 241002, China} %xwshu@ahnu.edu.cn}
% zhmzhang@ahnu.edu.cn

\author{Chichuan Jin}
\affil{National Astronomical Observatories, Chinese Academy of Sciences, Beijing 100101, China}
\affil{School of Astronomy and Space Science, University of Chinese Academy of Sciences, Beijing 100049, China}
\affil{Institute for Frontier in Astronomy and Astrophysics, Beijing Normal University, Beijing 102206, China}
% ccjin@nao.cas.cn 

\author{Arne Rau}
\affil{Max-Planck-Institut für extraterrestrische Physik, Giessenbachstrasse 1, 85748 Garching, Germany}
% arau@mpe.mpg.de 

\author{Tinggui Wang}
\affil{Department of Astronomy, University of Science and Technology of China, Hefei, 230026, China} 
% twang@ustc.edu.cn

\author{Xue-feng Wu}
\affil{Purple Mountain Observatory, Chinese Academy of Sciences, Nanjing 210023, China} %xwshu@ahnu.edu.cn}
% xfwu@pmo.ac.cn

\author{Weimin Yuan}
\affil{National Astronomical Observatories, Chinese Academy of Sciences, Beijing 100101, China}
% wmy@nao.cas.cn

\author[0000-0002-9725-2524]{Bing Zhang}
\affil{Department of Physics, University of Hong Kong, Pokfulam Road, Hong Kong, China}
\affil{Nevada Center for Astrophysics and Department of Physics and Astronomy, University of Nevada, Las Vegas, NV 89154, USA}

%Nevada Center for Astrophysics, University of Nevada Las Vegas, NV 89154, USA.} %xwshu@ahnu.edu.cn}
%\affil{Department of Physics and Astronomy, University of Nevada Las Vegas, NV 89154, USA}
% bing.zhang@unlv.edu

\author{Kirpal Nandra}
\affil{Max-Planck-Institut für extraterrestrische Physik, Giessenbachstrasse 1, 85748 Garching, Germany}
%  	knandra@mpe.mpg.de 

\author{Alexei~V.~Filippenko} 
\affil{Department of Astronomy, University of California, Berkeley, CA 94720-3411, USA}
% afilippenko@berkeley.edu

\author{Fr\'ed\'erick Poidevin}
\affil{Instituto de Astrof\'isica de Canarias, E-38205 La Laguna, Tenerife, Spain}
\affil{Departamento de Astrof\'isica, Universidad de La Laguna, E-38206 La Laguna, Tenerife, Spain}
% frederick.poidevin@iac.es

\author{Roberto Soria}
\affil{INAF--Osservatorio Astrofisico di Torino, Strada Osservatorio 20, I-10025 Pino Torinese, Italy}
\affil{National Astronomical Observatories, Chinese Academy of Sciences, Beijing 100101, China} % %
% rsoria@nao.cas.cn

\author{Amit Kumar}
\affil{Department of Physics, Royal Holloway - University of London, Egham, TW20 0EX, UK}
%\affil{Department of Physics, University of Warwick, Gibbet Hill Road, Coventry CV4 7AL, UK}
% Amit.Kumar@rhul.ac.uk

\author{David S. Aguado}
\affil{Instituto de Astrof\'isica de Canarias, E-38205 La Laguna, Tenerife, Spain}
\affil{Departamento de Astrof\'isica, Universidad de La Laguna, E-38206 La Laguna, Tenerife, Spain}
% david.aguado@iac.es

\author{Fangxia An}
\affil{Yunnan Observatories, Chinese Academy of Sciences, Kunming 650216}
% anfangxia@ynao.ac.cn

\author{Tao An}
\affil{Shanghai Astronomical Observatory, Chinese Academy of Sciences, 80 Nandan Road, Shanghai 200030, China}
\affil{State Key Laboratory of Radio Astronomy and Technology, A20 Datun Road, Chaoyang District, Beijing, P. R. China}
% antao@shao.ac.cn 

\author{Jie An}
\affil{National Astronomical Observatories, Chinese Academy of Sciences, Beijing 100101, China}
\affil{School of Astronomy and Space Science, University of Chinese Academy of Sciences, Beijing 100049, China}
% anjie@nao.cas.cn

\author{Moira Andrews}
\affil{Las Cumbres Observatory, 6740 Cortona Drive, Suite 102, Goleta, CA 93117-5575, USA}
% mandrews@lco.global 

\author{Rungrit Anutarawiramkul}
\affil{National Astronomical Research Institute of Thailand, 260 Moo 4, Donkaew, Maerim, Chiang Mai 50180, Thailand} 
% rungrit@narit.or.th 

\author{Pietro Baldini}
\affil{Max-Planck-Institut für extraterrestrische Physik, Giessenbachstrasse 1, 85748 Garching, Germany}
% baldini@mpe.mpg.de

\author{Thomas~G.~Brink} 
\affil{Department of Astronomy, University of California, Berkeley, CA 94720-3411, USA}
% tgbrink@berkeley.edu

\author{Pathompong Butpa}
\affil{National Astronomical Research Institute of Thailand, 260 Moo 4, Donkaew, Maerim, Chiang Mai 50180, Thailand}
% pathompong@narit.or.th

\author{Zhiming Cai}
\affil{Innovation Academy for Microsatellites, Chinese Academy of Sciences, Shanghai 201210, China}

\author{Alberto J. Castro-Tirado}
\affil{Instituto de Astrof\'isica de Andaluc\'ia (IAA-CSIC), Glorieta de la Astronom\'ia s/n, 18008 Granada, Spain}
\affil{Ingeniería de Sistemas y Autom\'atica, Universidad de M\'alaga, Unidad Asociada al CSIC por el IAA, Escuela de Ingenier\'ias Industriales, Arquitecto Francisco Pe\~nalosa, 6, Campanillas, 29071 M\'alaga, Spain}

\author{Huaqing Cheng}
\affil{National Astronomical Observatories, Chinese Academy of Sciences, Beijing 100101, China}
% hqcheng@nao.cas.cn

\author{Weiwei Cui}
\affil{Key Laboratory of Particle Astrophysics, Institute of High Energy Physics, Chinese Academy of Sciences, Beijing 100049, China}

\author{Joseph Farah}
\affil{Las Cumbres Observatory, 6740 Cortona Drive, Suite 102, Goleta, CA 93117-5575, USA}
\affil{Department of Physics, University of California, Santa Barbara, CA 93106-9530, USA }
%  josephfarah@ucsb.edu

%\author{Alexei~V.~Filippenko} 
%\affil{Department of Astronomy, University of California, Berkeley, CA 94720-3411, USA}
% afilippenko@berkeley.edu

\author{Shaoyu Fu}
\affil{Department of Astronomy, School of Physics, Huazhong University of Science and Technology, Wuhan, 430074, China}
% syfu@nao.cas.cn

\author{Johan P.U. Fynbo}
\affil{Cosmic Dawn Center (DAWN), Copenhagen 2200, Denmark}
\affil{Niels Bohr Institute, University of Copenhagen, Copenhagen 2200, Denmark}
% jfynbo@nbi.ku.dk

\author{Xing Gao }
\affil{Xinjiang Astronomical Observatory, Chinese Academy of Sciences, Urumqi, Xinjiang 830011, China}
% gaoxing@nao.cas.cn

\author{Dawei Han}
\affil{Key Laboratory of Particle Astrophysics, Institute of High Energy Physics, Chinese Academy of Sciences, Beijing 100049, China}

\author{Xuhui~Han} 
\affil{National Astronomical Observatories, Chinese Academy of Sciences, Beijing 100101, China}
% hxh@nao.cas.cn

\author{D. Andrew Howell}
\affil{Las Cumbres Observatory, 6740 Cortona Drive, Suite 102, Goleta, CA 93117-5575, USA}
\affil{Department of Physics, University of California, Santa Barbara, CA 93106-9530, USA }
% dahowell@gmail.com 

\author{Jingwei Hu}
\affil{National Astronomical Observatories, Chinese Academy of Sciences, Beijing 100101, China} 
% hujingwei@nao.cas.cn 

\author{Shuaiqing Jiang}
\affil{National Astronomical Observatories, Chinese Academy of Sciences, Beijing 100101, China}
\affil{School of Astronomy and Space Science, University of Chinese Academy of Sciences, Beijing 100049, China}
% sqjiang@nao.cas.cn 

%\author{Amit Kumar}
%\affil{Department of Physics, Royal Holloway - University of London, Egham, TW20 0EX, UK}
%\affil{Department of Physics, University of Warwick, Gibbet Hill Road, Coventry CV4 7AL, UK}
% Amit.Kumar@rhul.ac.uk

\author[0000-0001-7225-2475]{Brajesh Kumar}
\affiliation{South-Western Institute for Astronomy Research (SWIFAR), Yunnan University, Kunming, Yunnan 650500, People's Republic of China}
\affiliation{Key Laboratory of Survey Science of Yunnan Province, Yunnan University, Kunming, Yunnan 650500, People's Republic of China}

\author{Weihua Lei}
\affil{Department of Astronomy, School of Physics, Huazhong University of Science and Technology, Wuhan, 430074, China}
%  	leiwh@hust.edu.cn 

\author{Dongyue Li}
\affil{National Astronomical Observatories, Chinese Academy of Sciences, Beijing 100101, China} 
% dyli@nao.cas.cn

\author{Chengkui Li}
\affil{Key Laboratory of Particle Astrophysics, Institute of High Energy Physics, Chinese Academy of Sciences, Beijing 100049, China}

\author{Huaqiu Liu}
\affil{Innovation Academy for Microsatellites, Chinese Academy of Sciences, Shanghai 201210, China}

\author{Xing Liu }
\affil{National Astronomical Observatories, Chinese Academy of Sciences, Beijing 100101, China}
\affil{School of Astronomy and Space Science, University of Chinese Academy of Sciences, Beijing 100049, China}
% liuxing@nao.cas.cn 

\author{Yuan Liu }
\affil{National Astronomical Observatories, Chinese Academy of Sciences, Beijing 100101, China}
% liuyuan@bao.ac.cn

\author[0000-0003-1295-2909]{Xiaowei Liu}
\affiliation{South-Western Institute for Astronomy Research (SWIFAR), Yunnan University, Kunming, Yunnan 650500, People's Republic of China}
\affiliation{Key Laboratory of Survey Science of Yunnan Province, Yunnan University, Kunming, Yunnan 650500, People's Republic of China}

\author{Alicia L\'opez-Oramas}
\affil{Instituto de Astrof\'isica de Canarias, E-38205 La Laguna, Tenerife, Spain}
\affil{Departamento de Astrof\'isica, Universidad de La Laguna, E-38206 La Laguna, Tenerife, Spain}
% alicia.lopez@iac.es

\author{David L\'opez Fern\'andez-Nespral}
\affil{Instituto de Astrof\'isica de Canarias, E-38205 La Laguna, Tenerife, Spain}
\affil{Departamento de Astrof\'isica, Universidad de La Laguna, E-38206 La Laguna, Tenerife, Spain}
% david.lopez@iac.es

\author{Justyn R. Maund}
\affil{Department of Physics, Royal Holloway - University of London, Egham, TW20 0EX, UK}
% Justyn.Maund@rhul.ac.uk

\author{Curtis McCully }
\affil{Las Cumbres Observatory, 6740 Cortona Drive, Suite 102, Goleta, CA 93117-5575, USA}
\affil{Department of Physics, University of California, Santa Barbara, CA 93106-9530, USA }
%  cmccully@lco.global  

\author{Zexi Niu}
\affil{School of Astronomy and Space Science, University of Chinese Academy of Sciences, Beijing 100049, China}
\affil{National Astronomical Observatories, Chinese Academy of Sciences, Beijing 100101, China}
% nzx@nao.cas.cn

\author{Megan Newsome }
\affil{Las Cumbres Observatory, 6740 Cortona Drive, Suite 102, Goleta, CA 93117-5575, USA}
\affil{Department of Physics, University of California, Santa Barbara, CA 93106-9530, USA }
% mnewsome@lco.global 

\author{Paul O'Brien}
\affil{School of Physics and Astronomy, University of Leicester, University Road, Leicester, LE1 7RH, UK} 
% paul.obrien@leicester.ac.uk

\author{Haiwu Pan}
\affil{National Astronomical Observatories, Chinese Academy of Sciences, Beijing 100101, China}
% panhaiwu@bao.ac.cn

\author[0009-0002-7625-2653]{Yu Pan}
\affiliation{South-Western Institute for Astronomy Research (SWIFAR), Yunnan University, Kunming, Yunnan 650500, People's Republic of China}
\affiliation{Key Laboratory of Survey Science of Yunnan Province, Yunnan University, Kunming, Yunnan 650500, People's Republic of China}

\author{Estefania Padilla Gonzalez}
\affil{Space Telescope Science Institute 3700 San Martin Drive, Baltimore, MD 21218, USA}
% epadill7@jh.edu 

\author{Ismael P\'erez-Fournon}
\affil{Instituto de Astrof\'isica de Canarias, E-38205 La Laguna, Tenerife, Spain}
\affil{Departamento de Astrof\'isica, Universidad de La Laguna, E-38206 La Laguna, Tenerife, Spain}
% ipf@iac.es

%\author{Fr\'ed\'erick Poidevin}
%\affil{Instituto de Astrof\'isica de Canarias, E-38205 La Laguna, Tenerife, Spain}
%\affil{Departamento de Astrof\'isica, Universidad de La Laguna, E-38206 La Laguna, Tenerife, Spain}
% frederick.poidevin@iac.es

\author{Walter Silima}
\affil{Inter-University Institute for Data Intensive Astronomy (IDIA), Department of Astronomy, University of Cape Town, 7701 Rondebosch, Cape Town, South Africa}
% pfunzowalter@gmail.com

%\author{Roberto Soria}
%\affil{INAF--Osservatorio Astrofisico di Torino, Strada Osservatorio 20, I-10025 Pino Torinese, Italy}
%\affil{National Astronomical Observatories, Chinese Academy of Sciences, Beijing 100101, China} % %
% rsoria@nao.cas.cn

\author{Hui Sun}
\affil{National Astronomical Observatories, Chinese Academy of Sciences, Beijing 100101, China} 
% hsun@nao.cas.cn  

\author{Shengli Sun}
\affil{Shanghai Institute of Technical Physics, Chinese Academy of Sciences, Shanghai 200083, China }

\author{Xiaojin Sun}
\affil{Shanghai Institute of Technical Physics, Chinese Academy of Sciences, Shanghai 200083, China }

\author{Giacomo Terreran}
\affil{Adler Planetarium 1300 S Dusable Lk Shr Dr, Chicago, IL 60605, USA}
%  gterreran@lco.global 

\author{Samaporn Tinyanont }
\affil{National Astronomical Research Institute of Thailand, 260 Moo 4, Donkaew, Maerim, Chiang Mai 50180, Thailand}
% samaporn@narit.or.th 

\author{Junxian Wang}
\affil{Department of Astronomy, University of Science and Technology of China, Hefei, 230026, China} %xwshu@ahnu.edu.cn}
% jxw@ustc.edu.cn

\author{Yanan Wang}
\affil{National Astronomical Observatories, Chinese Academy of Sciences, Beijing 100101, China}
% wangyn@bao.ac.cn

\author{Yun Wang}
\affil{Key Laboratory of Dark Matter and Space Astronomy, Purple Mountain Observatory, Chinese Academy of Sciences, Nanjing 210023, China} %xwshu@ahnu.edu.cn}

\author{Klaas Wiersema}
\affil{Centre for Astrophysics Research, University of Hertfordshire, Hatfield AL10 9AB, UK}
% k.wiersema@herts.ac.uk

\author{Yunfei Xu}
\affil{National Astronomical Observatories, Chinese Academy of Sciences, Beijing 100101, China} 
% xuyf@nao.cas.cn

\author{Yongquan Xue}
\affil{Department of Astronomy, University of Science and Technology of China, Hefei, 230026, China} %xwshu@ahnu.edu.cn}
% xuey@ustc.edu.cn

\author{Yi~Yang} 
\affil{Physics Department, Tsinghua University, Beijing, 100084, China}
\affil{Department of Astronomy, University of California, Berkeley, CA 94720-3411, USA}
%  yiyangtamu@gmail.com

\author{Fabao~Zhang }
\affil{Department of Physics, Anhui Normal University, Wuhu, Anhui, 241002, China} %xwshu@ahnu.edu.cn}
% zhang681@mail.ahnu.edu.cn

\author{Juan Zhang}
\affil{Key Laboratory of Particle Astrophysics, Institute of High Energy Physics, Chinese Academy of Sciences, Beijing 100049, China}

\author{Pinpin~Zhang} 
\affil{National Astronomical Observatories, Chinese Academy of Sciences, Beijing 100101, China}
% ppzhang@nao.cas.cn

\author{Wenda Zhang}
\affil{National Astronomical Observatories, Chinese Academy of Sciences, Beijing 100101, China} 
% wdzhang@nao.cas.cn

\author{Yonghe Zhang}
\affil{Innovation Academy for Microsatellites, Chinese Academy of Sciences, Shanghai 201210, China}

\author{Haisheng Zhao}
\affil{Key Laboratory of Particle Astrophysics, Institute of High Energy Physics, Chinese Academy of Sciences, Beijing 100049, China}

\author{Zipei Zhu}
\affil{National Astronomical Observatories, Chinese Academy of Sciences, Beijing 100101, China}
% zpzhu@nao.cas.cn

\author{Liping Xin} 
\affil{National Astronomical Observatories, Chinese Academy of Sciences, Beijing 100101, China}
 
\author{Zhuheng Yao} 
\affil{National Astronomical Observatories, Chinese Academy of Sciences, Beijing 100101, China}

\author{Bertrand Cordier} 
\affil{CEA Paris-Saclay, Irfu/D´epartement d’Astrophysique, 9111 Gif sur Yvette, France}

\author{Jianyan Wei}
\affil{National Astronomical Observatories, Chinese Academy of Sciences, Beijing 100101, China}

\author{Yulei Qiu} 
\affil{National Astronomical Observatories, Chinese Academy of Sciences, Beijing 100101, China}

\author{Frédéric Daigne} 
\affil{Sorbonne Universit´e, CNRS, UMR 7095, Institut d’Astrophysique de Paris, 98 bis bd Arago, F-75014 Paris, France}

%\author{Xinwen Shu}
%\affil{Department of Physics, Anhui Normal University, Wuhu, Anhui, 241002, China} %xwshu@ahnu.edu.cn}

\begin{abstract}
%Einstein Probe (EP) recently discovered a peculiar X-ray transient, \srcs, 
%by the Einstein Probe (EP) mission, 
%as well as 
We present the discovery of a peculiar X-ray transient, \srcs, by the Einstein Probe (EP) mission, and the results from multiwavelength follow-up observations. 
%of a peculiar X-ray transient, \srcs, recently discovered by the Einstein Probe (EP) mission. 
The transient was first detected with the Wide-field X-ray Telescope as an intense flare lasting for $\sim$100 s, reaching a luminosity of $L_{\rm 0.5-4 keV}\approx 10^{48}$\,\erg. % at $z=0.748$. 
Further observations with EP's Follow-up X-ray Telescope reveal a huge drop 
in the X-ray flux by a factor of $>$1000 within 1.5 days. 
After maintaining a nearly plateau phase for 
$\sim$7 days, the X-ray flux decreases $\propto t^{-1.2}$ over a period of $\sim$30 days, followed by a sudden decrease to an undetectable level by EP and XMM-Newton, making it the longest afterglow emission detected among known fast X-ray transients. 
%The transient shows a non-thermal X-ray spectrum in 0.5--10 keV, which can be described with a 
%power-law with photon index of $\approx 1.8$. 
A bright counterpart at optical and radio wavelengths was also detected, with high peak luminosities in excess of $10^{44}$\,\erg and $10^{41}$\,\erg respectively. 
%These properties, together 
%Combining with 
In addition, \src exhibits a nonthermal X-ray spectrum, red optical color, X-ray and optical rebrightenings in the light curves, and fast radio spectral evolution, suggesting that relativistic jets may have been launched. 
%, possibly driven by an active central engine. 
%We favor a merger-triggered magnetar or repeating partial tidal disruption event involving an intermediate-mass black hole as possible origins of \srcs, although neither can perfectly explain the multiwavelength properties. 
We discuss possible origins of EP241021a, including a choked jet with supernova shock breakout, a merger-triggered magnetar, a highly structured jet, and a repeating partial tidal disruption event involving an intermediate-mass black hole, but none can perfectly explain the multiwavelength properties. 
\src may represent a new type of X-ray transient with month-duration evolution timescales; 
 future EP detections and follow-up observations of similar systems will provide statistical samples to understand the underlying mechanisms at work.

\end{abstract}


\section{Introduction} \label{sec:intro}
%There now exists compelling evidence 

In the past decade, a few tens of {extragalactic fast X-ray transients (EFXTs)} 
have been discovered with 
{X-ray missions such as Swift, Chandra, and XMM-Newton,}  
either from serendipitious observations or through target searches of archival data 
\citep[e.g.,][]{Soderberg2008, Jonker2013, Glennie2015, Bauer2017, Alp2020, Quirola2022, Quirola2023, Lin2022, Eappachen2023}. 
These {EFXTs} are characterized by intense bursts of soft X-ray emission lasting tens to thousands of seconds %and exhibiting 
with a wide range of luminosities,  
%Based on the sparse prompt multi-wavelength follow-ups 
the nature of which remains enigmatic. 
%typically characterized by intense X-ray emission with timescales between hundreds of seconds to hours. 
%The sparse prompt multi-wavelength follow-ups  %since the archival discovery of FXTs make 
%make it challenging to characterize the physical properties and uncover the origin of FXTs. 
Several possibilities have been invoked as the origins of {EFXTs}, 
including %stellar flares \citep{Glennie2015}, 
supernova shock breakout \citep{Soderberg2008, Alp2020}, 
long gamma-ray bursts \citep[GRBs;][]{Bauer2017}, 
magnetars after binary neutron star mergers \citep{Lin2022, Quirola2024}, and
stellar tidal disruption events involving an intermediate-mass black hole 
\citep{Jonker2013, Shen2019, Peng2019}. 
%or {\bf a possibly new class of transient phenomena previously unseen.}
%an entirely new class of transient phenomena. 
{On the other hand, EFXTs may represent a new and heterogeneous class of 
transient phenomena not well explained by any single model.}
%the main obstacle to distinguish different mechanisms is the 
Owing to the lack of prompt multiwavelength follow-up observations in previous studies, it is challenging to explore and  
distinguish these different mechanisms in detail.  
%lack of made their characterizations difficult. 

%The observational paradigm for FXTs has recently undergone
%a drastic shift with the launch of the Einstein Probe (EP) X-ray
%telescope (W. Yuan et al. 2022).
%X-ray flashes intense bursts of soft X-ray emission first identified ago.

%On the other hand, some jetted TDE flares could be misidentified as long gamma-ray bursts (GRBs) or X-ray flashes 
%lasting $\sim$$10^{3}$s \citep{Tchekhovskoy..2014}, making {the true abundance of jetted TDEs largely uncertain. }
%Radio observations of optical and X-ray TDEs in their peaks are crucial to 

Following its successful launch on 2024 Jan. 9, the Einstein Probe (EP, also known as the Tianguan telescope) has started
monitoring the sky in the soft X-ray regime \citep{Yuan2025}\footnote{https://ep.bao.ac.cn/ep/}, 
opening a new avenue to characterize and study {EFXTs}. 
%The Einstein Probe (EP) is a time-domain X-ray mission aiming to discover cosmic 
%high-energy transients with unprecedented sensitivities (Yuan+2022). 
%It features a 
Operating in the 0.4--5 keV band, the wide-field X-ray telescope (WXT) onboard EP has a large instantaneous field of view ($\sim$3600 deg$^2$), and is capable of surveying the available night sky several times per day. 
%and a sensitivity of $2-3 \times10^{-11}$ $\rm erg\,cm^{-2}\,s^{-1}$ with 1\,ks exposure. 
EP also carries an X-ray telescope (FXT) in the 0.3--10 keV range, 
%WXT has a large instantaneous field-of-view ($\sim$3600 sq. deg.),
%that is realised via the lobster-eye micro-pore X-ray focusing optics. 
%FXT is conventional X-ray focusing telescope 
which has a larger effective area to perform follow-up observations and provide more precise localization of %newly-discovered 
new transients triggered by WXT.  %with a spatial resolution of $\sim$10--20$^{\prime\prime}$ and 
%with a typical sensitivity of $\sim$$10^{-13}$ $\rm erg\,cm^{-2}\,s^{-1}$. 
%will rapidly increase the sample of TDEs discovered in real time, including jetted ones.
%This unique capability is 
The prompt follow-up FXT observations are capable of determining the temporal evolution  of the afterglow X-ray emission (down to a typical sensitivity of $\sim 10^{-14}$\,$\rm erg\,cm^{-2}\,s^{-1}$) --- specifically the duration, light-curve shape, and spectral evolution, which are crucial to understanding the origin of FEXTs.  
%EP/FXT provides a unique capability to determine 

\begin{figure*}[htbp!]
\epsscale{0.9}
\plotone{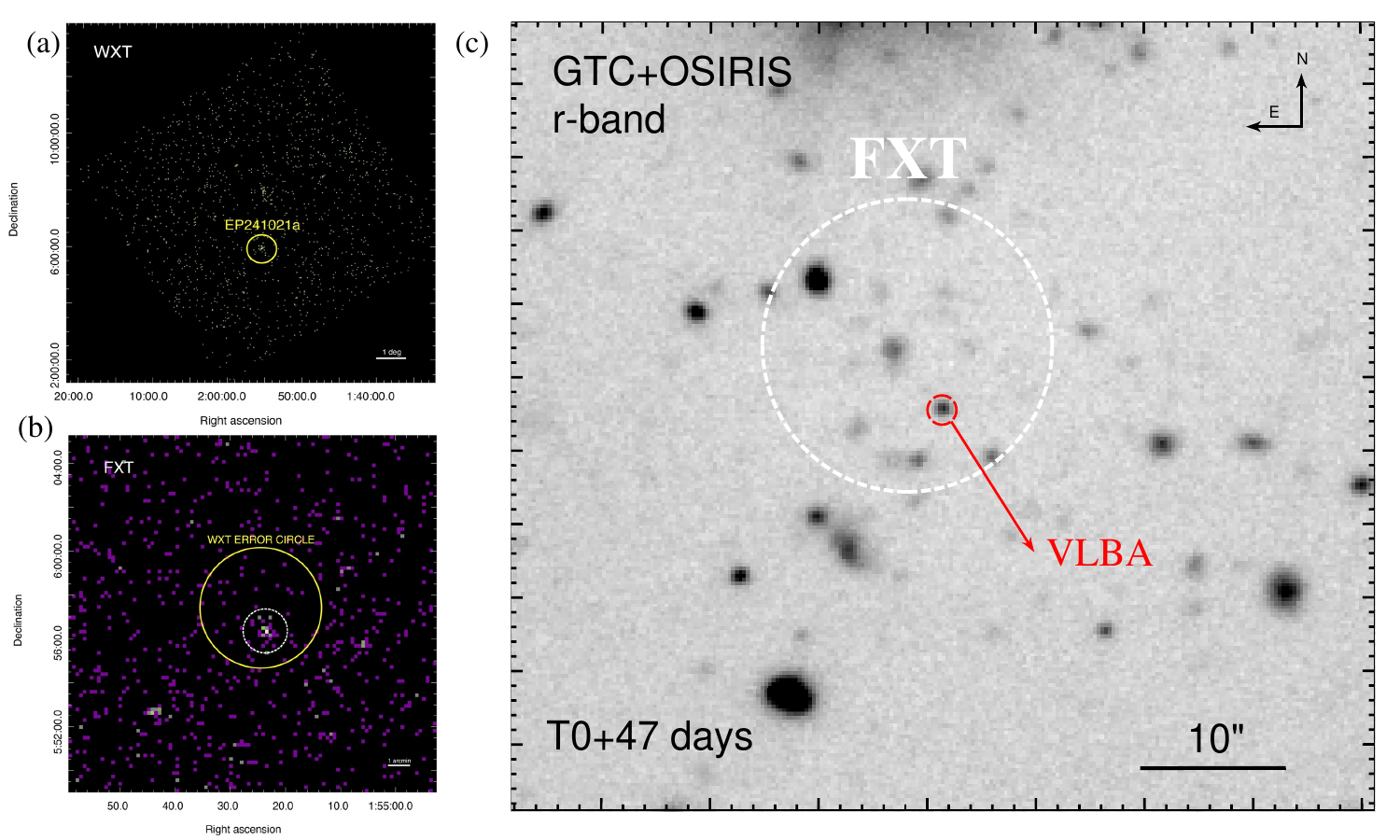}
\caption{
%Multiwavelength images and optical spectrum of \srcs. 
Multiwavelength images of \srcs. 
(a) The image of \src detected in one of the WXT CMOS detectors. 
(b) The image of \src taken by one of the follow-up FXT observations. 
The white circle represents the center of the X-ray source localized by FXT, 
with a radius of 1$^{\prime}$, while the yellow circle shows the localization 
of WXT with an error circle of 3$^{\prime}$. 
(c) The GTC $r$-band image observed on Dec. 7, 2024, in which 
the optical counterpart was found within the FXT error circle (radius $\sim 10^{\prime\prime}$, dashed white circle). 
The radio position obtained by VLBA is marked by the small red circle, 
which can localize the optical counterpart with a positional uncertainty of $\sim 1$ milliarcsecond. 
\label{fig:image}}
\end{figure*}

In its first year of scientific operations, EP/WXT has detected dozens of {EFXTs}  %more than 100 FEXTs 
in real time, most of which have received rapid follow-up observations at optical and radio bands. 
%to analyze %their multi-wavelength properties. 
%other wavelengths to  is capable of discovering
Analyzing the multiwavelength properties has revealed diversity in transient types and progenitor systems. 
%are diverse
%and likely represent a wide range of transient types and progenitor systems. 
%It is clear that at least some FEXT are linked to 
 A number of the {EFXTs} detected by EP have turned out to be GRBs, 
such as EP240219a \citep{Yin2024}, 
EP240315a \citep{Levan2024, Liu2025, Ricci2025}, and EP240801a \citep{Jiang2025}.
%there is a variety that has been linked to Galactic stellar flares %or X-ray binaries  
%\citep[e.g.,][]{M.J.Liu2025GCN}. 
%, there is a variety that has been linked to GRBs,  %shows clear GRB-like properties 
%Though not associated with GRBs, 
A few of the EP {EFXTs} have also been observed to show clear associations with Type Ic-BL supernovae,  
%which have no 
but without high-energy $\gamma$-ray counterparts,  
%but without associated $\gamma$-ray emission, 
including EP240414a \citep{Sun2024, vanDalen2025, Srivastav2025}, EP250108a \citep{Li2025, Srinivasaragavan2025, Eyles-Ferris2025, Rastinejad2025}, and EP250304a \citep{Izzo2025}, 
which could be explained by %a trapped or low-luminosity jet-driven explosion from a 
a jet-forming supernova trapped in a dense envelope. %inside a dense envelop. 
%The non-detections of $\gamma$-ray emission 
They have also been suggested {being linked to}
%resemble 
luminous fast blue optical transients (LFBOTs)
%due to the %optical bump 
%The luminous bump at 4 days has similar 
%according to the fast rise 
owing to the detection of a delayed optical bump with a fast rise time, although it is not as blue \citep{vanDalen2025}. 
%fast rise time in the optical rebrightening, though the colors are relatively red. 

\begin{figure*}[htbp!]
    \centering
    %	\epsscale{0.8}
    \includegraphics[width=0.48\textwidth]{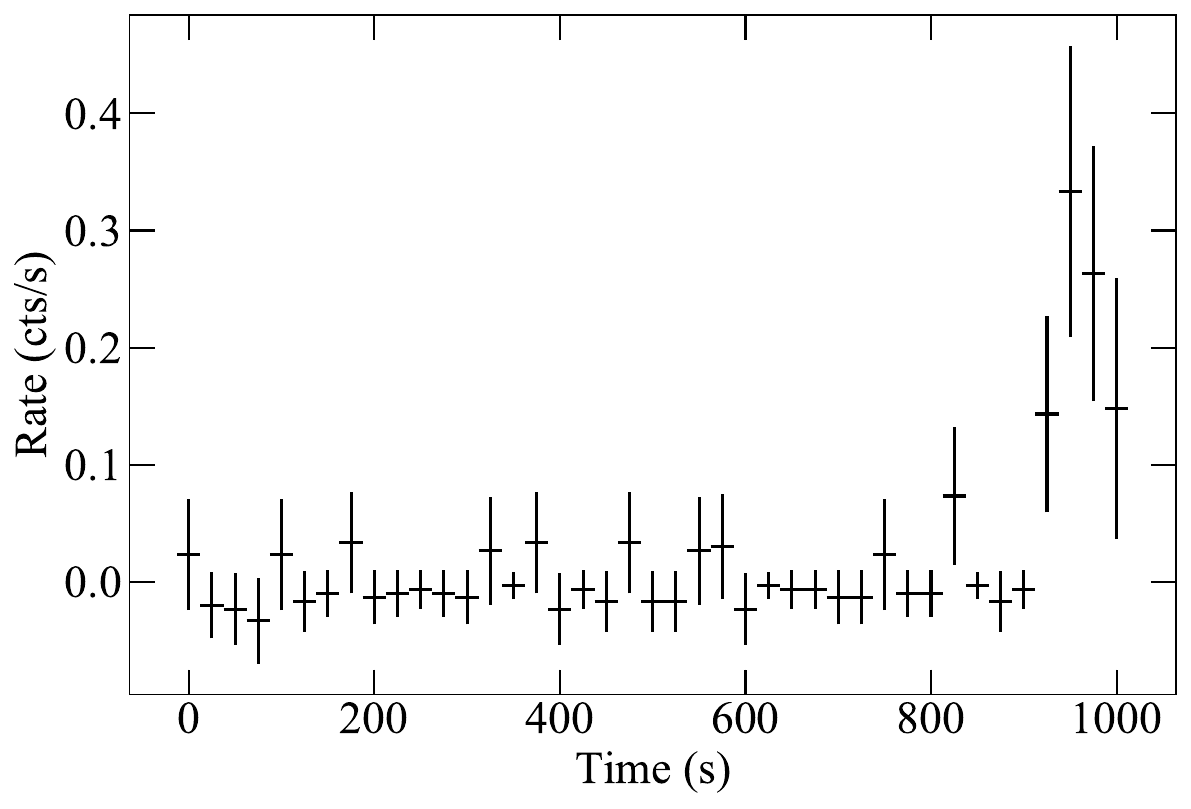}
    %\raisebox{0.1cm}
{\includegraphics[width=0.38\textwidth]{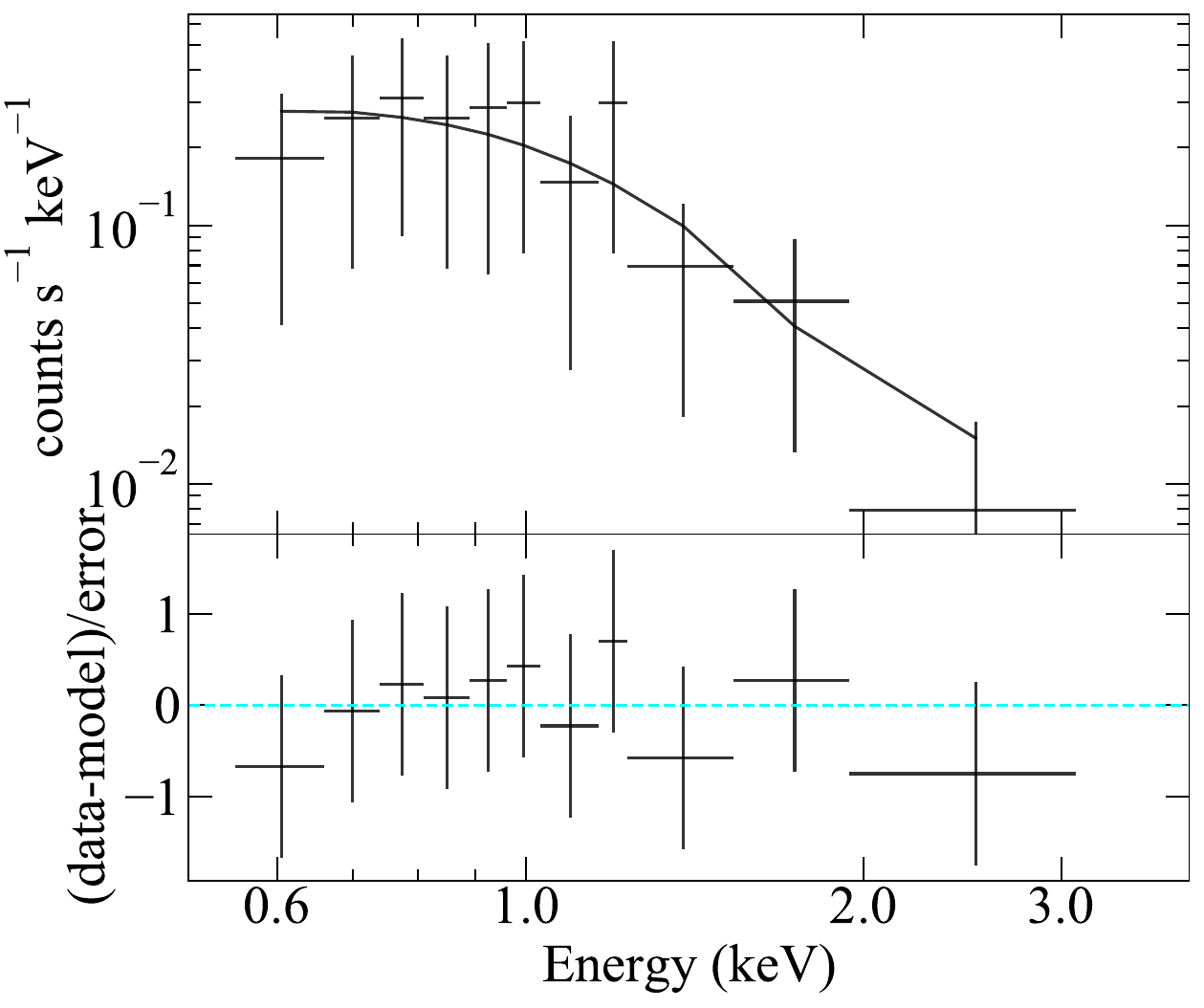}}
    \caption{
        \textbf{Left:} 
        The EP/WXT light curve of the count rate in the 0.5--4 keV band, in which an intense flare lasting $\sim 91.9$s was observed. 
        \textbf{Right:}  The WXT X-ray spectrum during the flare period, 
        %and EP/FXT spectrum by stacking the first six observations in the plateau phase (total exposure $\sim$35 ks), both of 
        which can be described by an absorbed single power-law model with photon index $\Gamma=1.8^{+0.57}_{-0.54}$. 
        %The photon index for the stacked EP/FXT spectrum is of $1.8\pm0.1$. Note that the X-ray spectrum is relatively flat and consistent with jet-dominated emission. 
            }
    \label{fig:wxtlc}
\end{figure*}

%Among these fast X-ray transients, \citet{Zhang2024} report 
In addition, EP has discovered %More recently, \citet{Zhang2025} reported the discovery 
a new type of X-ray transient with an intermediate evolutionary timescale. 
{One of the most intriguing EFXTs} is EP240408a, which was 
characterized by a post-flare plateau lasting for $\sim$4 d in the X-ray light curve, 
followed by a steep decay to an undetectable level about 10 d %since the initial trigger. 
after the initial detection \citep{Zhang2025}.  
No optical, near-infrared, or radio counterparts were detected, making EP240408a's temporal properties
inconsistent with any of the transient types known so far. 
Unfortunately, spectroscopic observations are not sensitive enough to confirm its redshift, 
preventing further investigations of the nature of EP240408a, although an exotic GRB or an abnormal jetted tidal disruption event (TDE) was suggested \citep{OConnor2025}. 
%. 
%but others remain unidentified due to the lack of sensitive spectroscopic observations to secure the redshift (Zhang et al. 2024; O'Connor et al. 2025).  
%since they have typical magnitudes of 20-21.5 in g band at peaks. 

\begin{table*}
\centering
\caption{Summary of X-ray observations obtained from EP and XMM-Newton}\label{tab:obs_xray}
\begin{tabular}{lclcccc}
\hline
{Obs ID} & 
{Exp. Time (s)} & 
{Obs. Start Time (UTC)} & 
{Phase (days)} & 
{$N_{\rm H}$} & 
{Flux (0.3--10\,keV)} & 
{Photon Index} \\
& & & &
($10^{20}{\mathrm{cm}^{-2}}$) &
($\mathrm{erg}\,\mathrm{s}^{-1}\,\mathrm{cm}^{-2}$) & \\
\hline

\multicolumn{6}{l}{WXT \footnote{The spectrum during the 91.9\,s flare is extracted and fitted at 0.5--4\,keV, while the flux at 0.3--10\,keV is reported based on the best-fit spectral model.}} \\
% 11908749577 & 2769 & 2024-10-21 00:07:20 & -- & $< 8.5 \times 10^{-12}$ & --  \\
% 11908750089 & 1308 & 2024-10-21 03:15:53 & -- & $< 3.0 \times 10^{-11}$ & --  \\
11900012322 & 1013 & 2024-10-21 04:51:25 & & 5 & $6.03_{-0.19}^{+0.42}\times 10^{-10}$ & $1.80_{-0.54}^{+0.57}$  \\
% 11900012325 & 843  & 2024-10-21 06:48:24 & -- & $< 7.3 \times 10^{-11}$ & --  \\
% 11900012326 & 1185 & 2024-10-21 08:05:41 & -- & $< 3.9 \times 10^{-11}$ & --  \\
\hline

\multicolumn{6}{l}{FXT} \\
06800000167 & 3024 & 2024-10-22 17:43:00 & 1.52 & 5 & $2.19_{-0.28}^{+0.32}\times 10^{-13}$ & $1.83\pm0.08$  \\
06800000168 & 6044 & 2024-10-23 17:46:00 & 2.52 & 5 & $2.00_{-0.18}^{+0.19}\times 10^{-13}$ & --  \\
06800000170 & 6037 & 2024-10-24 19:25:00 & 3.59 & 5 & $1.45_{-0.16}^{+0.18}\times 10^{-13}$ & --  \\
06800000173 & 8559 & 2024-10-25 17:51:00 & 4.52 & 5 & $1.41_{-0.15}^{+0.14}\times 10^{-13}$ & --  \\
06800000181 & 6122 & 2024-10-27 14:47:00 & 6.39 & 5 & $1.66_{-0.18}^{+0.20}\times 10^{-13}$ & --  \\
06800000186 & 5575 & 2024-10-29 08:28:00 & 8.13 & 5 & $2.09_{-0.18}^{+0.25}\times 10^{-13}$ & --  \\
06800000198 & 2093 & 2024-11-02 15:04:00 & 12.41 & 5 & $0.91_{-0.22}^{+0.26}\times 10^{-13}$ & --  \\
06800000202 & 6175 & 2024-11-04 07:08:00 & 14.08 & 5 & $0.79_{-0.13}^{+0.14}\times 10^{-13}$ & --  \\
06800000211 & 4791 & 2024-11-07 12:04:00 & 17.28 & 5 & $0.87_{-0.15}^{+0.18}\times 10^{-13}$ & --  \\
06800000250 & 5966 & 2024-11-20 15:41:00 & 30.43 & 5 & $0.47_{-0.11}^{+0.13}\times 10^{-13}$ & --  \\
06800000269\footnote{\src was only detected by FXT-B.} & 8943 & 2024-11-30 19:13:00 & 40.58 & 5 & $0.31_{-0.09}^{+0.10}\times 10^{-13}$ & --  \\
06800000356\footnote{\src was not detected, and the corresponding 3$\sigma$ upper limits on flux are given.} & 8706 & 2025-01-08 10:36:19 & 79.22 & -- & $< 1.82 \times 10^{-14}$ & -- \\ %$< 0.10 \times 10^{-13}$ & -- \\
\hline
\multicolumn{6}{l}{\textit{XMM-Newton}/MOS2} \\
% 0954190901$^{\rm c}$ & 39918 & 2025-01-18 13:03:43 & -- & $<2.83\times10^{-15}$ & --  \\
0954190901$^{\rm c}$ & 39918 & 2025-01-18 13:03:43 & 89.32 & -- & $<2.83\times10^{-15}$ & --  \\ %$<1.32\times10^{-15}$ & --  \\ 
\hline
\end{tabular}
\end{table*}

%In addition, the host galaxies of optically-selected TDEs are dominated by green valley galaxies, including post-starburst or E+A galaxies. 
%Such a preference is confirmed by the eROSITA sample of X-ray TDEs \citep{Sazonov2021}, 
%Such a preference can be attributed to a higher central stellar concentration following recent star-formation or galaxy mergers \citep{French+2020, Hammerstein2021}. 
%optical detections of dozens of TDEs.. 
%dwarf galaxies IMBHs are rare.. .
%While the optical-to-X-ray emission of TDEs  radio...delayed radio emission...
In this {\it Letter}, 
we report the discovery of a peculiar EP X-ray transient, \srcs, 
which exhibits long-lasting X-ray emission for more than one month, 
making it {possibly an EFXT} with the longest evolutionary timescale ever observed. 
%\src is one of few FEXTs that have a bright optical and radio counterpart and the results from X-ray and radio follow-up observations. 
%of a peculiar X-ray transient, \srcs, recently discovered by EP \citep{Hu2024}. 
Shortly after the trigger, it was detected by follow-up observations at
optical and radio bands.
Optical spectroscopy reveals that the redshift of \src is $z = 0.748 $,    %obtained by ESO’s Very Large Telescope
based on the detection of narrow emission and absorption lines from the faint host galaxy. % at the redshift of $z = 0.748 $.  %\citep{Pugliese2024}. 
%as well as the results from X-ray and radio follow-up observations. 
%we report the detection of delayed and rapidly rising radio emission from \srcs,  
%about 1105 days after the optical alert. 
%possibly originating from optically-thin jet ejections. 
%providing new insights into the jet production in accreting SMBHs. 
%challenge canonical theories of jet formation well-established
%in XRBs,
The multiwavelength observations and data reduction are described in Section \ref{sec:observation}. 
In Section \ref{sec:analysis}, we present the detailed analysis of X-ray spectral and variability properties, 
and the modeling of the radio flux and the evolution of the spectral energy distribution (SED). %, and the modeling of the multi-band light curves. 
A discussion of possible interpretations for \src is given in Section \ref{sec:discussion}. 
We summarize the results in Section \ref{sec:conclusion}.
%In Section 4, we summarize our results and findings. 
We adopt a cosmology of $\Omega_M = 0.3$, $\Omega_{\Lambda} = 0.7$, and H$_0$ = 70\,km\,s$^{-1}$\,Mpc$^{-1}$ when computing luminosity distances.

\section{OBSERVATIONS} \label{sec:observation}

\subsection{X-ray trigger and observations} 
\subsubsection{EP/WXT}

The source \src was initially detected by CMOS detector 10 of the EP/WXT during the observation conducted on October 21, 2024, between 04:51:21 and 05:10:39 UTC \citep{Hu2024}. Data reduction was performed using the \texttt{wxtpipeline} software tool (version 0.1.0). The WXT image in which \src is detected ($\alpha = 01^{\rm hr}55^{\rm m}24.48^{\rm s}$, $\delta = +05^\circ 57' 25\farcs20$, J2000, with an uncertainty of 2\farcs4) is shown in Figure \ref{fig:image}(a). The light curve exhibits a significant flux increase from \src that started at 05:07:56 UTC and lasted $\sim 100$\,s (Figure \ref{fig:wxtlc}), with a time-averaged X-ray flux of $3.31^{+1.26}_{-0.86}\times 10^{-10}$ \ergs in the 0.5--4.0 keV band, % to avoid confusion, the EP / WXT flux in Table~\ref{tab:obs_xray} is estimated in the 0.3--10 keV), 
corresponding to an isotropic equivalent luminosity as high as $8.5\times10^{47}$ \erg at $z=0.748$. 
The transient has a relatively hard X-ray spectrum which can be fitted by a power law with photon index $1.8^{+0.57}_{-0.54}$.  
We will refer to the time of the EP/WXT trigger as $T_{\rm 0}$ { (MJD = 60604.214) for the phases (in days)} of multiwavelength follow-up observations.

\subsubsection{EP/FXT}

Following the initial detection of \src by WXT, an observation was performed with the FXT onboard EP about 36.96\,hr later. The exposure time was 3024\,s, during which a previously uncataloged source was detected within the WXT's error circle, at J2000 coordinates $\alpha = 01^{\rm hr}55^{\rm m}23.59^{\rm s}$, $\delta = +05^\circ 56' 22\farcs20$ with an uncertainty of 10$^{\prime\prime}$ \citep{Wang2024}. Subsequent monitoring of \src continued until January 8, 2025, consisting of a total of 12 observations (Table 1). The FXT was configured in Full Frame mode in all the observations. The data were reducted using the \texttt{fxtchain} tool in the FXT Data Analysis Software (\texttt{FXTDAS}).
Detailed analysis of the X-ray spectra and light curve will be presented in Section \ref{subsec:xray spectra} and \ref{subsec:xray lc}, respectively.

\subsubsection{XMM-Newton} 

To fully constrain the late-time decay of \src, we also obtained a target-of-opportunity (ToO) observation with \textit{XMM-Newton} (Obsid 0954190901; PI Eyles-Ferris). This was performed on January 18, 2025, $\sim$90 days after the first detection of \src. We obtained the resulting Pipeline Processing System (PPS) files from the XMM-Newton Science Archive\footnote{\url{https://www.cosmos.esa.int/web/xmm-newton/xsa}} (XSA). After filtering for background flares, exposure times were 38.2, 39.9, and 30.7\,ks with the EPIC MOS1, MOS2, and pn detectors respectively.

No source was detected at the position of \src in any of the EPIC instruments. To derive the flux upper limits, %90\% upper limits, 
we employed the method of \citet{Kraft1991} using a circular aperture of radius 10\arcsec~centered on the source position. We estimated the background using a 60\arcsec~radius aperture placed on the same chip as the source position and applied the PPS exposure maps to correct for vignetting. This was repeated for all three detectors, finding that MOS2 gives the deepest limit. 
We then converted the 0.2--12\,keV count rate to 0.3--10\,keV flux using \texttt{Pimms v4.15} and a photon index of 1.83. The resulting unabsorbed flux upper limit is 
%$<1.32\times10^{-15}$ $\mathrm{erg}\ \mathrm{s}^{-1} \mathrm{cm}^{-2}$ (at 90\% confidence level).
$<2.83\times10^{-15}\,\mathrm{erg}\,\mathrm{s}^{-1}\,\mathrm{cm}^{-2}$ (at 3$\sigma$ confidence level).

\subsection{Nondetections from Gamma-Ray Observations}

%\src was not detected by any high-energy gamma-ray observations. 
There was no Fermi-GBM onboard trigger of \srcs. An automated, blind search for short GRBs below the onboard triggering threshold of Fermi-GBM also identified no counterpart candidates. 
The GBM targeted search for GRB-like signals was run from 30\,s before the EP trigger time until 300\,s after, 
but again
%seeking signals between 64 ms and 32.768 s in duration . 
no signal was identified \citep[GCN 37855]{Burns..2024}.  %., as confirmed by visual inspection of the data. 
Konus-Wind (KW) was observing the position of EP241021a (Svinkin et al., GCN 38034) 
for 1000\,s before and after the trigger time and did not detect 
the source, with 
%Using waiting-mode data within the interval T0 +/- 1000 s,
an upper limit (90\% confidence level) for the 20--1500\,keV peak flux
of $2.5\times10^{-7}$\,\ergs for a typical long GRB spectrum. 
%(the Band function with $alpha=-1, beta=-2.5, and Ep=300 keV) 
%and 2.944 s timescale.
%GECam-B?? 
{This is an order of magnitude lower than the peak flux of 
long GRBs observed by Konus-Wind \citep{Tsvetkova2017}.} 
%Finally, 
Therefore, \src was not detected by any high-energy gamma-ray observations, 
{ and appears to fall below the lower end of the gamma-ray flux distribution of cosmological GRBs \citep{Yadav2025}. 
}

\begin{figure}[!t] 
\epsscale{1.1}
\plotone{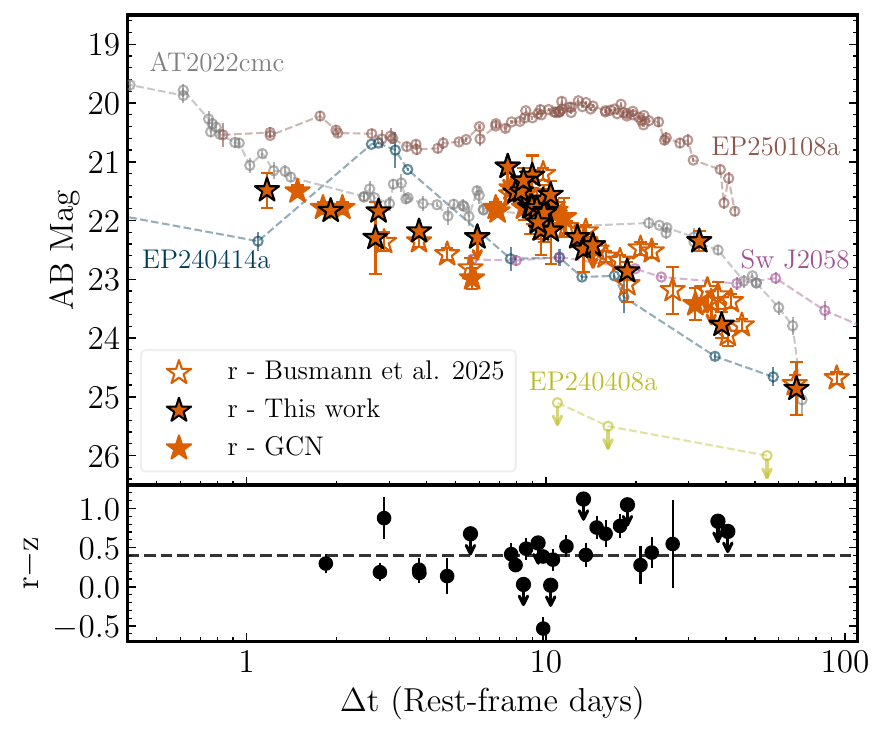}
\caption{
{ Optical ($r$-band) light curve of \srcs. 
The photometric data are taken from our own observations (Section 2.3), and those reported in GCN Circulars and \citet{Busmann..2025}. 
Note that the flux excess observed by SVOM/VT (at $\Delta t\approx$18.6 days) is due to the use of a broader $r$-band filter, and hence should be treated with caution.  
For comparison, we present the $r$ light curves of EP240414a \citep{vanDalen2025} and EP250108a \citep{Li2025} --- the two EFXTs that are associated 
with Type Ic-BL supernovae --- 
as well as jetted TDEs Sw J2058 \citep{Cenko2012} 
and AT2022cmc \citep{Andreoni2022}. 
The peculiar EFXT EP240408a with an intermediate timescale but without optical counterpart (down to 26 mag) is also plotted for comparison, assuming it at $z=0.5$ \citep{Zhang2025, OConnor2025}. 
%also show optical brightenings. 
The lower panel shows the $r-z$ color evolution of \src as a function of time. 
It is clear that the color is persistently red up to at least $\sim 30$ rest-frame days since discovery. 
}
%Keck spectrum of EP241021a taken on October 30, 2024 and the power-law model that can describe the continuum likely dominated by the transient. The wavelength was converted into rest-frame according to emission line redshift of $z$ = 0.7478, which was obtained from narrow emission lines we labeled. We show in the insert panel the single Gaussian models that fit the H$\beta$ and [\ion{O}{3}] emission lines.
\label{fig:optphot}}
\end{figure}

\subsection{{ Optical Photometry}}

The optical counterpart of EP241021a was first reported by the Nordic Optical Telescope (NOT) equipped with the Alhambra Faint Object Spectrograph and Camera (ALFOSC) at coordinates
$\alpha = 01^{\rm hr}55^{\rm m}23.41^{\rm s}$, $\delta = +05^\circ 56' 18.01''$ with an uncertainty of 0.5$^{\prime\prime}$ \citep{2024GCN.37840....1F}. This result was confirmed by earlier observations of the Thai Robotic Telescope network located at Fresno, California, USA (TRT-SRO) \citep{2024GCN.37842....1F}, the 0.76\,m Katzman Automatic Imaging Telescope (KAIT; \citealt{Filippenko2001}) located at Lick Observatory \citep{2024GCN.37849....1Z}, and the Liverpool Telescope (LT) located at the Observatorio del Roque de los Muchachos, La Palma, Spain \citep{Li2024GCN, Kumar2024GCN}. After the first detection of the optical counterpart, several optical photometry follow-up campaigns were carried out, such as NOT, Thai Robotic Telescope network located at New South Wales, Australia (TRT-SBO), 50\,cm-A and 100\,cm-C telescopes of the JinShan project, located at Altay, Xinjiang, China (ALT-50A\&-100C), HMT-0.5\,m telescope located at Nanshan, Xinjiang, China (HMT), 1.5\,m telescope at Observatorio de Sierra Nevada, Spain (OSN), LT, Gran Telescopio Canarias (GTC) located at Roque de los Muchachos Observatory, La Palma, Spain, {LCO}, {and the 1.6m Multi-channel Photometric Survey Telescope (Mephisto) located at Lijiang Observatory of Yunnan Astronomical Observatories, Chinese Academy of Sciences, and operated by the South-Western Institute for Astronomy Research, Yunnan University \citep{Pan2024GCN37968}}. %and the Visible Telescope of Space Variable Objects Monitor (SVOM/VT).

{The 1.6m Mephisto observed EP241021a on 2024 October 29 in the u$_m$, v$_m$, g$_m$, r$_m$, i$_m$, and z$_m$ bands, with two frames of 300-second exposures in each band. After stacking, the source was detected in the v$_m$ and r$_m$ bands. Follow-up observations in the v$_m$, r$_m$, and z$_m$ bands were conducted on October 30, October 31, November 3, and November 4. Except for November 4 (one exposure per band), three exposures were obtained each night. Stacked images from November 3 show marginal detections in all three bands, while clear detections on the other nights were limited to the v$_m$ and r$_m$ bands. There was no detection on November 4. The Mephisto data reduction and calibration were performed following the method discussed in \citet{Chen2024ApJ}.}

{For the images of NOT, Nanshan-HMT, and ALT-50A/100C, the data were processed using standard procedures with the Image Reduction and Analysis Facility (IRAF) v2.16 \citep{1986SPIE..627..733T}, including bias subtraction, flat fielding, and image combination. Aperture photometry was conducted on the stacked frames, and the resulting flux was calibrated by nearby Pan-STARRS1 field stars \citep{2016arXiv161205560C}.} {For the LT images and for the LCO images obtained in the program IAC2024B-004, we performed point-spread-function (PSF) photometry with the \texttt{autophot} package \citep{autophot.ref}, using stars in the SDSS catalog for photometric calibration.}
The log of photometric observations and results is listed in Table \ref{tab:obs_optical} of  Appendix A.
{Figure~\ref{fig:optphot} displays the $r$-band light curve and $r-z$ color evolution of \srcs. 
%, and the comparison with other EFXTs 
}

\subsection{{Optical Spectroscopy}}

%%%Weikang added Keck spec figure
\begin{figure*}[!t] 
\epsscale{1.1}
\plotone{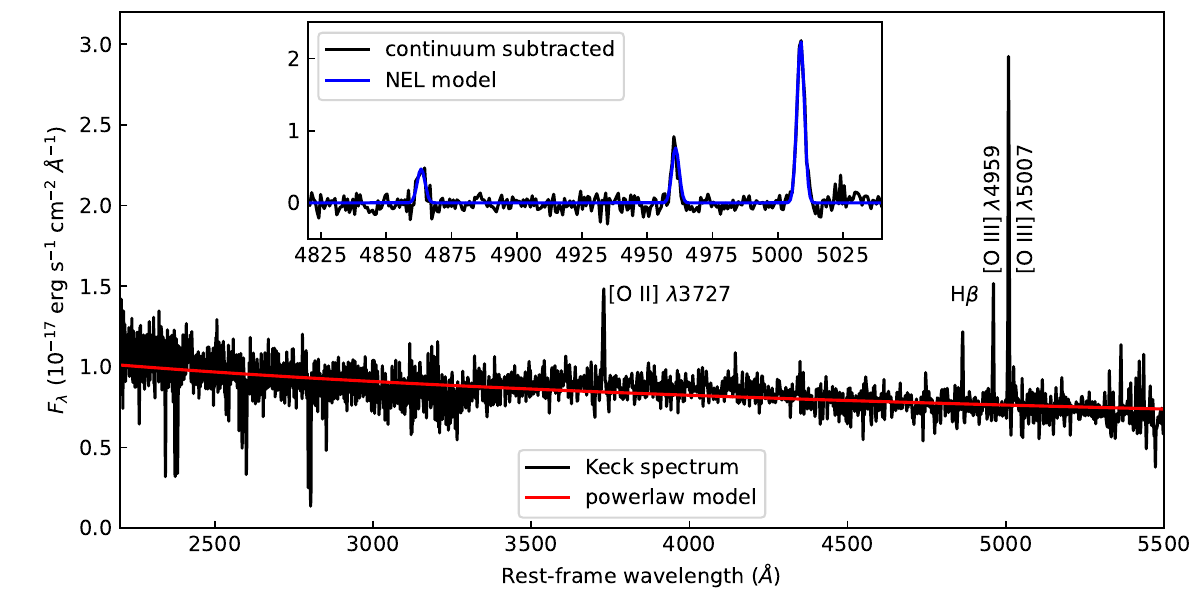}
\caption{Keck spectrum of EP241021a taken on October 30, 2024, and the power-law model that can describe the continuum likely dominated by the transient. Wavelengths were converted into the rest frame according to the redshift $z = 0.7478$, which was obtained from the labeled narrow emission lines. 
%{\bf The spectrum was binned by 3 pixels for clarity}. 
We show in the inset the single Gaussian models that fit the H$\beta$ and [\ion{O}{3}] emission lines.
\label{fig:keck_spec}}
\end{figure*}

{We acquired long-slit spectroscopy of EP241021a with OSIRIS+ on the GTC on Oct. 23, 2024 and Oct. 29, 2024 \citep{PF2024GCN}, each with an exposure time of 1800\,s. The R500R grism was used, corresponding to a wavelength coverage of 4800--10000\,\AA\ and a spectral resolution of $R \approx 500$. The data were reduced with the \texttt{pypeit} package \citep{pypeit.ref} and flux calibrated with standard stars observed on the same night.}

EP241021a was observed with the Low Resolution Imaging Spectrometer (LRIS; \citealp{oke1995}) on the Keck~I 10\,m telescope at the W. M. Keck Observatory, on October 30, 2024 \citep{Zheng24}. 
{The exposure time was $2 \times 1200$\,s with the 600/4000 grism on the blue side ($R \approx 1100$; $\sim 3000$--5600\AA) and (simultaneously) the 400/8500 grating on the red side ($R \approx 1100$; $\sim 5400$--9500\AA). }The spectrum was acquired with the $1''$-wide slit oriented at or near the parallactic angle to minimize slit losses caused by atmospheric dispersion \citep{filippenko1982}.
Data were processed using the \texttt{LPipe} \citep{Perley2019} data-reduction pipeline.
The spectrum was flux calibrated using observations of appropriate spectrophotometric standard stars observed on the same night, at similar airmasses, and with an identical instrument configuration. 

As shown in Figure~\ref{fig:keck_spec}, the continuum of the Keck spectrum can be well fitted with a power law, $f_\lambda \propto \lambda^{-0.34}$.
We clearly detect narrow emission lines of [\ion{O}{2}] $\lambda$3727, H$\beta$, and [\ion{O}{3}] $\lambda\lambda 4959$, 5007 at a common redshift of $z = 0.7478$, which likely originate from the host galaxy.
We also detect an absorption-line system containing the \ion{Mg}{2} $\lambda\lambda 2796$, 2803 doublet and \ion{Fe}{2} lines at a similar redshift, making the redshift more reliable.
The [\ion{O}{3}] $\lambda$5007 and [\ion{O}{2}] $\lambda$3727 lines have luminosities of $3.6\pm0.1\times10^{41}$ and $1.7\pm0.2\times10^{41}$\,erg\,s$^{-1}$, respectively.
The measured full width at half-maximum intensity (FWHM) of [\ion{O}{3}] is $200\pm6$\,km\,s$^{-1}$, consistent with the line-spread function of $\sim210$ km s$^{-1}$ (measured using nigh-sky emission lines), indicating that the emission lines are narrow and unresolved.
The line intensity ratio [\ion{O}{3}]/H$\beta = 4.4\pm0.7$ indicates that the emission lines could be produced either by star formation or an active galactic nucleus \citep[AGNs;][]{Kewley2006}.
{ No [Ne~V] $\lambda$3426 or [Ne~III] $\lambda$3869 are detected, and the intensity ratios  [Ne~III]/[\ion{O}{2}] $< 0.17$ and [\ion{O}{3}]/[\ion{O}{2}] $= 2.15\pm0.19$ are consistent with star-forming galaxies in the diagnostic diagram of \cite{Feuillet2024}.
Thus, the host is more likely to be a star-forming galaxy than an AGN.}
If so, the [\ion{O}{2}] luminosity corresponds to a star-formation rate of 2.4\,M$_\odot$\,yr$^{-1}$ according to the relation of \citet{Kennicutt1998}. 
{ While the narrow emission lines such as [\ion{O}{3}], H$\beta$, and [\ion{O}{2}] 
were detected in the GTC spectrum as well, confirming that \src is at $z = 0.7478$, 
the signal-to-noise ratios (S/N) are too low for meaningful spectral fittings (Figure A1 in  Appendix A). 
Therefore, we report only the spectral fitting results from the Keck data. 
} 

\begin{deluxetable*}{cccccc}
  \centering
\tablewidth{0pt}
\tablehead{
\colhead{Observatory} & 
\colhead{Project} & 
\colhead{$\nu$} & 
\colhead{UTC Date} & 
\colhead{Phase} & 
\colhead{$F_\nu$}\\
\colhead{} & \colhead{} & \colhead{(GHz)} & \colhead{} & \colhead{(days)} & \colhead{(mJy)}}
\tablecaption{Summary of the radio observations of \srcs \label{tab:obs_radio}}
\setlength{\tabcolsep}{3mm}{
\startdata
ATCA & CX585 & 5.0 & 2024 Oct 29 & 8 & 0.350 $\pm$ 0.029\\
     &       & 5.5 & 2024 Oct 29 & 8 & 0.382 $\pm$ 0.024\\
     &       & 6.0 & 2024 Oct 29 & 8 & 0.407 $\pm$ 0.029\\
     &       & 8.5 & 2024 Oct 29 & 8 & 0.434 $\pm$ 0.028\\
     &       & 9.0 & 2024 Oct 29 & 8 & 0.453 $\pm$ 0.026\\
     &       & 9.5 & 2024 Oct 29 & 8 & 0.467 $\pm$ 0.028\\
\hline
VLBA & BS346 & 6.2 & 2024 Nov 28 & 37 & 0.774 $\pm$ 0.072\\
     &       & 8.4 & 2024 Nov 30 & 39 & 0.762 $\pm$ 0.081\\
\hline
VLA & 24B-487 & 5.0  & 2025 Jan 28 & 98 & 0.544 $\pm$ 0.038\\
    &         & 7.0 & 2025 Jan 28 & 98 & 0.374 $\pm$ 0.038\\
    &         & 15.0 & 2025 Jan 28 & 98 & 0.110 $\pm$ 0.020\\
    &         & 10.0 & 2025 Feb 14 & 115 & 0.225 $\pm$ 0.074\\
    &         & 6.0  & 2025 Feb 21 & 122 & 0.319 $\pm$ 0.025\\
\hline
MeerKAT & DDT-20250123-XS-01 & 2.2 & 2025 Jan 29 & 99 & 0.522 $\pm$ 0.027\\
% & & VAST & 804 & 0.89 & $\dots$ \\
        &                    & 3.1 & 2025 Jan 29 & 99 & 0.478 $\pm$ 0.025\\
        &                    & 1.3 & 2025 Feb 15 & 116 & 0.375 $\pm$ 0.020\\
        &                    & 2.2 & 2025 Feb 16 & 117 & 0.367 $\pm$ 0.020\\
        &                    & 3.1 & 2025 Feb 16 & 117 & 0.413 $\pm$ 0.022\\
 % & & 0.89 & 2023 Jul 06 & 1846 & 3.744 $\pm$ 0.076 \\
 % & & 0.86 & 2023 Aug 11 & 1882 & 7.132 $\pm$ 0.050 \\
 % & & 0.89 & 2023 Aug 30 & 1901 & 8.110 $\pm$ 0.180 \\
 % & & 0.89 & 2023 Oct 29 & 1961 & 9.010 $\pm$ 0.170 \\
\enddata}
%\tablenotetext{beam size of VLA are from the cleaned image}
%\tablecomments{
%\tablenotetext{
%$^{\dag}$For VLASS non-detections, the corresponding 3$\sigma$ upper limits on peak flux density are given. \\
%$^{\ddag}$The upper limit was measured by stacking the VAST images observed between 2019 Aug 27 and 2020 Aug 28.  
%Due to the poor imaging quality, the beam size cannot be measured using CASA. 
%$^{\ddag}$VLASS consists of three-epoch observations, each separated by approximately a period of 32 months \citep{Lacy2020}. 
%The last three rows represent the archival VLA and ASKAP data. 
%}
\end{deluxetable*}

\subsection{Radio observations} 

\subsubsection{VLA}

While \src was not detected by e-MERLIN on 2024 Oct. 24 \citep{Gianfagna..2024}, three days after discovery, 
 a radio brightening was identified in both ATCA and AMI-LA observations on 2024 Oct. 29--30 \citep{Ricci..2024, Carotenuto..2024}, with a radio luminosity at 5\,GHz of $\simgt 5\times10^{40}$\,\erg. 
 %comparable to the jetted TDEs at similar evolution epochs (Fig. 1b). 
To further study the radio properties, we triggered two-epoch VLA observations through a DDT program (project code VLA/24B-487) at central frequencies of 6\,GHz (C vand), 10\,GHz (X band), and 15\,GHz (Ku band).
%covering a frequency range 6--15 GHz. 
The VLA observations were carried out 98 and 115 days after discovery. 

Data were reduced using the Common Astronomy Software Applications \citep[CASA;][version 6.6.1]{McMullin2007} and the standard VLA calibration pipeline (version 2024.1.1). Where the S/N allowed, we enhanced frequency resolution by splitting the Measurement Set (MS) from groups of the spectral windows. We reran the pipeline after manually inspecting and additional flagging.
%By examining the pipeline product, 
%For the reduced data product, we inspected each spectral window and 
%manually flagged channels affected by radio frequency interference %(RFI). 
%flagged the abnormal data for frequencies due to RFI issue. 
The calibrated data were imaged using the {\tt CLEAN} algorithm with 
Briggs weighting and ROBUST parameter of 0, which helps to reduce 
side lobes and achieve good sensitivity. 
\src was clearly detected in all observations. 
We used the {\tt IMFIT} task in CASA to fit the radio emission with a two-dimensional 
elliptical Gaussian model to determine the position, integrated flux density, and peak flux density. 
%There is no extended emission 
The radio emission at all bands is unresolved and no extended emission is detected. 
%The compactness of the radio emission is confirmed by the ratios of integrated and peak flux density, 
%which are in the range 0.92--1.35, with a median of 1.09. 
For consistency, only peak flux densities are used in our following analysis.  
The VLA observation log and flux-density measurements are presented in Table \ref{tab:obs_radio}.

We also searched for radio emission at 3\,GHz using
the archival data from the Very Large Array Sky Survey \citep[][]{Lacy2020}, 
but found that \src remains undetected over its three-epoch observations between 2017 Oct. and 2023 Mar., 
with a 5$\sigma$ upper limit in the peak flux of $\sim 0.8$\,mJy\,beam$^{-1}$.

\subsubsection{MeerKAT}

We conducted multifrequency radio observations with the MeerKAT telescope over two epochs (project code DDT-20250123-XS-01). 
The first-epoch observations were performed in the S0 and S4 bands (centered at 2.2\,GHz and 3.1\,GHz, respectively), 
while the L-band observations (with central frequency 1.3\,GHz) were included in the second epoch. 
The MeerKAT observations were observed quasisimultaneously as VLA, to measure the broad-band SED 
and its evolution. 
We used the ``4K'' wideband coarse mode to ensure high sensitivity.   
For the first epoch, the total time was $\sim 1.9$\,hr in each band, of which 1.5\,hr were spent on source and 0.4\,hr on the flux and phase calibrators (J0408-6545 and J0149+0555), 
resulting in a root-mean-square (RMS) of $\sim 10$\,$\mu$Jy. 
A shorter on-source exposure (0.7\,hr) in each band was used for the second-epoch observations, 
yielding a larger RMS of $\sim 20$\,$\mu$Jy. 
%For
%these observations % (performed between 2023 Oct 20 and Nov 7) 
%we used J0408-6545 as the flux calibrator
%and J0906-6829 as the phase calibrator. 
%The MeerKAT data were reduced using the OxKAT software \citep{Heywood2020}, 
%and the final images were cleaned with the the WSClean algorithm \citep{Offringa2017}. 

We reduced the MeerKAT data using the OxKAT software \citep{Heywood2020}, 
and the final images were cleaned with the WSClean algorithm \citep{Offringa2017}. 
We then measured the integrated and peak flux in CASA, following the same procedures
described above. 
%Thanks to the high sensitivity of the MeerKAT observations, 
The source is clearly detected in all three bands. %, even if the radio SED has evolved to peak at lower flux and frequencies. 
The ratio of the integrated flux to the peak flux is in the range of 0.92 to 1.06, with a median value of 0.99,
suggesting that most, if not all, of the radio emission is
unresolved, consistent with observations from other telescopes, although the spatial resolution
is very different. 
%Note that we 
%To check the data quality of the pipeline products, 
%We retrieved the primary beam corrected continuum images from the SARAO Science Data Processor Pipeline products. %, and the source was clearly detected in all bands. W}
% then used OxKAT software \citep{Heywood2020} to check the quality of the pipeline products above.) 
%We find the source's flux densities obtained with OxKAT and pipeline products 
%are consistent with each other (the difference is $\sim$10\%). %3\%--15\%). 
%Hence, we report mainly t
The flux densities obtained from MeerKAT observations are %measured from the MeerKAT pipeline products, which are 
listed in Table \ref{tab:obs_radio}.

\subsubsection{VLBA}

We carried out Very Long Baseline Array (VLBA) observations at the location of \src 
on 2024 Nov. 28 at 6.186\,GHz and on 2024 Nov. 30 at 8.368\,GHz with its 10 antennas (project code BS346). 
The observing frequency was centered at 6.186\,GHz in the C band and 8.368\,GHz in the X band. The observation was 
performed in the phase-referencing mode to the nearby strong compact radio source J0149+0555. 
%Total time assigned for this project was 10 hours. 
Phase-reference cycle times were 4.5\,min, with 3.5\,min on-target and 1.0\,min for the phase calibrator. 
We also inserted several scans of the bright radio source 3C~84 for fringe and bandpass calibration with an integration time of 3.0\,min 
for each scan. 
The resulting total on-source time on both bands is 5\,hr. 
%The data were recored at 2 Gbps, and 
To achieve sufficiently high imaging sensitivity, we adopted the observational mode RDBE/DDC to 
use the largest recording rate of 2\,Gbps, corresponding to a recording bandwidth of 256\,MHz in each of the dual circular polarizations. 
%The data from the VLBA experiment were correlated with the DiFX software correlator \citep{Deller2011}. 
%each cycle is three minutes for the target J1302 and one and a half minutes for the phase calibrator J1300+28 
%(RA:13$^h$00$^m$28$^s$.5300, Dec:+28${\degr}$30$\arcmin$10$\arcsec$.189). 
%The BS255 VLBA experiment data were correlated at the DiFX correlator in Socorro, USA \citep{Deller2011}. 
We used the NRAO AIPS software to calibrate the amplitudes and phases of the visibility data, following the standard procedure 
from the AIPS Cookbook\footnote{\url{http://www.aips.nrao.edu/cook.html}}. 
The calibrated data were imported into the Caltech DIFMAP package \citep{Shepherd1997} for imaging and model fitting. 
During imaging, we noted that the gain solutions were bad for the MK, SC, and HN antennas at 6.2\,GHz, and for the HN antenna at 8.4\,GHz. Therefore, we flagged data from these antennas.

VLBA detects a compact source in the final cleaned image, which has a deconvolved size of 3.77\,mas $\times$ 1.12\,mas. 
% The integrated and peak flux density for the source is 695 $\pm$ 59 $\mu$Jy and 406 $\pm$ 23 $\mu$Jy/beam, respectively. 
%The source appears to be marginally resolved according to the ratio of integrated and peak flux density which is 1.7. However, by using 
To further investigate whether the source is resolved or not, we used the task {\tt Modelfit} in DIFMAP to fit the radio emission, but found no additional emission components in the residual map. 
Therefore, \src remains compact and unresolved at the resolution of VLBA observations, with an upper 
limit on its size of $< 8.22$\,pc. 
{ The position of the radio source observed by VLBA is (J2000)
$\alpha = 01^{\rm hr}55^{\rm m}23.4323^{\rm s}$, $\delta = +05^\circ 56' 17\farcs7978$.
The VLBA observation log and flux density measurements are presented in Table \ref{tab:obs_radio}.}

\subsubsection{ATCA}
\src was observed with the Australia Telescope Compact Array (ATCA) in the 16\,cm and 4\,cm bands. The observations were made in the 6A configuration on 2024 Oct. 29 \citep[Program ID CX585, GCN 37949]{Ricci..2024} and were carried out with two 2\,GHz-wide intermediate frequencies (IFs) of 4.5--6.5\,GHz (centered at 5.5\,GHz) and 8--10\,GHz (centered at 9\,GHz). PKS B1934-638 was used as a bandpass and flux-density calibrator and 0146+056 was used as a complex-gain calibrator. Data reduction was carried out with the software \textit{Miriad} \citep{RJSault..1995}, following standard procedures.

First, we performed the automatic radio frequency interference (RFI) flagging by using the task \textit{pgflag} before calibration to reduce the influence of RFI. Second, standard calibration involved bandpass and flux-density calibration on PKS B1934-638 using the \textit{Miriad} tasks \textit{mfcal} and \textit{gpcal}, and this was applied to the gain calibrator 0146+056. Time-varying gains and polarization leakage calibration were done using the task \textit{gpcal} on 0146+056, and these gains were transferred and applied to the target before imaging.

Imaging was performed using the tasks \textit{invert}, \textit{cgdisp}, \textit{mfclean}, and \textit{restor} to create six continuum images (three for each of the IFs). Images were made in Stokes \textit{I}, with a Briggs visibility weighting robustness parameter of 0.5. Images were restored with a Gaussian synthesized beam of $27.21'' \times  1.36''$ with a position angle of $-1.5^circ$  for the 5.5\,GHz image and $16.78'' \times 0.83''$ with a position angle of $-1.5^circ$  for the 9\,GHz image. \src was clearly detected in all bands; the radio emission is unresolved and no extended emission is detected. The central region of the 5.5\,GHz image has an RMS noise level of $15.1\,\mu$Jy\,beam$^{-1}$, while the 9\,GHz image has an RMS noise level of $11.8\,\mu$Jy\,beam$^{-1}$. The flux density of the target was fitted by using the task \textit{imfit}.  The ATCA observation log and flux-density measurements are presented in Table \ref{tab:obs_radio}.

\section{Analysis and Results} \label{sec:analysis}

\subsection{X-ray Spectra}\label{subsec:xray spectra}
\begin{figure*}[htbp!]
    %	\epsscale{0.8}
    \centering
    \includegraphics[width=0.46\textwidth]{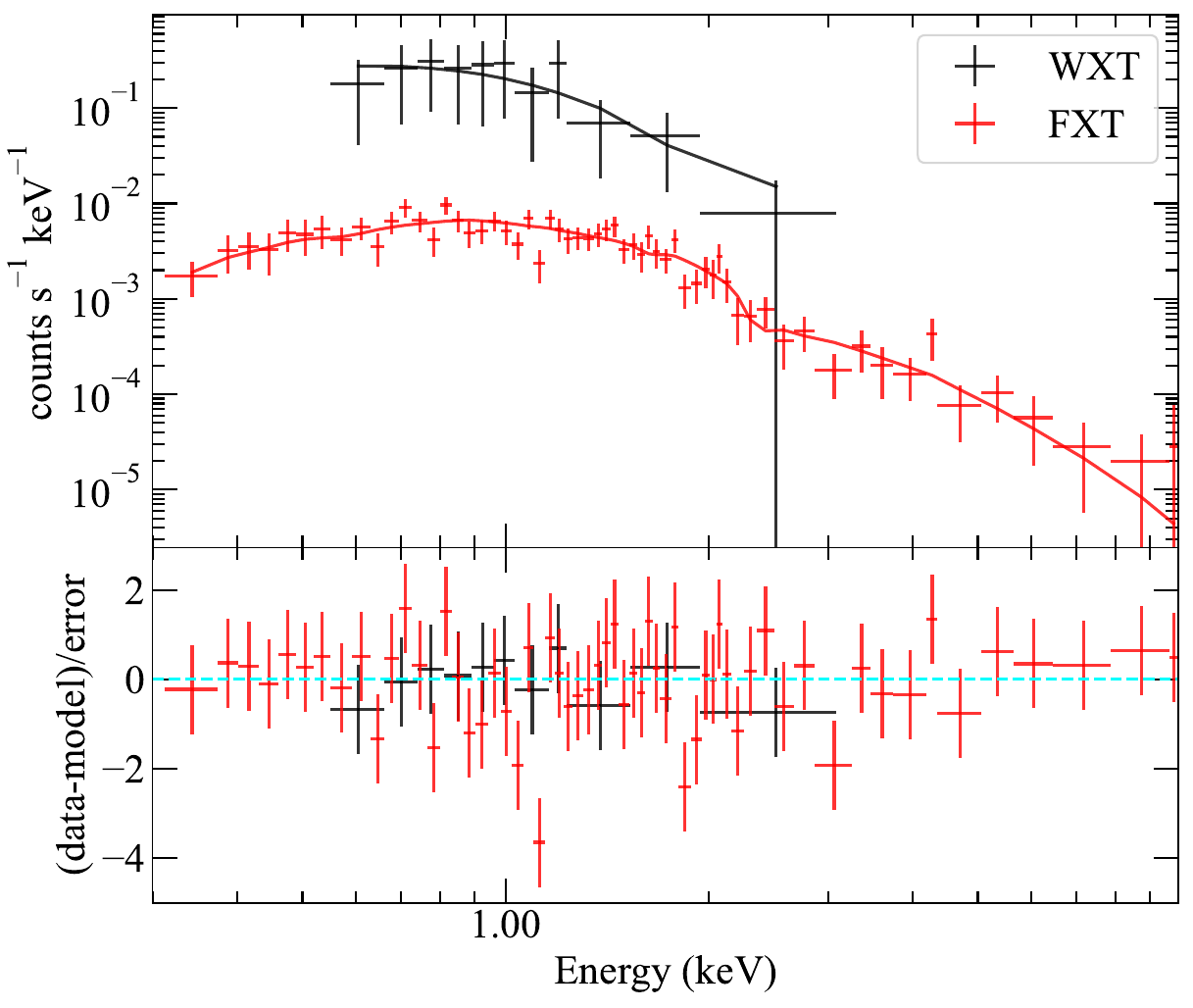}
    \raisebox{0cm}{\includegraphics[width=0.52\textwidth]{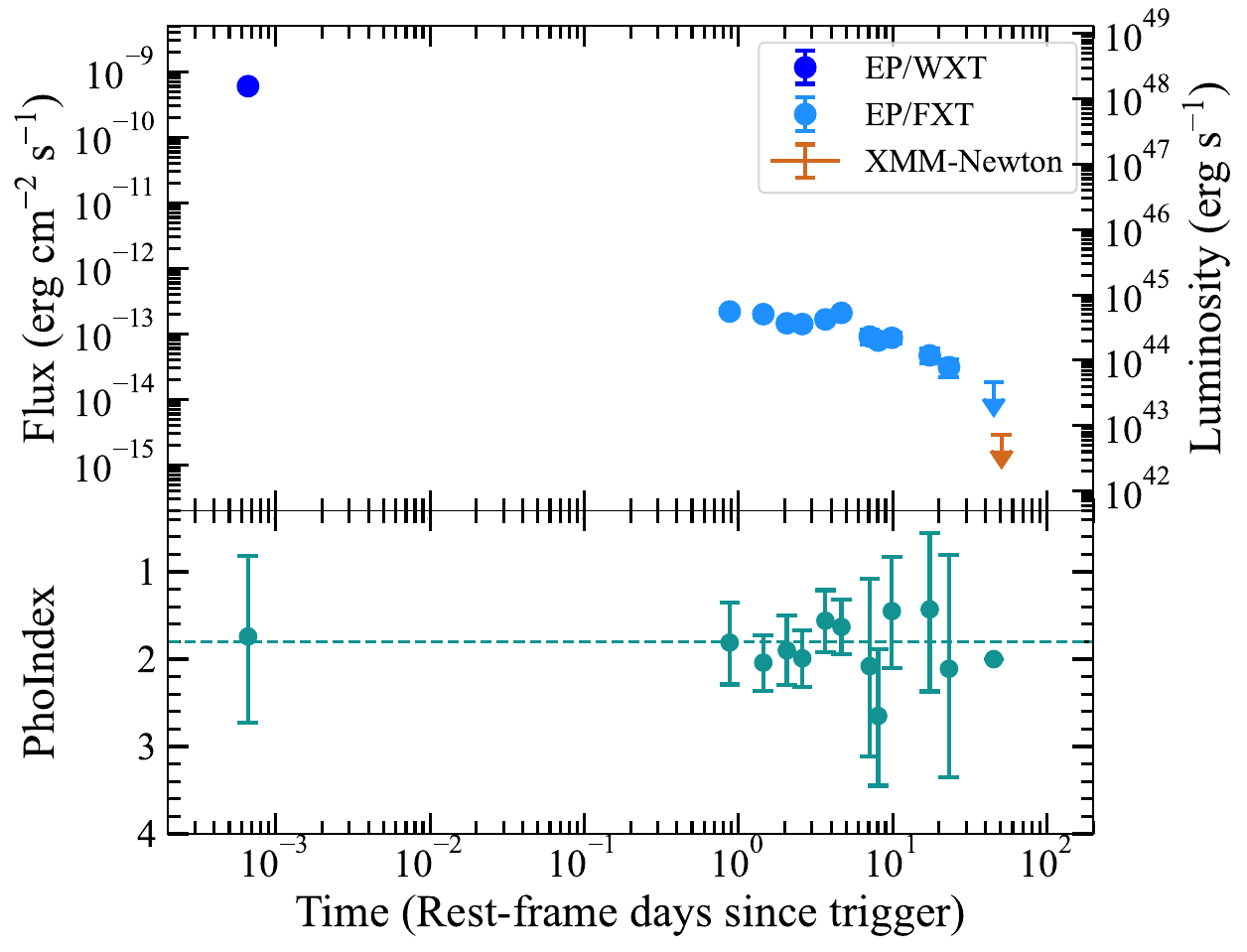}}
    \caption{
        \textbf{\it Left:} 
        The EP/FXT spectrum in the 0.3--10\,keV band by stacking the first six observations in the plateau phase (total exposure $\sim 35$\,ks). %both of which can be described by an absorbed single powerlaw model. 
        The photon index for the stacked EP/FXT spectrum is $1.8\pm0.1$ if fitted by an absorbed power law, 
        consistent with that of EP/WXT. 
        Note that the X-ray spectrum is relatively flat and consistent with nonthermal emission. 
        \textbf{\it Right:}  X-ray light curves of \src observed by EP/WXT, EP/FXT, and XMM-Newton. 
The EP/FXT fluxes in the upper panel were obtained with the photon index tied during fitting. 
The lower panel shows the distribution of photon index for the power-law component which was allowed to vary in the individual spectral fittings, and is consistent with the value $\Gamma=1.8$ within the errors. 
More details of modeling the light-curve evolution can be found in Appendix B.    
            }
    \label{fig:fxtlc}
\end{figure*}

% Since the spectral signal-to-noise ratios
% (S/Ns) for most of the individual EP/FXT spectra are not sufficient
% to perform meaningful fits, we combined the spectra from individual
% observations using the FTOOLS task???. 
We grouped the spectra to have at least 1 count in each bin so as to adopt the $C$-statistic for the spectral fits. 
%All statistical errors given hereafter correspond to 90\% confidence for one interesting parameter ($\Delta C$ = 2.076), unless stated otherwise. 
The following spectral analysis was performed with the software BXA \citep{Buchner2014},
which connects the nested sampling algorithm UltraNest\footnote{\url{https://johannesbuchner.github.io/UltraNest/}} \citep{Buchner2021}, and integrates the conventional spectral fitting software XSPEC (version 12.14.1).
%performed in the 0.3–10 keV range using XSPEC version 12.14.1. 
For the EP/WXT spectral fittings, a simple absorbed power-law model provides an acceptable fit (Figure \ref{fig:wxtlc}), yielding a best-fit photon index of $\Gamma=1.8^{+0.57}_{-0.54}$. 
Given the limited spectral S/N for the EP/FXT data, we performed a joint fit of all the EP/FXT spectra in the 0.3–-10\,keV band using the data from epochs when \src was detected; the photon indices were tied and $N_\mathrm{H}$ was frozen at the Galactic value of $5\times 10^{20}$\,cm$^{-2}$. A single uniform prior is defined for each parameter, and the best-fit values with 68.3\% uncertainty margins are given by the posterior distributions.
The resulting photon index is $1.83\pm0.08$, consistent with that obtained by WXT, 
suggesting marginal spectral evolution during the FXT observations. 
To secure the result, we used \texttt{addspec} 1.4.0 to stack the exposure-weighted X-ray spectra 
from the first 6 FXT observations which exhibit relatively small flux variations.  
The stacked spectrum can also be described by an absorbed power-law model with a photon index of $1.8\pm 0.1$, as shown in Figure~\ref{fig:fxtlc} (left). 
%the exposure was considered as weights. 
We used  this best-fit model to extrapolate the flux in the 0.3--10\,keV band 
for all the EP/FXT data as well as that observed by XMM-Newton.  
For the last FXT observation, \src was faint and not detected. Based on the aperture-photometry method \citep{Kraft1991,Ruiz2022}, we estimated the flux upper limit at a 3$\sigma$ confidence level. 
All the X-ray flux measurements are shown in Table \ref{tab:obs_xray}. 

%The first 6 FXT observations which exhibit relatively small flux variations can be considered to be in a plateau phase. We used \texttt{addspec} 1.4.0 to stack their spectra assuming exposure as weights. The stacked spectrum can also be described by an absorbed power-law model with the photon index of $1.8\pm 0.1$, as shown in Figure~\ref{fig:x-rayspec}.

\subsection{X-ray Light Curve}\label{subsec:xray lc}

During the 91.9\,s flare period, EP/WXT collected a total of 26 photon counts from \srcs, with no significant short-timescale features being discernible. \src could not be distinctly separated from the background during nonflare intervals.
Following the X-ray flare detected by EP/WXT, 
EP/FXT observed \src between 1 and 79 days since trigger. 
%Figure \ref{fig:fxtlc} (right) shows the X-ray light curve of \srcs. 
As shown in Figure \ref{fig:fxtlc} (right), the X-ray light curve exhibits a nearly plateau phase for the first 7 days, 
followed by a steep decline up to 79 days, after which \src becomes undetectable by EP/FXT, 
with a 3$\sigma$ upper limit on the flux of $1.82\times10^{-14}$\,\ergs. 
\src remains undetected by more sensitive XMM-Newton observations at $t_{\rm 0}+89$ days, 
with a 3$\sigma$ confidence upper limit on the flux of $2.83 \times 10^{-15}$\,\ergs (Section 2.1.3). 

To quantify the evolution of the X-ray emission, we first fitted a smoothed broken 
power law to the light curve in the rest frame (Appendix \ref{appeddix:xray lc}), excluding the EP/WXT data. 
Although the model can describe the EP/FXT data, it seems to overpredict the flux if compared to 
the upper limit obtained by XMM-Newton, indicating a steep break in the light curve. 
Therefore, we included a third power law to take into account the steep flux break, 
and found that it provides a statistically better fitting result over the previous broken power-law model, 
according to the Bayes factor of $K=4.12$. 
Therefore, we favor the three piecewise power-law function as the model to describe 
the light-curve evolution. 
We used an MCMC fitting technique to determine the best-fitting 
parameters and uncertainties (Appendix \ref{appeddix:xray lc}). 
Based on this, we found a slope of $L_{\rm X}\propto t^{-0.28^{+0.17}_{-0.13}}$ 
for the plateau's decline, which breaks to $t^{-1.16^{+0.30}_{-1.11}}$ at a rest-frame time 
of $t_{\rm rest}=6.1^{+8.58}_{-1.39}$ days post-trigger, followed by a steeper drop at $t_{\rm rest}=33.49^{+10.67}_{-9.72}$ days, with $t^{-9.64^{+6.92}_{-7.04}}$. 
We note that the power-law slope for the final drop phase in the luminosity evolution 
was poorly constrained, 
%a degeneracy
%between 
%the jet shut-off time $t_{\rm off}$ and the flux normalization
%FX given 
owing to the sparse temporal coverage in the light curve
between the last EP/FXT detection and the late-time limit provided by XMM-Newton. 

%that can be described by a powerlaw of $t^{-1.14}$
%The X-ray light curve is peculiar among known transients. 

\subsection{Radio Flux and SED Evolution}

\begin{figure}[t!]
\epsscale{1}
\plotone{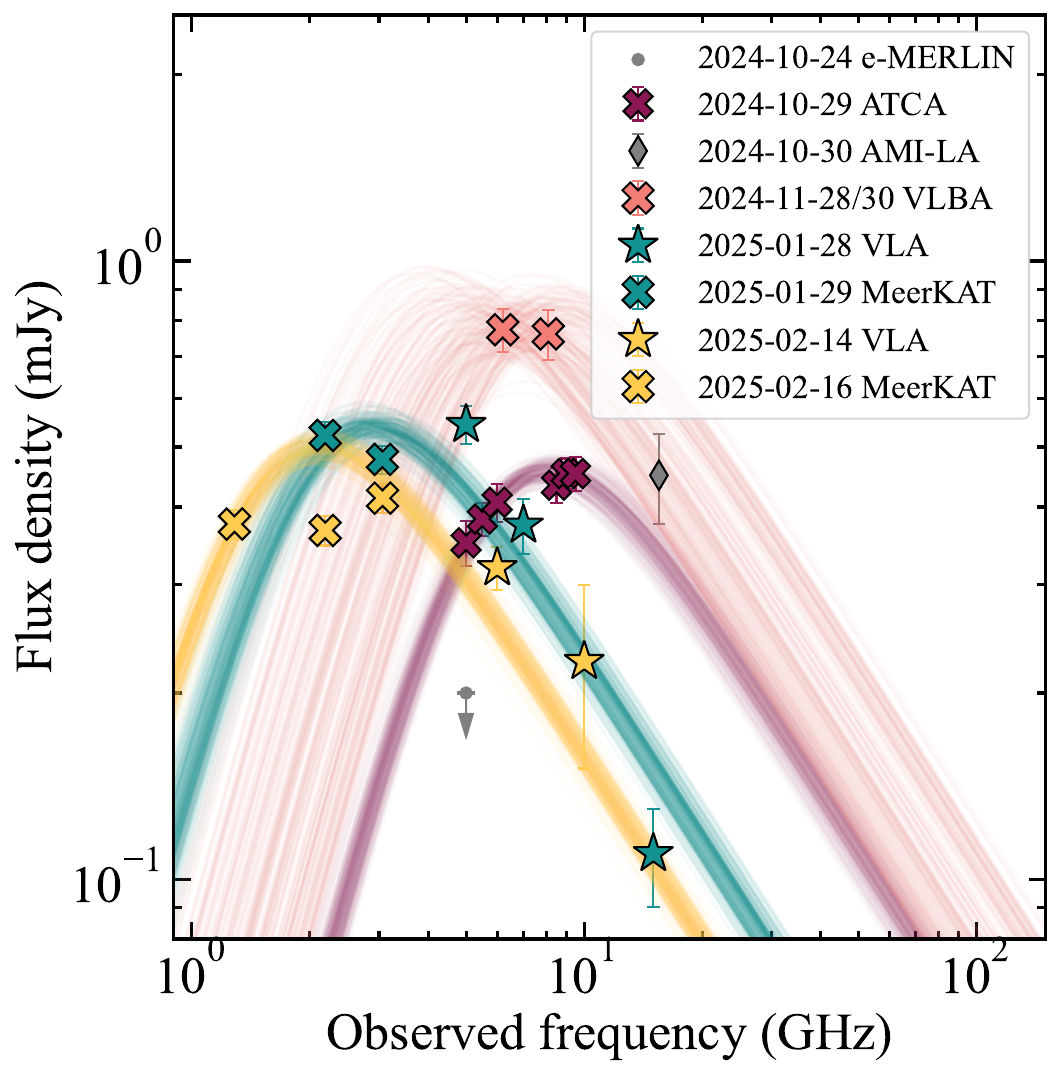}
\caption{The radio SEDs over four epochs that have quasi-simultaneous observations at different frequencies. { For non-detections, the corresponding 5$\sigma$ upper limits on} flux density are shown. The colored lines represent the best fit to each SED from our MCMC modeling, which are the model realizations on a basis of 200 random samples from the MCMC chains. 
%There is the peak flux of the four epochs has been increasing, while the value of peak frequency has been fluctuating.
\label{fig:radiosed}}
\end{figure}

{ As mentioned in Section 2.5.1, a radio transient was detected $\sim 8$ days after the discovery 
of \srcs. 
The radio luminosity at 5\,GHz is as high as $\simgt 5\times 10^{40}$\,\erg, comparable to jetted TDEs and long GRBs at similar evolution epochs \citep[e.g.,][]{Gillanders2024}. 
Combining with the detections of fast X-ray variability, nonthermal X-ray spectrum (Section 2.1), and red optical color (Section 2.3), 
this indicates that a relativistic outflow may have been launched. 
}
Figure \ref{fig:radiosed} shows the radio SED and its evolution over four epochs, 
which was constructed using the publicly available ATCA data \citep[GCN 37949]{Ricci..2024}, 
as well as our own data obtained with MeerKAT, VLA, and VLBA. 
Upon the radio detection, the SED peaks at $\sim$10\,GHz, but exhibits a shift toward higher flux density through rest-frame 18 days. The joint MeerKAT and VLA observations, taken between 57 and 67 rest-frame days post-trigger, revealed that the peak of the SED shifts downward to $\sim 2$\,GHz, with a flux of $\sim$0.5\,mJy. 

To quantify the temporal evolution, we fit the SED with a synchrotron emission
model in the context of a {relativistic} outflow expanding
%into and shocking the surrounding medium, %outflow–CNM interaction, 
%following the same approach outlined in \cite{Goodwin..2022} and \cite{Zhang..2024}. 
{into the surrounding medium. 
The interaction of outflow with the surrounding medium 
leads to synchrotron emission owing to the acceleration of electrons 
and amplification of magnetic fields.  
%generating blastwave that amplifies the 
%magnetic field and accelerates the electrons in surrounding medium into a power law distribution, 
%$N(\gamma_{e})\propto\gamma_{e}^{-p}$ for $\gamma_{e}\geq\gamma_{\rm m}$, where $\gamma_{e}$ is the electron Lorentz factor, 
%$\gamma_{\rm m}$ is the minimum Lorentz factor of the distribution, and $p$ is the powerlaw index. 
This model has been widely used to explain the radio emission from GRBs \citep[e.g.,][]{Granot2002, Granot2014} and TDEs \citep[e.g.,][]{Berger2012, Alexander..2016}, as the basic framework is quite general.  
}.

% As in Goodwin et al. (2022), due to the sparse sampling of radio spectrum, 
Owing to the sparse sampling of the radio spectrum and the lack of high-frequency observations, we fit the radio spectrum using the synchrotron spectrum 2 model described by \cite{Granot2002}, assuming $\nu_m \ll \nu_a$, where $\nu_m$ is the characteristic synchrotron frequency of the emitting electrons with the least energy and $\nu_a$ 
is the self-absorption frequency,  %As in Goodwin et al. (2022), 
and fixing the synchrotron energy index in the optically thin regime to $p=3$ \citep[e.g.,][]{Cendes..2021}. 
We use an MCMC fitting technique \citep[python module emcee;][]{Foreman-Mackey..2013} to marginalize over the synchrotron model parameters to determine the best-fitting parameters and uncertainties. 
%Due to the limited data points, we fix the synchrotron energy index to $p=3$ \citep[e.g.,][]{Alexander..2016, Cendes..2021}. 
In Figure \ref{fig:radiosed}, we show the resulting SED models, which provide reasonable fits to the data. From the best-fitting SED models, we determine the peak flux density and frequency, $F_{\nu,p}$ and $\nu_p$, respectively. 
%We find that $\nu_p$ decreased over time, while $F_{\nu, p}$ first increased and then decreased.

Using the inferred values of $F_{\nu,p}$ and $\nu_p$, we can calculate the equipartition radius and energy in the relativistic regime assuming the outflow is viewed on-axis 
\citep{BarniolDuran..2013}, which was derived as %. We computed the equipartition radius 

\[ R_{\rm eq} \approx (1.7 \times 10^{17} \, \text{cm}) 
\begin{bmatrix}
F_{p, \text{mJy}}^{8/17}d_{L,28}^{16/17}\eta^{35/51} \\
\hline
\nu_{p,10}(1+z)^{25/17}
\end{bmatrix}
\begin{array}{c}
\Gamma^{10/17} \\
\hline
f_A^{7/17}f_V^{1/17}
\end{array} ,
\]

%\noindent and the minimal total energy

\[ E_{\rm eq} \approx (2.5 \times 10^{49} \, \text{erg}) 
\begin{bmatrix}
F_{p, \text{mJy}}^{20/17}d_{L,28}^{40/17}\eta^{15/17} \\
\hline
\nu_{p,10}(1+z)^{37/17}
\end{bmatrix}
\begin{array}{c}
f_V^{6/17} \\
\hline
\Gamma^{26/17}f_A^{9/17}
\end{array} .
\]

Here $f_V$ and $f_A$ are geometric factors, $\eta =1$ for $\nu_m \ll \nu_a$, and $\Gamma$ is the Lorentz factor of the outflow. Considering the exceptionally high X-ray and radio luminosity likely due to a strong beaming effect, we assume a narrow jet with a half-opening angle of $\theta_j = 0.1 < 1/\Gamma$, such that $f_A = f_V = (\theta_j\Gamma)^2$.

%Following Barniol Duran $\&$ Piran (2013), Barniol Duran et al. (2013), Eftekhari et al. (2018) and Yao et al. (2024), 
%We further assumed microphysical parameters $\epsilon_e = 0.1$, $\epsilon_B = 10^{-3}$, and that kinetic energy of the outflow is dominated by hot protons \citep{BarniolDuran&Piran2013, BarniolDuran..2013, Eftekhari..2018, Yao2024}. The equipartition energy will increase by a factor of $\xi^{11/17}$ and the radius will increase by a factor of $\xi^{1/17}$, where $\xi = 1 + \epsilon_e^{-1} \approx 11$. 
% Moreover, the deviation from equipartition is parameterized by the factor $\epsilon = (\epsilon_B/\epsilon_e)(11/6)$, which introduces a multiplicative factor of $\epsilon^{1/17}$ for the radius and $(11/17)\epsilon^{-6/17} + (11/17)\epsilon^{7/12}$ for the energy. 
%The actual radius R corresponding to the minimal energy is different from $R_{eq}$ by a multiplicative factor of $\epsilon^{1/17}$ and the total energy E is greater than $E_{eq}$ by a multiplicative factor of $(11/17)\epsilon^{-6/17} + (6/17)\epsilon^{11/17}$. 

To find the mimimal energy using equipartition arguments, we need another equation that relates the time since the onset of the relativistic outflow $t$, the radius $R$, and the Lorentz factor ($\Gamma$):  

\[t \approx 
\begin{array}{c}
R(1 - \beta)(1+z) \\
\hline
\beta c
\end{array}.
\]
\noindent
where $\beta$ is the velocity of the outflow. 
%In the right panel of Figure \ref{fig:radiosed}, we show the evolution of outflow kinetic energy ($E_{\rm K}$), radius of radio-emitting region ($R$), $\Gamma$, and $\Gamma\beta$.
%The kinetic energy increases by a factor of $\sim$10 from $E_{\rm K}\approx 5.35\times10^{49}$ 
%to $E_{\rm K}\approx 6.09\times10^{50}$ erg over 62 rest-frame days. 
%The rapid increase in kinetic energy indicates that we were observing the deceleration of the outflow, 
%which is consistent with 
The equipartition analysis suggests a modest Lorentz factor $\Gamma\approx3$, 
%to $\Gamma\approx2.1$ 
and an equipartition energy $E_{\rm eq}\approx 10^{50}$\,\erg. 
Note that the total minimal energy will be increased by a factor of $(1 + \epsilon_e^{-1})^{7/12}$, if considering that the hot protons carry a significant portion of the total energy \citep{BarniolDuran&Piran2013}. 
%over the same epochs. 
%However, the equipartition analysis suggests that 
%while the outflow velocity ($\Gamma\beta$) 
%decreased from $\approx3.3$ to $\approx$1.8 over the same epochs, which is still in the relativistic expansion phase ($\Gamma\beta>1$). 
Furthermore, if considering the case that we are away from equipartition, 
the estimate on the Lorentz factor is still valid. 
%the change in the changes in the microphysical parameters ($\epsilon_e$ and $\epsilon_B$), i.e., the case away from 
In this case, the total energy will be larger by a factor of $\sim 0.5\epsilon_e^{-0.6}\epsilon_B^{-0.4}$ if $\epsilon_e + \epsilon_B < 1$,  
where $\epsilon_e$ and $\epsilon_B$ are the microphysical parameters, respectively
the fractions of the total energy in electrons and magnetic field. 

%The peak of the radio SED is around ?? GHz, and 
%the lack of detection of high-frequency emission at $>$100 GHz (Aryan et al., GCN 38924) 
%indicates that the observed radio emission is not originating from a compact region. 
%Given the shape of the observed SEDs, we associate the peak frequency $\nu_{\rm p}$ 
%with the synchrotron self-absorption frequency $\nu_{\rm a}$, 
%and assume that the 

%As shown in Figure \ref{fig:radiosed} (left), 
We note that the flux in the S0 band (2.2\,GHz) observed by MeerKAT decreased from 0.51\,mJy to 0.39\,mJy over two weeks,  % from 2025 Jan 29 to 2025 Feb 16. 
while the variability amplitude observed in other MeerKAT bands was not as significant as that in the S0 band. 
We consider whether the observed variability might be affected by %was induced by the effect of refractive 
interstellar scintillation (ISS). This process occurs when radio waves propagate through an inhomogeneous plasma in our Galaxy, which could cause intraday variability in some AGNs with compact radio emission \citep{Lovell..2003, Rickett2007}. The amount and timescale of radio variation caused by ISS depend on the Galactic electron column density along the line of sight and the observing frequency. Using the NE2001 free electron density model developed by \cite{Cordes..2002}\footnote{\url{https://pypi.org/project/pyne2001/}}, and the Galactic dispersion measure (DM)\footnote{EP241021a has  Galactic coordinates of ($l,~b = 150^\circ.3520610,~-53^\circ.4941773$). We used
the DM of 27.5\,cm$^{-3}$\,pc for EP241021a, which was derived from the pulsar J0156+0402 ($l,~b = 151^\circ.9997139,~-55^\circ.1844179$) in the ATNF Pulsar Catalogue.}, % along the line of sight to \srcs, 
we find that, at the position of \srcs, the transition frequency between the strong- and weak-scattering regime is $\nu_0 \approx 7.93$\,GHz and the angular size limit of the first Fresnel zone at the transition is 6.2\,$\mu$as. Adopting the formalism of \cite{Walker1998}, we estimate that 
%the radio emission from \src will be in the strong, refractive scintillation regime, 
%until the source reaches an angular size of 284 $\mu$as. Radii of $10^{17} \sbond 10^{18}$cm at $D_A$ = 1512 Mpc would correspond to angular diameters of 4.4 $\sbond$ 44 $\mu$as. {\bf The emission from \srcs is expected to be affected by ISS with a time-scale of and a modulation fraction of .} 
% Hence \srcs is in the strong-scattering regime at S0-band. In this case, we computed the fractional modulation index due to ISS (Walker 1998), \footnote{In Walker(1998),the modulation index m is used to calculate the variability amplitude caused by ISS, which is defined as the ratio of the rms deviation to the mean value of the observed flux densities, $m = \sqrt{S^2 / [F_\nu]^2}$.} which is m = 0.42 at S0-band, comparable to that observed in \srcs.
%In order to account for this flux density variation, and any flux density variation due to ISS, we introduced an additional error on each radio flux density measurement. We calculated the appropriate error due to ISS for each frequency depending on the expected modulation fraction of ISS at that frequency, where the errors varied 
the level of frequency- and time-dependent random flux variations induced by ISS is from 59\% at 3.1\,GHz to 2.8\,\% at 15.0\,GHz for \srcs. 
%To inspect the effect of the ISS-induced flux variations, we repeated the SED fittings 
%We verify that the 
%When considering the effect of the ISS-induced flux variations affect the 
When adding the ISS-induced flux variations to the measurement uncertainties, 
the radio SED fittings in the Epoch II and III observations cannot converge. 
This is because the SED sampling is either too sparse or peaks at lower frequency where the expected ISS effect is large. 
On the other hand, we find that while the ISS effect 
increases the uncertainty in the derived physical parameters for 
the Epoch I and IV observations, %the best-fit values changed by $\sim$ 3 percent, 
%which do 
it does not bias the above estimation of Lorentz factor and total minimal energy of the outflow. 
{ Note that the ISS effect on the radio variability is also explored by \citep{Yadav2025}, 
in which the size of the radio-emitting region can be constrained. 
By estimating the brightness temperature of the radio emission, \citet{Yadav2025} derived 
a lower limit for the Lorentz factor $\Gamma \simgt 5.5$ at $T_{0}+1.5$\,d ($T_{\rm 0}$ is the time of trigger), consistent with the value constrained by the SED fittings. 
}

\section{Discussion} \label{sec:discussion}

Here we summarize the multiwavelength properties of \srcs,  
%traced by its evolution in the X-ray emission: 
highlighting the uniqueness in its temporal evolution.
%Our analysis of the multi-wavelength data suggests that \src exhibits several 
%unique properties: 

(1) \src was triggered by EP/WXT and had a duration of $\sim$100\,s, with a time-averaged X-ray flux (0.5--4 keV) of $3.31\times10^{-10}$\,\ergs, corresponding to an isotropic-equivalent energy of the prompt emission $E_{\rm iso}\approx 8.5\times10^{49}$\,erg. 
%The emission is predominatnly in the soft X-ray bands 
The flare's X-ray spectrum is relatively hard %which can be fitted by an absorbed power-law model 
with a photon index $\Gamma=1.8^{+0.57}_{-0.54}$, likely associated with nonthermal emission.  %indicating a non-thermal emission. 

(2) An X-ray counterpart was detected by EP/FXT about 36.96\,hr after the trigger,  
with a flux (0.3--10 keV) of $2.19\times10^{-13}$\,\ergs,  
implying a rapid flux decline by more than three orders of magnitude within
about 1.5\,d. % after the detection. 
During the phase from $T_{\rm 0}+1.5$\,d to 8\,d, the evolution of the X-ray emission enters a nearly plateau phase, with a sign of late-time rebrightening. 
The FXT observations confirmed the hard X-ray spectrum, with no significant spectral evolution.  
Then the X-ray emission shows a power-law decline with $t^{-1.2}$ 
over $\sim 30$\,d, followed by rapidly dropping below detection. 

(3) Optical emission associated with \src was detected within 1.8\,d 
after the X-ray trigger, possibly related to the afterglow of the prompt X-ray emission. 
At $T_{\rm 0}+6$\,d, an optical brightening was observed peaking at $\sim -21.5$ mag in the $r$ band, with a fast rise time of only $\sim 2$\, d. 
Such a fast timescale of the optical evolution and high peak luminosity resemble those of luminous fast blue optical transients \citep[FBOTs,][]{Ho2023}. 
During the optical peak, Keck spectroscopy was  performed, revealing narrow emission lines from the host galaxy and providing the redshift of \src at $z=0.748$. 
%which represents the
%most luminous optical emission ever recorded to be associated with an X-ray transient. 
A late-time optical rebrightening appears at $T_{\rm 0}+20.7$\,d, though not as significant and bright as the second one, with a peak of $\sim-20.2$\,mag.  
\src exhibits a persistent red color ($r-z\approx 0.4$\,mag) over a period of at least 40\,d since discovery, which is unusual among known optical transients. 
%The phases of rebrightening are potentially coincident with the X-ray rebrightening. 
There is a potential coincidence between the phases of optical and X-ray brightenings. 

(4) Follow-up radio observations reveal the appearance of the radio counterpart at $T_{\rm 0}+8.4$\,d. Over the period of $\sim 110$ days of post-trigger radio observations, the peak flux density increases for the first month, followed by a decrease in the subsequent two months.  
During this time range, the self-absorption frequency decreases. 
%at 9 GHz peaks at $\sim$$T_{\rm 0}+8.4$ d, with a 
The radio spectral luminosity is as high as $\sim 10^{31}$\,erg\,s$^{-1}$\,Hz$^{-1}$, comparable to that of long GRBs and jetted TDEs at similar evolution epochs. An equipartition analysis suggests that a relativistic outflow with a moderate bulk Lorentz factor ($\Gamma\approx 3$) and a minimumn energy of $10^{50}$\,\erg may have been launched. 

%The exceptionally high apparent luminosity across X-ray, optical and radio, red optical color, 
%extreme X-ray luminosity drop (a factor of $>$$1000$ over a month), 
%and fast X-ray variability on a short time-scale of $\sim$100 s, 
The X-ray, optical, and radio observations suggest that \src is an extremely unusual transient, and multiple emission components might 
be required to 
%We will discuss below possible scenarios to 
account for its multiwavelength properties, 
%possibly related to a long-lived central engine activity and
%or multiple outflows. 
such as various mechanisms to produce the rebrightening emission 
or episodic energy injections from multiple outflows. 
In the next section, we will explore the possible scenarios that could accommodate the 
multiwavelength behavior of \srcs. 

\subsection{A Choked/Weak Jet and a Supernova Shock Breakout}

Both the isotropic energy and the peak energy of the prompt emission are significantly lower than
those of conventional long GRBs. 
While the luminosity of \src falls into the range of low-luminosity GRBs, it stands as an outlier from the Amati-Yonetoku relation owing to its lower peak energy, 
similar to the EP { EFXTs} EP240414a and EP250108a, both of which are associated with broad-lined Type Ic supernovae \citep[SNe Ic-BL;][]{Sun2024, Li2025}. 
%which was believed to be generally satisfied by LGRBs, X-ray flashes, or low-luminosity GRBs \citep{}. 
%In addition to \srcs, there are two extragalactic FEXTs, 
%Note that both EP240414a and EP250108a also exhibited optical brightening in their light curves \citep{Sun2024, Li2025}.  
Together with the nondetection of high-energy gamma-ray emission, 
it has been proposed that a collapsar-driven low-energy jet 
choked in extended circumstellar material (CSM) can explain their multiwavelength properties \citep{Hamidani2025, Eyles-Ferris2025}. 
In this scenario, multiple emission components are invoked to account for the evolution in different phases, including the early-phase afterglow emission from the choked/weak jet, the late-phase emission from a Type Ic-BL supernova, and the middle rebrightening phase due to supernova shock breakout (SBO). In addition, the cocoon produced by the interaction of the jet with the dense CSM could contribute to the emission \citep{Hamidani2025}.
%The jet afterflow emission and cooling emission from outer cocoon 

Generally speaking, no supernova-like features were identified in the late-time optical spectra
 of \src \citep{Busmann..2025}. However, such a negative supernova observation could just be due to the relatively large distance of the source. 
One cannot fully rule out that the second optical bump is produced by the breakout of the supernova shock from the CSM. In order to account for the fast evolution of the optical bump and its bolometric luminosity of $L_{\rm opt}\approx 10^{44}\,\rm{erg\,s^{-1}}$, the mass of the CSM would be required to be about $10\,{\rm M}_{\odot}$, which spreads within a range of several thousands of solar radii \citep{Khatami2024}. However, it could still be not easy to account for the large radius of the emitting photosphere, which can be constrained to 
\begin{equation}
    \begin{aligned}
        &R_{\mathrm{ph}}=\left(\frac{L_{\rm p}}{4\pi \sigma_{\mathrm{SB}}T^4}\right)^{1/2}\\
        &\approx 3.8\times 10^{15}\left(\frac{L_{\rm p}}{10^{44}\,\mathrm{erg\,s^{-1}}}\right)^{1/2}\left(\frac{T}{10^{4}\,\mathrm{K}}\right)^{-2}\mathrm{cm}, 
    \end{aligned}  \label{Eq:Rph}
\end{equation}
where $\sigma_\mathrm{SB}$ is the Stefan-Boltzmann constant and the reference value of the blackbody temperature $T$ is taken according to its relatively red color. Furthermore, the existence of the dense CSM could effectively hinder the propagation of the jet/cocoon and thus the generation of the afterglow emission. Finally, very different from EP240414a and EP250108a, \src displays long-lasting X-ray afterglow emission ($\Delta T\approx 39$\,d), 
%which can even brighten up too with the optical bump. 
{ with a flux brightening that appears to coincide with the optical bump. 
This X-ray brightening} obviously cannot be contributed by the cooling cocoons or generated by a radiation-mediated shock propagating into the dense CSM \citep{Margalit2022}. 
%While \srcs's X-ray evolution is 
%EP240408a was observed to show 
%similar to EP240408a, the peculiar FEXT with an intermediate evolution timescale of $\approx$10 d \citep{Zhang..2024}, 
%the lack of optical and radio counterparts for the latter makes %its origin still obscured. 
%it hard to explore whether they are driven by the same mechanisms. 
%In Figure \ref{fig:LC}, it is evident that the model involving two discrete jets with a time interval of $\sim$5 days adequately explains the optical and radio lightcurves. However, the model under predicts the X-ray emission during the plateau phase, suggesting that an additional component is required, such as an active central engine. One possible explanation is energy injection from the central engine of the GRB \citep{Dai..1998,Zhang..2001}. Energy injection is commonly invoked to explain the X-ray plateau observed in GRBs \citep{Zhang..2006,Rowlinson..2013}. 

\begin{figure}[htbp!]
    	\epsscale{1.2}
    \plotone{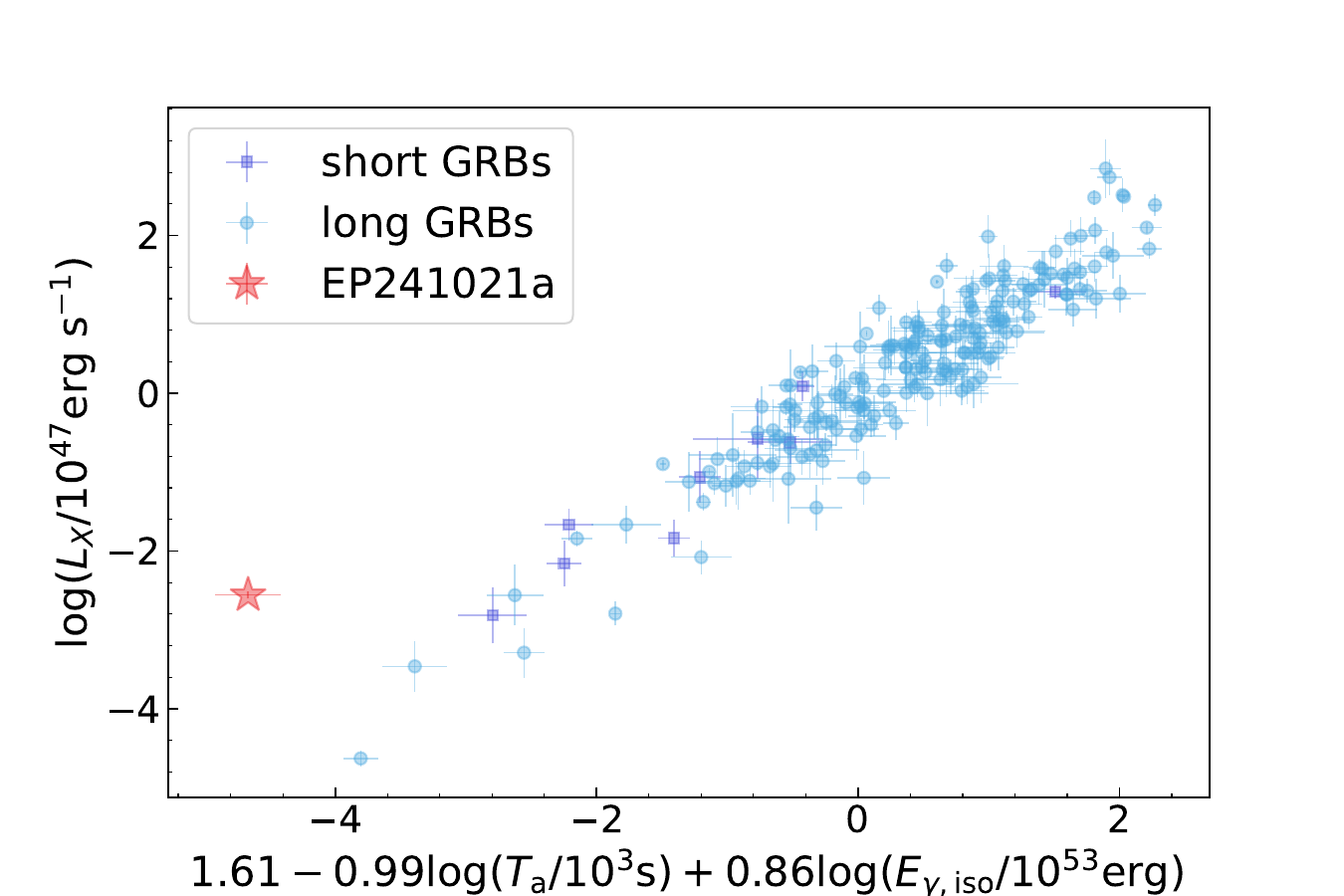}
    \caption{
        EP241021a and the observed \emph{L-T-E} relation for 210 GRBs \citep{Xu..2021,Deng..2023}. Long GRBs and short GRBs are marked with blue circles and purple squares, respectively. EP241021a is an outlier of this relation, indicating that its plateau originates from a different mechanism compared to the typical GRB plateau.
            }
    \label{fig:LTE}
\end{figure}

\subsection{A Magnetar Engine and its Possible Origin}
{Even if the rebrightening is not due to a supernova SBO, could it instead originate from another type of supernova-related emission?} Using the photospheric radius {given in Eq. (\ref{Eq:Rph})} to represent the outer radius of the explosion ejecta, which can be correct when the ejecta are optically thick at early times, we can constrain the expansion velocity of the ejecta to be $v_{\mathrm{ej}}\approx 0.25c$. This value is obviously higher than the typical velocity of supernovae {which is no more than $\sim0.1c$ \citep[e.g.,][]{Barbarino2021,Liu2017}.} Meanwhile, according to the prompt emission energy of $\sim10^{50}$\,erg and assuming a reference radiation efficiency of $\sim10\%$, we can further take the initial kinetic energy of the explosion ejecta to be  $E_{\mathrm{k}}\approx10^{51}\,\mathrm{erg}$, which is typical for both normal supernova explosions and compact binary mergers. Combining the estimates of $v_{\mathrm{ej}}$ and $E_{\mathrm{k}}$, the mass of the ejecta can be calculated as  $M_{\mathrm{ej}}\approx 0.02\,{\rm M}_{\odot}$. Then, we can express the photon diffusion timescale of the expanding ejecta by 
 \begin{equation}
    \begin{aligned}
       & t_{\mathrm{diff}}\approx\left(\frac{3\kappa_{\rm opt} M_{\mathrm{ej}}}{4\pi c v_{\mathrm{ej}}}\right)^{1/2}\\
        &\approx 6.5\left(\frac{\kappa}{10~\mathrm{{cm}^2g^{-1}}}\right)^{1/2}\left(\frac{M_{\mathrm{ej}}}{0.02\,{\rm M}_{\odot}}\right)^{1/2}\left(\frac{v_{\mathrm{ej}}}{0.25c}\right)^{-1/2}~\mathrm{day},
    \end{aligned}
 \end{equation}
 which indicates that a relatively high opacity $(\kappa_{\rm opt}\approx 10\,\mathrm{{cm}^2\,g^{-1}})$ would be invoked to match the observed peak time of the optical rebrightening. 

The large speed, low mass, and high opacity of the ejecta point to \src possibly originating from the merger of a binary neutron star or the collapse of a super-Chandrasekhar-mass white dwarf. In the former case, it is widely believed that an amount of lanthanides can be synthesized in the neutron-rich ejecta, which can give a natural explanation for the ultrahigh opacity of 10--100\,cm$^2$\,g$^{-1}$ \citep{Kasen2013}. The low mass of the ejecta further indicates that the prompt emission energy of \src, $\sim10^{50}$\,erg, cannot be provided by the decay of radioactive elements in the ejecta, and instead requires a more powerful energy source. As widely suggested, such an extra energy supply can be provided naturally by the remnant compact object, which is most likely a millisecond magnetar \citep{Yu2013,Metzger2014,Yu2019a,Yu2019b}. In this case, the magnetar not only enhances the optical thermal emission of the ejecta, but  also generates X-rays directly through its wind nebula emission. So, the X-ray bump associated with the optical transient can be explained by the leakage of the nebula emission from the ejecta \citep{Yu2019b}. For detailed modeling of \src in this scenario, please see \citet{Wu2025}. In particular, if this millisecond magnetar is specifically formed from the accretion-induced collapse (AIC) of a white dwarf, then the interaction between the magnetar and the residual companion star could further lead to intermittent energy injection into the ejecta. As a result, some post-peak light-curve bumps could be produced \citep{Zhu2024}, just as observed in \src.

In such magnetar engine cases, the $\sim 100$\,s bright X-ray emission detected by WXT may be attributed to the magnetar wind dissipation during the initial spin-down period \citep{zhang2013,sun2017,chen2025}. Gravitational-wave signals may accompany the event but are too faint to be detected with the current gravitational-wave detectors. {The dissipation of the released spin-down energy could occur through magnetic reconnection that converts a Poynting flux to an ultrarelativistic wind, which is most likely detected from a viewing angle away from the central jet. Additionally, X-ray emission can be produced by the collision of this wind with the preceding jet or with the CSM. In any case, similar to normal GRBs, we cannot rule out that the prompt X-ray emission is contributed by internal dissipation in the mildly relativistic jet.} 

\subsection{ A Structured GRB Jet}

%likely produced by a relativistic outflow that 
%To model the multi-wavelength light curves, we collect the optical $r$-band data from GCN Circulars and the dataset presented by \cite{Busmann2025}, with corrections for Galactic extinction.
%\cite{Busmann2025}, corrected for the Galactic extinction. 
%For the radio light curves, we utilize mainly the flux densities observed 
%at 5 GHz and 9 GHz, as they are less affected by the ISS effect and are relatively well sampled. 
%In addition to data provided in Table 2, we retrieve the relevant data observed by ATCA (Program ID: CX585), consisting of totally 8 observations. 
%We then produced the  measured the peaked , 
%We then produce the calibrated clean images, and measure the flux densities at 5 GHz and 9 GHz, %were then measured 
%following the same procedures described in Section 2.3.4. 
%Given that the early-time radio emission is likely produced by a relativistic outflow (Section 3.3), %we modelled \srcs's data under the jet paradigm. Here, 

The general synchronous evolution of the optical and X-ray emission could also indicate that the multiwavelength brightenings are the result of the dynamical evolution of a structured jet, { as observed in some GRBs \cite[e.g.,][]{Salafia2022}}. On the one hand, the jet could be radially structured and its external shock can be continuously refreshed because of energy injection from the catching-up and more energized jet material \citep{Rees1998,Yu2007,Busmann..2025}. However, in this scenario, it could not be easy to produce a flux jump of $\delta t/t\approx 0.3$, unless the jet actually consists of a series of discrete segments that have very different Lorentz factors (see Sec. \ref{sec:TDE} for a specific example). On the other hand, the jet could be angularly structured and the line of sight (LOS) somewhat deviates from the symmetry axis of the jet \citep{rossi2002,zhang2002}. 
{ This is possible if the nondetection of high-energy emission in \src (Section 2.2) is due to an off-axis viewing angle.}
Then, as the emission from the core of the jet gradually enters the view, a rising afterglow can be obtained. Here, in order to explain the fast jump of the emission, the jet should consist of two or more separated angular components, which could specifically correspond to the core, wing, and cocoons of the jet (see \cite{Gianfagna2025} for detailed modeling). In this case, since the LOS initially pointed to the wide slow component, optical and X-ray bumps can be produced when the narrow fast component suddenly emerges in the view at a later time \citep{Granot..2002, Yamazaki..2004, Xu..2023a, OConnor..2024}. Such a nonthermal origin of both optical and X-ray brightenings is further beneficial for explaining the hard X-ray spectrum and the relatively red color of the optical emission.
%to near-infrared colors point to a non-thermal origin for the X-ray and the optical emission as well as its rebrightening. 
%Using this setup, we fit the multi-wavelength light curves. 

To model the multiwavelength light curves in the two-component jet scenario, we use mainly the optical 
$r$-band data taken from our follow-up observations, GCN Circulars, and the dataset presented by \cite{Busmann..2025}, in which the bump features are more clearly presented. 
All the fluxes have been corrected for the Galactic extinction. 
%with corrections for Galactic extinction. 
%we collect the optical $r$-band data from GCN Circulars and the dataset presented by \cite{Busmann2025}, with corrections for Galactic extinction.
%\cite{Busmann2025}, corrected for the Galactic extinction. 
For the radio data, we utilize mainly the flux densities observed 
at 5\,GHz and 9\,GHz, as they are less affected by the ISS and are relatively well sampled. 
In addition to data provided in Table 2, we retrieve the relevant data observed by ATCA (Program ID CX585), consisting of 8 observations. 
%We then produced the  measured the peaked , 
We then produce the calibrated clean images, and measure the flux densities at 5\,GHz and 9\,GHz, %were then measured 
following the procedures described in Section 2.3.4. 
However, the fitting to the multiwavelength light curves with a structured jet yields an unphysically high Lorentz factor (exceeding 1000) for the fast, narrow jet. Furthermore, this model fails to account for the third optical bump observed around 20 days after the burst. The optical flux observed after this bump is also significantly higher than that predicted by the model.

{If \src is related to a GRB, the missing prompt gamma-rays might imply a low-luminosity burst. 
In this case, the absence of a supernova disfavors a GRB 980425-like event that is associated with a ``hypernova''  \citep{Iwamoto..1998}, and its nonthermal spectrum differs from the thermal shock-breakout source GRB 060218 \citep{Campana..2006}, pointing to a distinct subclass.}
In any case, \cite{Xu..2012} identified a tight relation linking the luminosity and duration of the plateau to the isotropic-equivalent gamma-ray energies. As shown in Figure \ref{fig:LTE}, EP241021a deviates significantly from this relation, suggesting a possibly different origin for the X-ray emission as observed in conventional GRBs.

\begin{figure*}[htbp!]
    %	\epsscale{0.8}
    \includegraphics[width=0.5\textwidth]{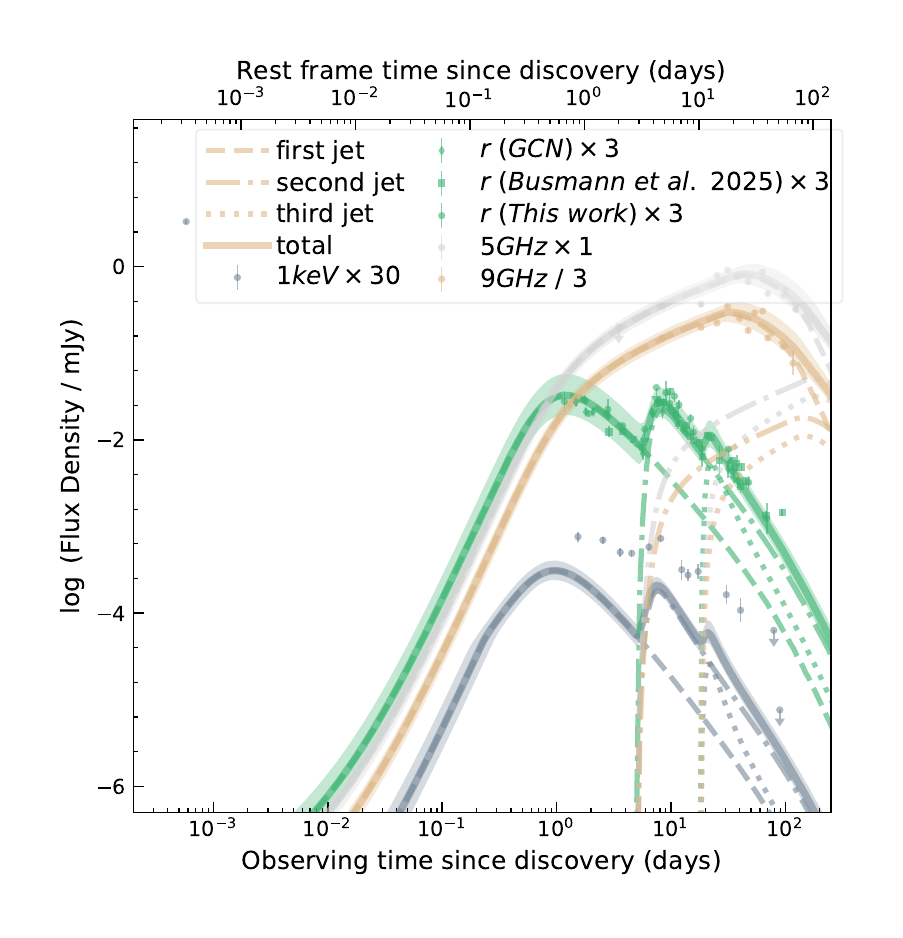}
    \raisebox{0.14cm}{\includegraphics[width=0.5\textwidth]{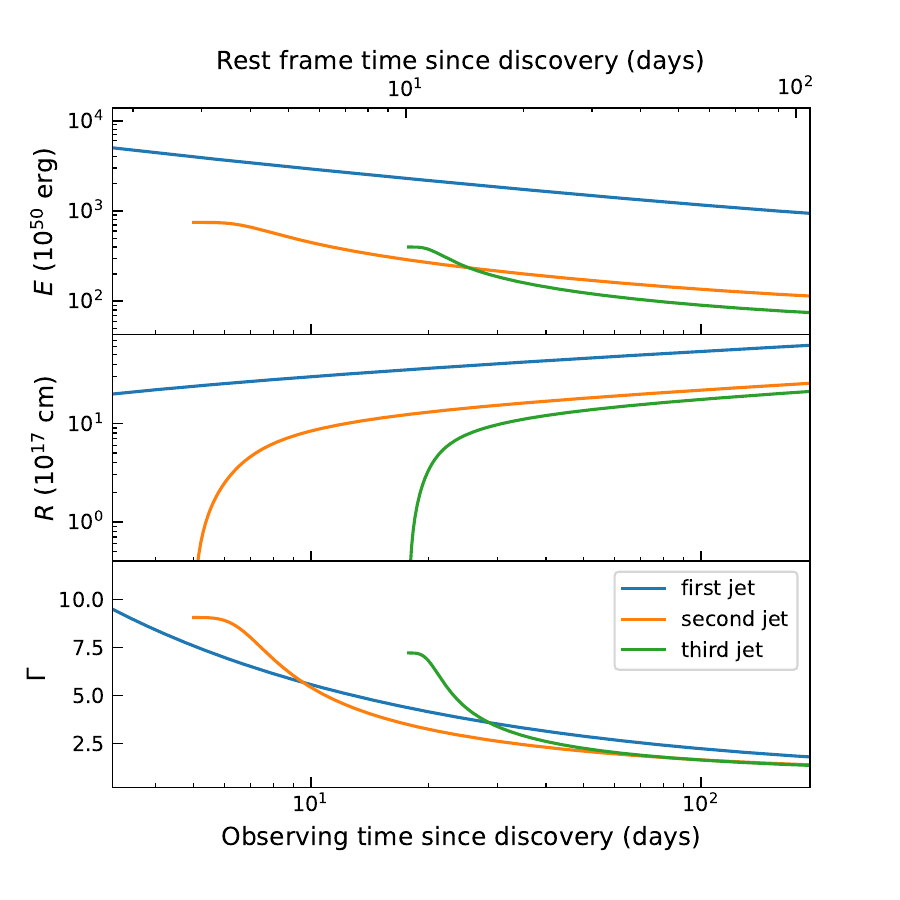}}
    \caption{
        \textbf{\it Left:} Observed multiwavelength afterglow of EP241021a and best-fit results from the three-jet model (solid curves). Shaded regions exhibit the $1\sigma$ confidence intervals of the model light curves derived from posterior distributions of model parameters. Both optical and radio light curves are well explained by this model, while the X-ray emission is underpredicted. A possible solution to this inconsistency is the inclusion of a jetted TDE (see Section \ref{sec:TDE}). \textbf{\it Right:}  The evolution of kinetic energy, radius, and Lorentz factor for three jets. The first, second, and third jets are represented by the blue, orange, and green curves, respectively. 
            }
    \label{fig:LC}
\end{figure*}

\subsection{ Multiple Jet Ejections from Repeating Partial TDEs}\label{sec:TDE}
%Partial tidal disruptions by an intermediate-mass black hole? }
\subsubsection{Modeling the multiwavelength light curves with three jets launched at separate times}

The discrepancies of modeling with a two-component structured jet motivate us to explore an alternative explanation. We thus turn to the scenario involving three jet ejections, 
%launched at separate times, 
in which the two late-time optical bumps arise from the interactions of two additional jets launched a few days after the initial one.
%Determining the opening angle of the slow wide jet is challenging, as no clear jet break is observed prior to the optical rebrightening. However, we find that variations in the opening angle of the wide jet do not significantly impact the multi-wavelength lightcurve. As a result, we fix the opening angle of the slow wide jet at $\theta_{\rm{j,s}} = 0.3$ rad. In contrast, the opening angle of the fast jet and the viewing angle can be well constrained within this model.
%To model the multi-wavelength light curves in optical, X-ray 
Following the approach outlined by \cite{Huang..2000}, we calculate the jet dynamics and associated synchrotron emission separately for each of the three jet components \citep[see also][]{Xu..2023}. The MCMC algorithm is employed to constrain the model parameters. The subscripts ``1,'', ``2,'' and ``3'' denote the first, second, and third jets, respectively. 
Determining the opening angles of all three jets is challenging owing to the absence of a clear jet break in the light curve. As a result, the opening angles of all three jets are fixed at 0.1\,rad, and assumed to be viewed on-axis. 
%with a viewing angle of $0$ rad.  

The best-fit models to match the multiwavelength light curves are plotted in the left panel of Figure \ref{fig:LC}. In the right panel of Figure \ref{fig:LC}, we illustrate the evolution of the kinetic energy, radius, and Lorentz factor for the three jets.
The best-fit results are presented in Table \ref{tab:fit}, while the corresponding corner plot is shown in Figure \ref{fig:corner}. The inferred initial Lorentz factors are $\Gamma_1 \approx 20$ for the first jet, $\Gamma_2 \approx 9$ for the second, and $\Gamma_3 \approx 7$ for the third, indicating that all three outflows are mildly relativistic. The Lorentz factors are not at odds with that constrained by the radio SED fittings and the equipartition analysis (Section 3.3), though only one jet shocking synchrotron emission component was assumed for the latter. The isotropic kinetic energies are estimated to be $1.1 \times 10^{54}$\,erg for the first jet, $7.5 \times 10^{52}$\,erg for the second, and  $5.0 \times 10^{52}$\,erg for the third. 
%Notably, the first jet has four times the kinetic energy and twice the Lorentz factor of the second, implying that the ejecta mass of the first jet is twice that of the second. 
The time interval between the first and second jets is constrained to $\sim 2.9$\,days in the rest frame, while the interval between the second and third jets spans about $7$ days in the rest frame. 
We note that the model underpredicts the X-ray emission, suggesting that an additional emission component is required, 
perhaps from the internal dissipation process within the jet powered by an active central engine \citep{Andreoni2022, Yao2024}. 
%internal dissipation within the jet Andreoni et al. 2022; Yao et al. 2024).
%such as an active central engine. 
We will discuss the interesting implications of the three-jet model in the next section. 
%The parameter constraints are shown in Figure \ref{fig:corner}, while the best-fit lightcurves are plotted in Figure \ref{fig:LC}. The best-fit Lorentz factor for the slow wide jet is found to be approximately $91$, indicating that the wide jet is too weak to produce detectable gamma-ray emission. This naturally explains the non-detection of EP241021a by Fermi-GBM and Konus-Wind \citep{Burns..2024, Svinkin..2024}. Conversely, the Lorentz factor of the fast narrow jet, observed off-axis, is about $500$. The opening angle of this jet is constrained to be $\theta_{\rm{j,f}} \sim 0.04$ rad, with a best-fit viewing angle of $\theta_{\rm{obs}} \sim 0.2$ rad. The strong relativistic beaming effect significantly suppresses its early afterglow emission, resulting in a steep late-time bump.

\subsubsection{Constraints on the black hole mass in the context of TDEs}

The long-lived X-ray emission may be attributed to a stellar tidal disruption, 
in which an extreme accretion episode onto a black hole produces relativistic jet ejections. 
Thus, \src could be a rare jetted TDE. 
So far only four TDEs with relativistic jets have been disclosed, 
with the latest being AT2022cmc \citep{Andreoni2022}. 
Jetted TDEs are generally characterized by luminous X-ray and radio emission, 
%as well as rapid short-term variability on a timescale of hours 
as well as rapid variability in the early X-ray emission on timescales as short as $\sim 100$\,s \citep{Mangano2016, Pasham2023}. 
In Swift J1644+57, the prototype of jetted TDEs, the X-ray flux can vary by more than two orders of magnitude in the first several days since trigger, with multiple brief flares lasting for $\sim 10^3$--$10^4$\,s \citep{Krolik2011}. %The short-duration flares have been explained by erratic 
The large-amplitude flares could result from geometric effects due to 
erratic { ``wobbling'' of the jet during the alignment process between the jet and black hole spin axis  \citep{Tchekhovskoy..2014, Teboul2023, Lu2024}. 
In this case, the prompt X-ray emission of \src may be related to a misaligned precessing jet, 
which can temporarily point along our line of sight, causing bright X-ray flares with durations of 
$\sim10^2$--$10^3$\,s \citep[e.g.,][]{Tchekhovskoy..2014}. }
It is possible that other flares (if present) may have lower luminosities and hence escaped detection by EP/WXT \citep{Yuan2025}. 
{ Note that in the context of TDEs, precessing jets are initially choked by an accretion-disk wind, 
which can only break out of the wind confinement if the misalignment angle is sufficiently small, 
such as $\theta_{\rm LS}\simlt15^{\circ}-20^{\circ}$ \citep{Lu2024}. 
}
The observed duration for the prompt emission ($\sim100$\,s) of \src can be used to place 
an upper limit on the size of the X-ray-emitting region, $R_{\rm X}\simlt c \Gamma_{\rm j}^2 t_{\rm var}/(1+z)=1.7\times10^{14}(\Gamma_{\rm j}/10)^2$\,cm.  
%If equating the size of the X-ray-emitting region 
By requiring that $t_{\rm var}/(1+z)$ exceeds the light-crossing time of the Schwarzschild radius of a black hole with mass $M_{\rm BH}$, we find 
%an upper limit on the mass can be derived as 
$M_{\rm BH}\simlt5.7\times10^6$\,\msun.     

%\begin{equation}
%\begin{equation}

%mass constraints from the variability time-scale?? 
%the tidal disruption occurred earlier than the WXT trigger, 
%and the non-detection of other high-luminosity flares could be due to the lower flux hence 

%Another remarkable feature of jetted TDEs is a sudden drop in their X-ray light curves 
%[100--400 days in the rest frame][]\citep{Eftekhari2024}. 

%on top of a powerlawdecay. % of $t^{-2}$. 
%\srcs's apparent X-ray luminosityand evolution are only comparable to Sw J1644+57 
%The long-lived X-ray emission can be naturally explained by an 
%extreme accretion episode onto a black hole, 
%stellar tidal disruption event, in which the extreme accretion 
%\srcs's high apparent X-ray and radio luminosity, extreme X-ray luminosity variations  and fast va 
%The scenario of \src may be similar to that of AT2022cmc, where the optical and radio emissions arise from shock-accelerated electrons as jet propagated into the surrounding gas, and the X-ray 
%lightcurve at late time 
%emission is dominated by internal dissipation within the jet %an optically thick outflow 
%associated with the TDE 
%\citep{Andreoni2022, Yao2024}, 
%though for the latter a single steady jet was invoked. 
%The X-ray plateau %is also plausible in the framework of TDE 
%can also be explained in the framework of TDE if the accretion timescale is much longer than the fallback timescale, preventing the returned mass from being digested promptly by the black hole \citep{Peng..2019}. 

%\subsection{Estimating the mass of the disrupting }
Our deep XMM-Newton limit at $t_{\rm rest}\approx 50$\,days indicates a decrease in the X-ray flux by a factor of $\simgt10$ and a deviation from the earlier power-law decline. 
Such a sudden decrease could be interpreted as an
accretion-state transition from super- to sub-Eddington, 
%where the jet activity cease
%attributed 
leading to the cessation of jet ejections, 
as observed in other jetted TDEs \citep{Eftekhari2024}.
In the scenario of a TDE, we can estimate the black hole mass ($M_{\rm BH}$) of \src 
by equating the X-ray luminosity at shut-off to the Eddington luminosity \citep{Eftekhari2024}, 
such that 
\begin{equation}
    M_{\rm BH}=8.4\times10^4\frac{L_{\rm jet, off}}{10^{45}\,\rm erg\,s^{-1}}\left(\frac{\epsilon_{\rm disk}}{\epsilon_{\rm jet}}\right) f^{-1}_{\rm beam, 200}f_{\rm bol, 3}\,{\rm M}_\odot\, ,
\end{equation}
\noindent
where ${L_{\rm jet, off}}$ is the breaking luminosity 
at the time of jet shut-off ($\sim2.5\times10^{43}$\,\erg). 
The terms
$\epsilon_{\rm disk}$ and $\epsilon_{\rm jet}$ represent the radiative efficiencies of the
disk and jet, respectively. 
As before, we adopt a jet half-opening angle $\theta_c=0.1$\,rad, yielding the jet beaming factor 
$f_{\rm b}=200$. 
Following \citet{Eftekhari2024}, by assuming $\epsilon_{\rm jet}/\epsilon_{\rm disk}\simgt0.1$ and a bolometric correction to the jet luminosity $f_{\rm bol}=3f_{\rm bol, 3}$, 
%the breaking luminosity from our MCMC fit ($L_{\rm X}\approx??$) suggests 
we obtain a black hole mass $M_{\rm BH}\simlt10^4$\,\msun, making \src  
a possibly jetted TDE involving an intermediate-mass black hole (IMBH). 
%Comparison between \srcs and AGN/other transients???
%Separate from the luminosity at jet shut-off, 
%Besides the luminosity, 
%we can also place constraint on the black hole mass from the timescale of jet shut-off, 
%which is given by 
On the other hand, if the fast rise time ($t_{\rm rest}\approx 1.3$\,d) in the second optical bump is related to the fallback timescale 
of the disrupted stellar material, which is given by 

\begin{equation}
 %   t_{\rm jet, off}=t_{\rm fb}\lambda^{1/\alpha}_{\rm peak, BH}
 t_{\rm fb}=(3.5\times10^5)\, \left(\frac{M_{BH}}{10^{4}\,{\rm M}_{\odot}}\right)^{1/2}\left(\frac{m_{*}}{{\rm M}_{\odot}}\right)^{-1}\left(\frac{r_{*}}{{\rm R}_{\odot}}\right)^{3/2}\,{\rm s},
\end{equation}
where $m_{\rm *}$ is the mass of the disrupted star and $r_{*}$ represents its radius for a main-sequence star,  
%the peak Eddington ratio, which is an order of $\approx100$ for %super-Eddington accretion onto 
%a tidal disruption of a main-sequence star by a black hole with mass $M_{\rm BH}=10^6$\msun  \citep{Stone2013}. 
%With the powerlaw decay index $\alpha\simlt2$ as
%constrained from our MCMC analysis, the inferred jet shut-off
%time ($t_{\rm jet, off}\sim30$ days) requires a short fallback
%time of a few days, 
the implied black hole mass is $\sim 10^3$\,\msun\ for a solar-type star, 
supporting the scenario of a TDE by an IMBH. 
%which can be accommodated by a lower mass black hole of 
%$M_{\rm BH}\approx10^4$\msun. 

Finally, at the redshift of \srcs, the host galaxy is not detected in deep imaging 
with the DESI-Legacy survey down to a limiting $r$-band magnitude of 24.5. 
%More sensitive limit to the host has been obtained 
The host may be fainter than 24.9 mag as observed by the VLT \citep{Busmann..2025}, 
corresponding to an upper limit on the stellar mass of $<3\times10^9$\,\msun.  
%({\bf need check by JN}). 
We use a galaxy bulge vs. black hole mass relation and the upper limit
on \srcs's host-galaxy mass to obtain an upper limit on the BH mass,
$M_{\rm BH} < 10^6$\,\msun, which is 
not at odds with the mass estimates 
inferred from the timescale and luminosity of the jet shut-off. 
%probe the low-mass regime where our models are favored. 
%Recurring TDEs may serve as the central engine powering the two jets. The kinetic energy of the jets is approximately $\sim E_{\rm k, iso} \Gamma^{-2} \simeq 10^{51}\ {\rm erg}$, where $E_{\rm k, iso}$ and $\Gamma$ denote the isotropic kinetic energy and Lorentz factor, respectively. The accretion of the stellar debris by an IMBH can generate energy of $\sim 10^{52}\ (\Delta M/0.01\ M_{\odot}) c^2\ {\rm erg}$, which is sufficient to fuel the jets.

%The radiative light curve, resulting from the central accretion disk or jet radiation, is believed to approximated the mass fallback rate of a TDE \citep{Rees1988,Ramirez-Ruiz_THE_2009,Chen_Tidal_2018,Mockler_Weighing_2018}. 

%It is noteworthy that there is a possible brightening in the X-ray emission around $t_{\rm rest}\approx 5$ days, 
%after which the light curve breaks into a steeper decay (Figure \ref{fig:x-raylc} \& Figure \ref{fig:LC}). 
%This X-ray brightening appears to be coincident with the optical rebrightening, %with the latter being better sampled. 
%which displays a fast rise to peak on a timescale of only $\sim $1 day. 
%One scenario which can explain 

\subsubsection{Energy and timescale considerations for repeating partial TDEs}

To account for the rebrightening of both the X-ray and optical emission, we invoke a scenario that a star is tidally disrupted by the IMBH  more than once --- repeating partial TDEs. %not only once but twice.
The timescale of X-ray and optical rebrightenings will then correspond to the orbital period of the survived stellar core after the first disruption. 
%Given the short timescale ($\sim 4.5$ day in rest frame), the disrupted star is likely a white dwarf \citep{Zalamea_White_2010,Chen_tidal_2023} 
We consider a main-sequence star initially on a slightly bound orbit. 
Following each disruption, the accretion of stellar debris onto the IMBH powers a relativistic jet, 
%This picture is also consistent with the two-jet model discussed previously. 
which accommodates the model of multiple jet ejections discussed previously. 
The kinetic energy of the jets is $E_{\rm k, iso} \Gamma^{-2} \approx 10^{51}\,{\rm erg}$, where $E_{\rm k, iso}$ and $\Gamma$ denote the isotropic kinetic energy and Lorentz factor, respectively. The accretion of the stellar debris by an IMBH can generate energy of $\sim 10^{52}\,(\Delta M/0.01\ {\rm M}_{\odot})\,c^2\ {\rm erg}$, which suggests that stellar material of mass as low as $\sim 0.01\, {\rm M}_{\odot}$ needs to be tidally stripped off each time in order to fuel the jet (with a jet energy efficiency of $\sim10\%$.)

Based on this picture, we conduct theoretical calculations to match three key constraints placed by observations. (1) The orbital timescale of the stellar core, after the first disruption, should match the timescales of the X-ray and optical rebrightenings. (2) The overall debris fallback rate $\dot{M}_{\rm fb}$  should decay from super-Eddington to sub-Eddington around the time that the jet is observed to shut off. (3) The accretion of the stellar materials is sufficient to power the observed jet energy. 

For the calculation of $\dot{M}_{\rm fb}$ for stars in a bound orbit, we modified the parabolic fallback rate that is given by \cite{Guillochon_Hydrodynamical_2013} (assuming a $\gamma=5/3$ polytropic star), by shifting the debris specific energy $E$ to $E + E_0$, where $E_0$ is the specific orbital energy of the star. Now, $\dot{M}_{\rm fb}$ is given by $\dot{M}_{\rm fb} = dM/dE \times dE/dt$ with shifted $E$, aligning with the simulations of eccentric TDEs \citep{Hayasaki_Finite_2013}. We also account for the change of orbital energy of the stellar remnant during a partial disruption following the approach of \citet{chen_fate_2024}. 

We find that the observational constraints  mentioned above can be reasonably well satisfied using the model of a $10^4\,{\rm M}_{\odot}$ IMBH partially disrupting a solar-like star. The star initially approaches the IMBH along a bound orbit with eccentricity $e \approx 0.897$ and impact factor $\beta \approx 0.82$. After the first encounter, the star goes through a partial TDE, and the surviving core returns for a second encounter with $e \approx 0.9$ and impact factor $\beta \approx 0.826$, followed by a second partial TDE. After the second TDE, the suviving core returns with $e \approx 0.903$ and impact factor $\beta \approx 0.93$, leading to a complete disruption. To reconcile the energy ratio of the three jets,  $(\Delta M_1 c^2)/(\Delta M_2 c^2) \approx (E_{\rm k, iso, 1}/E_{\rm k, iso, 2})(\Gamma_2/\Gamma_1)^2 \approx 2.4$ and $(\Delta M_2 c^2)/(\Delta M_3 c^2) \approx (E_{\rm k, iso, 2}/E_{\rm k, iso, 3})(\Gamma_3/\Gamma_2)^2 \approx 1.6$, the accreted stellar masses in these three encounters are $\sim 0.6 \, {\rm M}_{\odot}$, $\sim 0.25 \, {\rm M}_{\odot}$, and $0.15\, {\rm M}_{\odot}$, respectively. We note that the stellar mass-radius relation should deviate from that of a main-sequence star after the encounter. In our calculation, the stellar density has a slight decrease after each encounter, consistent with tidal effects such as tidal heating and tidal spin-up causing the star to expand.
%We find that the observational constraints as mentioned above can be reasonably well satisfied using the model of a $10^4~M_{\odot}$ IMBH partially disrupting a main-sequence star with a mass of $1.6 M_{\odot}$. The star initially approaches the IMBH along a slightly bound orbit with eccentricity $e \simeq 0.9936$ and impact factor $\beta \simeq 0.8$. After the first encounter, the star goes through a partial TDE, and the survived core orbits back for the second time with $e \simeq 0.994$ and impact factor $\beta \simeq 1.34$ and gets fully disrupted. The resultant accreted stellar masses in these two encounters are $\sim 0.053 \ M_{\odot}$ and $0.047\ M_{\odot}$, respectively. This mass ratio reconciles the energy ratio of the two jets, i.e., $(\Delta M_1 c^2)/(\Delta M_2 c^2) \simeq (E_{\rm k, iso, 1}/E_{\rm k, iso, 2})(\Gamma_2/\Gamma_1)^2 \simeq 1.1$.

\begin{figure}[t!]
    \epsscale{1.1}
    \plotone{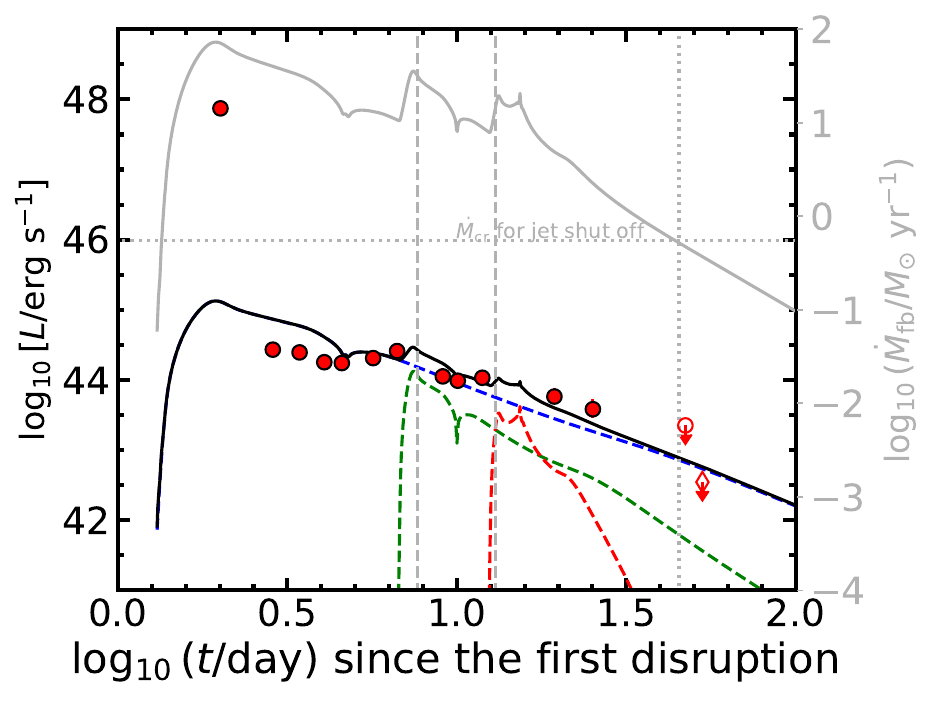}
    \caption{
        Isotropic luminosity of repeating TDEs resulting from a relativistic jet, and the X-ray luminosity in the 0.3--10\,keV band. Circular markers denote EP observations, while diamond markers represent XMM-Newton data. Open markers indicate $3\sigma$ upper limits. The blue, green, and red dashed lines show the isotropic light curves of the first, second, and third TDEs, respectively. The total light curve (sum of three events, Eq. \ref{eq:L_TDE}) is illustrated by the solid black line. The gray solid line represents the mass-fallback rate $\dot{M}_{\rm fb}$. The horizontal gray line indicates the critical value of $\dot{M}_{\rm cr}$, marking the transition from super-Eddington to sub-Eddington accretion state. The two vertical dashed lines denote the rebrightening time, and the dotted gray line indicates the time at which the jet shuts off.
            }
    \label{fig:repeating_TDEs}
\end{figure}

We also show the comparison between $\dot{M}_{\rm fb}$ and the observed data in Figure \ref{fig:repeating_TDEs}. One can see that the overall fallback rate behavior can well explain the observed rebrightening at $\sim 5$ and $12$ days, as well as the accretion-state transition observed at $\sim 50$\,day after the initial trigger. Subsequent to this transition, $\dot{M}_{\rm fb}$ falls below the critical value $\dot{M}_{\rm cr}$, leading to the jet shut-off. The value of $\dot{M}_{\rm cr}$ is given by $\dot{M}_{\rm cr} \approx L_{\rm Edd}/c^2 (R_{\rm disk}/R_{\rm Sch})$ \citep{shen_evolution_2014}, where $L_{\rm Edd}$, $R_{\rm disk}$, and $R_{\rm Sch}$ are respectively the Eddington luminosity, disk radius, and Schwarzchild radius of the IMBH. To match the timing of the X-ray sharp decay, we find that the disk size should be $R_{\rm disk} \approx 20\,R_{\rm p}$, where $R_{\rm p}$ is the pericenter radius of the stellar orbit. This value is reasonable for a typical TDE, where the initial disk experiences modest expansion due to viscous shear.

The inferred isotropic luminosity is given by 
\begin{equation} \label{eq:L_TDE}
  L \approx \eta \dot M_{\rm fb}c^2 \Gamma^2,
\end{equation}
\noindent
where $\eta$ is the efficiency of converting accretion power into jet radiative energy, and the values of $\Gamma$ for the three jets are taken to be $18$, $9.3$, and $6.8$, as derived from previous calculations. For matching the overall level of the X-ray flux, a very small $\eta$ value ($\sim 10^{-6}$) is needed, suggesting that the jet energy is mostly in the form of the kinetic energy %(and electromagnetic energy) 
instead of radiative energy, which is consistent with recent simulations of super-Eddington jets \citep{dai_unified_2018,curd_grrmhd_2019}.

%\subsection{Active galactic nucleus}
%Some bright active galactic nuclei (AGN) have X-ray luminosities as high as $L_{\rm X} = 10^{46} $\erg，
%and display rapid X-ray variability on timescales of hours to
%days \citep{Uttley..2005}. However, few AGN are known to show the similar variability of more than a factor of 1000 
%and X-ray flare with a duration 
%of $\sim$100 s 
%to that observed in \srcs.  
%clearly disfavorthe interpretation of the source as an AGN. 
%Moreover, the most
%X-ray luminous AGN also display bright optical luminosities
%of $\approx$--28 to --30 AB mag, corresponding to an 
%apparent magnitude of $\approx$13-15 mag for \srcs, which %would be detected for \src by
%is inconsistent with the non-detections by the pre-flare DESI legacy survey down to $\sim$24 mag. 
%We therefore strongly disfavor an AGN flare as the cause of \srcs. 

%Although luminous fast blue optical transients (LFBOTs) display slow fading X-ray emission before a sharp decay that is similar to \src \citep{Margutti..2019}, this scenario is also disfavored due to the extremely high plateau X-ray luminosity of \src as well as its red optical color in the peak, 
%which is unlike the prototypical LFBOT AT2018cow. 

\section{Conclusion} \label{sec:conclusion}

We report the discovery of a peculiar X-ray transient, \srcs, by EP/WXT on Oct. 21, 2024, 
and the results of follow-up multiwavelength observations including X-ray, optical, and radio. 
Our sensitive X-ray observations with EP/FXT and XMM-Newton reveal 
long-duration X-ray emission lasting for at least one month, 
which is unprecedented among known extragalactic fast X-ray transients. 
%that after 1.5 days since trigger, the X-ray emission drops in flux significantly by a factor of more than three orders of magnitude, settling into a nearly plateau phase of $\approx$7 days. 
%Further X-ray decline is found lasting for $\approx$30 days, followed by a remarkable 
%drop to a faint flux of $<$$2.8\times10^{-15}$\ergs.  
%Such a long-duration X-ray afterglow emission associated with an extragalactic fast X-ray transient is unprecedented, possibly representing a new type of X-ray transients with
%intermediate timescales. 
Both optical and radio counterparts are detected within several days after the trigger, 
with high peak luminosities in excess of $10^{44}$\,\erg and $10^{41}$\,\erg, respectively. 
These properties, together with the nonthermal X-ray spectrum, red optical color, X-ray and optical rebrightenings, and fast radio spectral evolution, suggest that relativistic jets may have been launched, possibly driven by an active central engine. 
We have considered a variety of scenarios as origins of \srcs, and
favor a magnetar engine or jetted TDE, although neither can perfectly explain the multiwavelength properties. 
%explains the observations.
%with peak luminosities as high as $10^{44}$\erg and $10^{41}$\erg respectively.
%in line with the results from the above 
%For a main-sequence star, 
%jetted TDE; 
%mass from host galaxy ??
%and light curve... 

Observations with the Hubble Space Telescope, and possibly with the James Webb Space
Telescope, are encouraged; they should have sufficient sensitivity to potentially unveil the faint host galaxy, once the transient emission has { fully} disappeared. 
%, allow for better constraining the host hence the black hole mass.  
%If a host galaxy is eventually detected, 
The detection of the host will enable further exploration of whether  
the transient position (for example, the radio position from our high-resolution VLBA observations) is spatially coincident with the host light centroid. 
This is crucial for understanding the nature of \srcs, from a rare jetted TDE involving an IMBH or an off-nuclear extremely unusual stellar explosion. 
%We note that the observed peak X-ray flux of the initial trigger of \src is
%a factor of higher than the brightest soft X-ray detection of Sw J1644+57.
%Finally, due to is large distance, 
%Future observations of FEXTs discovered by EP 
%\section{Conclusion} \label{sec:style}
With its unique capability of a wide-field survey and rapid X-ray follow-up observations, 
EP will discover more transients similar to \src with intermediate evolution timescales of months. 
Timely and deep multiwavelength monitoring are essential for further characterizing and uncovering their intriguing properties as a potentially new phenomenon. % and uncover their physical origins. 

%\src was first discovered as a bright X-ray transient by SRG/eROSITA \citep{Bykov2023}.  
%As mentioned in \citet{Bykov2023}, the object has also been named as an optical transient by Gaia Alerts Team (AT2018cqh) prior to the X-ray discovery. 
%By examining the optical light curves obtained by Zwicky Transient Facility \citep[ZTF,][]{Bellm2019}, \citet{Bykov2023} 
%found provides almost continuous data for this object from mid 201818 and
%is shown in Fig. 10 along with X-ray light curve.

%Detailed analysis of the host properties of \src will be presented in Section 3.1
  %All spectra are shown in Figure 2. 

\acknowledgments{
%The authors thank the VLA operations staff for their
%assistance in scheduling and performing the observations. 
%understand their physical origins.
This work was based on data obtained with Einstein Probe, a space mission
supported by the Strategic Priority Program on Space Science of the Chinese
Academy of Sciences, in collaboration with ESA, MPE, and CNES (grant
 XDA15310000), the Strategic Priority Research Program of the Chinese
Academy of Sciences (grant XDB0550200), and the National
Key R\&D Program of China (grant  2022YFF0711500).

%The data presented in this paper are based on observations 
The data presented in this paper are based in part on observations
made with the Karl G. Jansky Very Large Array, % from the program VLA/4B-487,  
the Very Long Baseline Array, % from the project VLBA/24B-491, 
the Australia Telescope Compact Array, % from the project CX585, 
 MeerKAT, % from the project DDT-20250123-XS-01, 
and XMM-Newton. % (Obsid: 095419090).
%We thank the staff of the VLA, VLBA, ATCA and MeerKAT that made these observations possible. 
%made with the Karl G. Jansky Very Large Array, the Australia Telescope Compact Array, the , and the European Space Agency space mission XMM-Newton. 
 The National Radio Astronomy Observatory is a facility of the U.S.
National Science Foundation operated under cooperative agreement
by Associated Universities, Inc.  
The Australia Telescope Compact Array is part of the Australia
Telescope National Facility (https://ror.org/05qajvd42) which is funded by
the Australian Government for operation as a National Facility managed by
CSIRO. 
The MeerKAT telescope is operated by the South African Radio Astronomy Observatory, which is a facility of the National Research Foundation, an agency of the Department of Science and Innovation.
We acknowledge the use of the ilifu cloud computing facility (\url{www.ilifu.ac.za}) and the Inter-University Institute for Data Intensive Astronomy (IDIA), a partnership of the University of Cape Town, the University of Pretoria, and the University of the Western Cape. 
 This work has made use of the ``MPIfR S-band receiver system'' designed, constructed, and maintained by funding of the MPI f{\''u}r Radioastronomy and the Max-Planck-Society. 
XMM-Newton is an ESA science mission with instruments and
contributions directly funded by ESA Member States and NASA.

Some of the data presented herein were obtained at the W. M. Keck
Observatory, which is operated as a scientific partnership among the
California Institute of Technology, the University of California, and
NASA; the observatory was made possible by the generous financial
support of the W. M. Keck Foundation.
KAIT and its ongoing operation were made possible by donations from Sun Microsystems, Inc., the Hewlett-Packard Company, AutoScope Corporation, Lick Observatory, the U.S. NSF, the University of California, the Sylvia \& Jim Katzman Foundation, and the TABASGO Foundation. Research at Lick Observatory is partially supported by a generous gift from Google, Inc. 
Mephisto is developed at and operated by the South-Western Institute for Astronomy Research of Yunnan University (SWIFAR-YNU), funded by the ``Yunnan University Development Plan for World-Class University" and ``Yunnan University Development Plan for World-Class Astronomy Discipline". 

 X.S. is supported by the National SKA Program of China (2022SKA0130102), and the National Science Foundation of China (NSFC) through grants 12192220 and 12192221. 
 J.H.C. and L.D. acknowledge the support from the Hong Kong Research Grants Council (HKU106220154, 109000722).
 C.J. acknowledges the National Natural Science Foundation of China through grant 12473016.
 Y.X. acknowledges support of NSFC 12025303. 
 P.O.B. acknowledges support from
UKRI under grant ST/W000857/1.
R.S. acknowledges INAF grant 1.05.23.04.04.
A.K. and J.R.M. are supported by the UK Science and Technology Facilities Council (STFC) Consolidated grant ST/V000853/1.
F.P. acknowledges support from the Spanish Ministerio de Ciencia, Innovación y Universidades (MICINN) under grant PID2022-141915NB-C21. 
F.X.A. acknowledgessupport from the National Natural Science Foundation of China (12303016) and the Natural Science Foundation of Jiangsu Province (BK20242115). 
T.A. acknowledges the support of the Xinjiang Tianchi Talent Program and FAST Special Program (NSFC 12041301).
A.V.F.’s research group at UC Berkeley acknowledges financial assistance from the Christopher R. Redlich Fund, Gary and Cynthia Bengier, Clark and Sharon Winslow, Alan Eustace and Kathy Kwan, William Draper, Timothy and Melissa Draper, Briggs and Kathleen Wood, Sanford Robertson (W.Z. is a Bengier-Winslow-Eustace Specialist in Astronomy, T.G.B. is a Draper-Wood-Robertson Specialist in Astronomy, Y.Y. was a Bengier-Winslow-Robertson Fellow in Astronomy), and numerous other donors. 
BK acknowledges the support from the ``Special Project for High-End Foreign Experts", Xingdian Funding from Yunnan Province, the Key Laboratory of Survey Science of Yunnan Province (202449CE340002), and the National Key Research and Development Program of China (Grant No. 2024YFA1611603). 
}
%\end{acknowledgments}

\software{ 
%MOSFiT \citep{Guillochon2018}, 
Autophot \citep{autophot.ref}, 
Astropy \citep{Astropy2013, Astropy2018, Astropy2022}, 
%DiFX software correlator \citep{Deller2011}, DIFMAP \citep{Shepherd1997}, GILDAS\footnote{\url{https://www.iram.fr/IRAMFR/GILDAS}}
AIPS \citep{Greisen2003}, 
CASA \citep[v6.6.1; ][]{McMullin2007}, %AIPS \citep{Greisen2003}, 
DiFX software correlator \citep{Deller2011}, 
DIFMAP \citep{Shepherd1997}, 
IRAF \citep{1986SPIE..627..733T}, 
LPipe \citep{Perley2019}, 
Miriad \citep{RJSault..1995},
OxKAT \citep{Heywood2020},
Pypeit \citep{pypeit.ref}. 
 }

\clearpage

 \bibliographystyle{aasjournal}
\bibliography{ep241021a_ms_v1.bib}

\clearpage

\appendix   %仅一个附录时用appendix，否则\appendices

%\section{Log of optical photometry observations}
\section{Summary of results from optical photometry and spectroscopy}
\setcounter{table}{0}   %从0开始编号，显示出来表会A1开始编号
\renewcommand{\thetable}{A\arabic{table}}
\setcounter{figure}{0}
\renewcommand{\thefigure}{A\arabic{figure}}

\begin{figure*}[htbp!]
%	\epsscale{0.8}
    \plotone{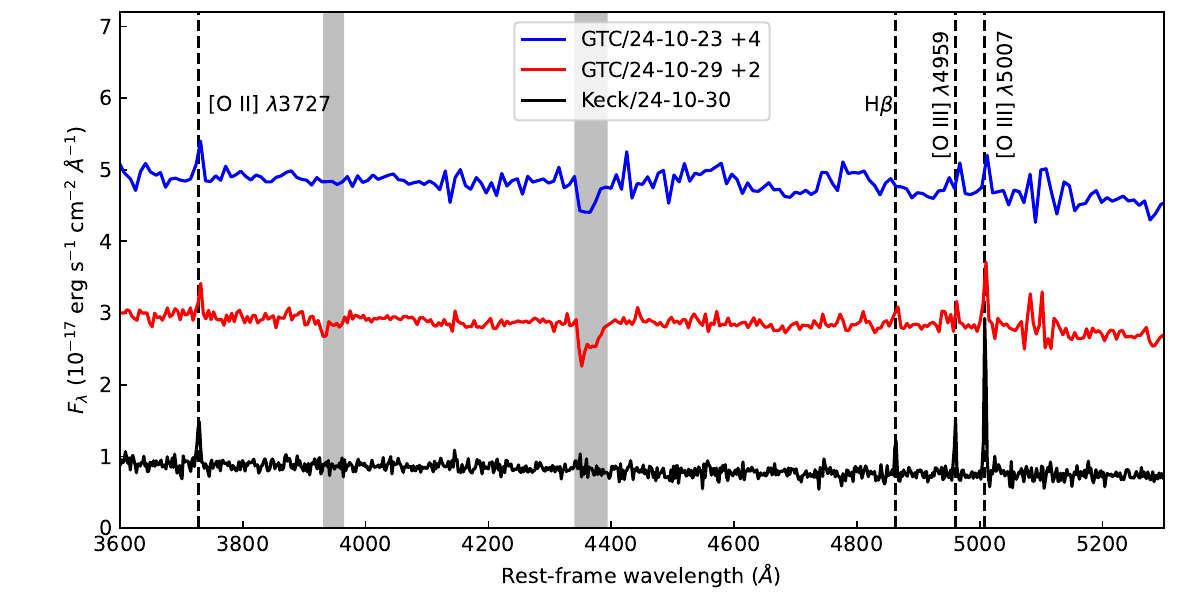}
    \caption{
        {\bf The optical spectrum observed by GTC and Keck on $\Delta t=2$, 8, and 9 days since the discovery of \srcs. 
        %While the GTC spectral quality is relatively poor, 
        While the GTC spectra have lower resolution and lower S/N than the Keck spectrum,
        %due to differences in the observation strategy, 
        the narrow emission lines of {\sc [O iii]} $\lambda5007$, {\sc [O iii]} $\lambda$4959, and {\sc [O ii]} $\lambda$3727 are detected, 
        confirming the Keck redshift of $z=0.7478$. }
        %with no obvious broad components. }
        %Parameters derived for EP241021a by using the three-jet model. The contour curves represent the $1\sigma-2\sigma-3\sigma$ confidence levels. The best-fit parameter values, along with their $1\sigma$ uncertainties, are indicated above the panels of their respective posterior distributions.		
            }
    \label{fig:gtc}
\end{figure*}

The log of optical photometry observations and results is given in Table \ref{tab:obs_optical}. 
{ Figure \ref{fig:gtc} shows the comparison of the optical spectra observed by GTC and Keck (Section 2.4). }

\begin{longtable}{cccc}
\caption{Summary of optical observations}\label{tab:obs_optical} \\
\hline
\hline
\colhead{Time (days)} & 
\colhead{Band} & 
\colhead{Magnitude\textsuperscript{a}} & 
\colhead{Telescope} \\
\hline
\endfirsthead

\multicolumn{4}{c}{{\bfseries Table~\thetable{} (continued)}} \\
\hline
\hline
\colhead{Time (days)} & 
\colhead{Band} & 
\colhead{Magnitude} & 
\colhead{Telescope} \\
\hline
\endhead

\hline
\multicolumn{4}{r}{{Continued on next page}} \\
\endfoot

\hline
\multicolumn{4}{l}{\textsuperscript{a}The magnitude has not been corrected for Galactic extinction.} \\
\endlastfoot
%\hline
%\multicolumn{4}{l}{
%\begin{flushleft}
%\textbf{Notes.}\\
%\tablenotetext{a}{The magnitude has not been corrected for Galactic extinction.}
%\end{flushleft}
%} \\
%\endlastfoot
    1.13 & clear & $>18.3$ & BOOTES-7\\
    1.14 & g & $>17.5$ & BOOTES-7\\
    1.14  & clear & $21.20 \pm 0.20$  & KAIT(Vega)\\
    1.15 & r & $>18.8$ & BOOTES-7\\
    1.17 & V & $21.70 \pm 0.30$ & TRT-SRO \\ 
    1.17 & R & $21.60 \pm 0.30$ & TRT-SRO \\ 
    1.76 & r & 21.55 $\pm$ 0.13 & LT \\
    1.77 & g & 22.04 $\pm$ 0.22 & LT \\
    1.77 & z & $21.60 \pm 0.11$ & NOT \\ 
    1.91 & r & $21.95 \pm 0.06$ & NOT \\ 
    2.07 & r & 21.88 $\pm$ 0.11 & LCO \\
    2.08 & i & 21.73 $\pm$ 0.11 & LCO \\
    2.12  & clear & $21.70 \pm 0.30$  & KAIT(Vega)\\
    2.7&r& $22.29\pm0.61$& LCO \\
    2.71&g& $22.38\pm0.44$& LCO \\
    2.76 & r & $21.96 \pm 0.03$ & NOT \\ 
    2.79&i& $21.86\pm0.51$ & LCO \\
    2.80 	&	g	&	22.06 	$\pm$	0.18 	&	LT \\
    2.81 	&	r	&	21.83 	$\pm$	0.30 	&	LT \\
    2.82 	&	i	&	21.60 	$\pm$	0.18 	&	LT \\
    2.82 & z & $21.72 \pm 0.11$ & NOT \\ 
    2.83 	&	z	&	21.78 	$\pm$	0.20 	&	LT \\
    2.93 	&	R	&	21.92 	$\pm$	0.16 	&   1.5m-OSN \\
    3.72 	&	g	&	22.52 	$\pm$	0.35 	&	LT \\
    3.73 	&	i	&	21.92 	$\pm$	0.19 	&	LT \\
    3.77 & r & $22.30 \pm 0.04$ & NOT \\ 
    3.79 & z & $22.07 \pm 0.12$ & NOT \\ 
    5.90 & r & $>22.4$ & NOT \\ 
    6.55 & r & $>20.3$ & ALT-50A\\
    7.46 & R & $>21.2$  & TRT-SBO \\ 
    7.76 	&	g	&	21.73 	$\pm$	0.07 	&	LT \\
    7.78 	&	i	&	21.27 	$\pm$	0.06 	&	LT \\
    7.90 	&	r	&	21.58 	$\pm$	0.12 	&	LT \\
    7.92 & r & $21.64 \pm 0.03$ & NOT \\ 
    7.92 	&	z	&	20.93 	$\pm$	0.16 	&	LT \\
    7.93 & z & $21.31 \pm 0.08$ & NOT \\ 
    8.26&r& $21.5\pm0.27$ & LCO \\
    8.27&g& $21.82\pm0.25$& LCO \\
    8.28&i& $21.12\pm0.35$ & LCO \\
    { 8.42} & v$_m$ & $22.10 \pm 0.38$ & Mephisto \\
    { 8.42} & r$_m$ & $21.32 \pm 0.21$ & Mephisto \\
    { 8.42} & z$_m$ & $>21.29$ & Mephisto \\
    { 8.43} & u$_m$ & $>22.48$ & Mephisto \\
    { 8.43} & g$_m$ & $>22.92$ & Mephisto \\
    { 8.43} & i$_m$ & $>22.33$ & Mephisto \\
    8.86&r& $21.77\pm0.45$ & LCO \\
    8.9&g& $21.98\pm0.35 $& LCO \\
    8.96&g& $21.96\pm0.36 $& LCO \\
    8.93 	&	g	&	21.69 	$\pm$	0.20 	&	LT \\
    8.96 	&	z	&	21.04 	$\pm$	0.32 	&	LT \\
    8.96 & B & $ >22.10 $ & TRT-SRO \\ 
    8.89&i& $21.42\pm0.38$ & LCO \\
    8.98&r& $21.77\pm0.29 $& LCO \\
    8.99&i& $21.38\pm0.33 $& LCO \\
    8.99 & V & $21.14 \pm 0.25$ & TRT-SRO \\ 
    9.01 & R & $21.35 \pm 0.34$ & TRT-SRO \\ 
    9.03 & I & $>20.3$ & TRT-SRO \\ 
    9.3&g& $22.11\pm0.27$& LCO \\
    9.33&r&$21.79\pm0.29$& LCO \\
    9.33&i&$21.45\pm0.37$& LCO \\
    { 9.42} & v$_m$ & $22.12 \pm 0.30$ & Mephisto \\
    { 9.42} & r$_m$ & $22.03 \pm 0.28$ & Mephisto \\
    { 9.42} & z$_m$ & $>21.47$ & Mephisto \\
    9.62&g& $22.27\pm0.41$& LCO \\
    9.63&r&$22.15\pm0.44$& LCO \\
    9.63&i&$22.14\pm0.77$& LCO \\
9.64 & g & 22.03 $\pm$ 0.11 & LCO \\
9.65 & r & 21.67 $\pm$ 0.13 & LCO \\
9.71 	&	g	&	21.89 	$\pm$	0.21 	&	LT \\
9.72 	&	i	&	21.25 	$\pm$	0.08 	&	LT \\
9.73 	&	r	&	21.66 	$\pm$	0.31 	&	LT \\
9.73 	&	z	&	20.95 	$\pm$	0.22 	&	LT \\
    9.78 & r & $21.77 \pm 0.03$ & NOT \\ 
    9.80 & z & $21.33 \pm 0.06$ & NOT \\ 
9.86&r&$21.87\pm0.49$& LCO \\
9.88&i&$21.59\pm0.43$& LCO \\
9.88&g&$21.96\pm0.31$& LCO \\
9.96&g&$22.1\pm0.33$& LCO \\
9.96&i&$21.98\pm0.65$& LCO \\
9.97&r&$21.84\pm0.38$& LCO \\   
10.35&g&$22.34\pm0.45$& LCO \\
10.38&i&$21.65\pm0.57$& LCO \\
{\bf 10.38} & v$_m$ & $22.48 \pm 0.35$ & Mephisto \\
{\bf 10.38} & r$_m$ & $21.56 \pm 0.24$ & Mephisto \\
{\bf 10.38} & z$_m$ & $>21.54$ & Mephisto \\
10.39&r&$22.17\pm0.57$& LCO \\
10.74 	&	g	&	22.25 	$\pm$	0.22 	&	LT \\
10.74 	&	i	&	21.33 	$\pm$	0.11 	&	LT \\
10.75 	&	r	&	21.45 	$\pm$	0.16 	&	LT \\
10.76 	&	z	&	21.11 	$\pm$	0.17 	&	LT \\
11.72 	&	g	&	22.12 	$\pm$	0.16 	&	LT \\
11.73 	&	i	&	21.92 	$\pm$	0.14 	&	LT \\
11.74 	&	r	&	21.70 	$\pm$	0.13 	&	LT \\
11.74 	&	z	&	21.33 	$\pm$	0.49 	&	LT \\
12.74 & r & $22.40 \pm 0.07$ & NOT \\ 
{\bf 13.36} & v$_m$ & $22.54 \pm 0.40$ & Mephisto \\
{\bf 13.36} & r$_m$ & $22.49 \pm 0.38$ & Mephisto \\
{\bf 13.36} & z$_m$ & $>21.37$ & Mephisto \\
13.88 & z & $>21.8$ & NOT \\ 
13.97 & r & 22.53 $\pm$ 0.20 & LCO \\
13.98 & i & 21.86 $\pm$ 0.15 & LCO \\
{\bf 14.37} & v$_m$ & $>21.90$ & Mephisto \\
{\bf 14.37} & r$_m$ & $>22.42$ & Mephisto \\
{\bf 14.37} & z$_m$ & $>20.65$ & Mephisto \\
14.72 	&	g	&	22.54 	$\pm$	0.11 	&	LT \\
14.73 	&	i	&	21.86 	$\pm$	0.10 	&	LT \\
14.85 	&	g	&	22.53 	$\pm$	0.25 	&	LT \\
14.86 	&	r	&	22.09 	$\pm$	0.12 	&	LT \\
14.86 	&	i	&	22.12 	$\pm$	0.18 	&	LT \\
14.87 	&	z	&	21.58 	$\pm$	0.25 	&	LT \\
15.71 	&	g	&	22.86 	$\pm$	0.27 	&	LT \\
15.72 	&	r	&	22.49 	$\pm$	0.17 	&	LT \\
15.72 	&	i	&	22.19 	$\pm$	0.12 	&	LT \\
15.73 	&	z	&	21.89 	$\pm$	0.26 	&	LT \\
    18.70 & r & $22.98 \pm 0.15$ & NOT \\ 
    19.80 & z & $22.50 \pm 0.20$ & NOT \\ 
    27.39 & r & $>20.0$ & NanShan-HMT \\
    30.00 & r & $23.70 \pm 0.20$  & SVOM/VT\\
31.71 	&	r	&	$>$22.98			&	LT \\
31.72 	&	g	&	$>$23.01			&	LT \\
31.74 	&	i	&	22.35 	$\pm$	0.32 	&	LT \\
31.75 	&	z	&	$>$22.23			&	LT \\
    37.47 & r & $>22.9$ & ALT-100C\\
    37.52 & z & $>21.1$ & ALT-100C\\
    38.68 & r & $23.89 \pm 0.22$ & NOT \\ 
    39.41 & r & $>22.9$ & ALT-100C\\
    39.43 & z & $>21.4$ & ALT-100C\\
47.77   &   r   &   23.94   $\pm$   0.18    &   GTC \\
    48.76 & i & $23.50 \pm 0.23$ & NOT \\ 
    68.83 & r & $24.98 \pm 0.45$ & NOT \\ 
    73.67 & i & $23.87 \pm 0.20$ & NOT \\ 

\hline
\end{longtable}

%\begin{fzhe'geigure*}[htbp!]
%\epsscale{0.8}
%\plotone{Xray/lightcurve/EP241021a_lc_wxt.pdf}
%\caption{Light curve of EP241021a when it is detected by EP-WXT.
%\label{fig:lc_wxt}}
%\end{figure*}

%\begin{figure*}[htbp!]
%    \epsscale{0.5}
%%    \plotone{GTCdeepimage.pdf}
%    \caption{Deep imaging of EP241021a taken by GTC/OSIRIS+ at $t$ = 47.77~day after trigger.}
%    \label{fig:deepimage}
%\end{figure*}

%\clear
%定义编号格式，在数字序号前加字符“A"
%\section{The VLBA high-resolution imaging of \srcs}

\section{Fittings to the X-ray light curve}\label{appeddix:xray lc}
\setcounter{figure}{0}
\renewcommand{\thefigure}{B\arabic{figure}}

\begin{figure*}[htbp!]
    %	\epsscale{0.8}
    \includegraphics[width=0.5\textwidth]{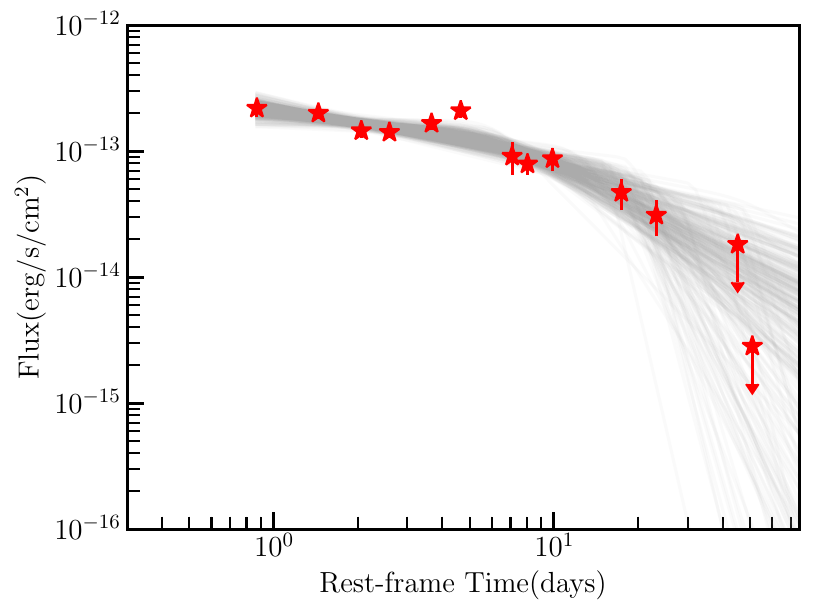}
    \raisebox{0.14cm}{\includegraphics[width=0.5\textwidth]{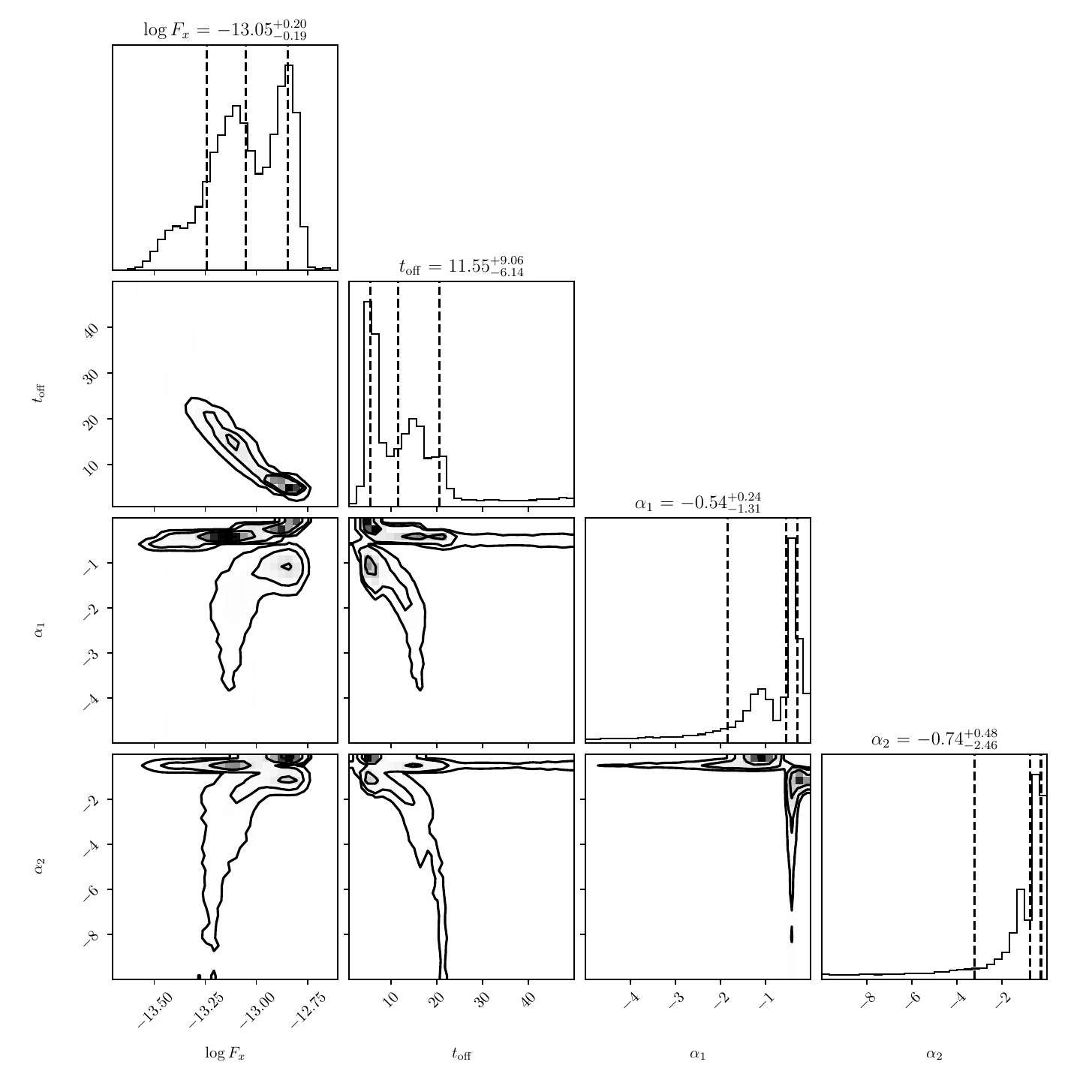}}
    \caption{
        {\bf \it Left:} 
 X-ray light curve in the 0.5--4\,keV range within $\sim 1$--50 rest-frame days since discovery. The data are taken from EP/WXT, EP/FXT, and XMM-Newton observations. 
%Bottom panel shows the host-subtracted light-curve comparison in absolute magnitude. 
% Inset panel shows the EP/WXT light curve that caught the X-ray flare. 
%The $r$-band optical light curve is shown in the lower panel. The optical transient was not detected in the pre-flare DESI Legacy Survey DR10, with an $r$-band magnitude of $>$24 mag.   
Our best-fit model with three power laws is shown as dark-gray lines, representing random samples from the MCMC chains. 
%Also shown for comparison are the X-ray light curves of four jetted TDEs known so far, including AT2022cmc (brown circles), Sw J1644+57 (light gray circles), 
%Sw J2058+05 (black squares), and Sw J1112-82 (yellow squares). The data are taken from Eftekhari et al. (2024, ApJ, 974, 149). 
%Vertical dashed lines indicate the time of jet shut-off for each event, except for 
%Note that the time of jet shut-off for 
%Sw J1112-82 as it cannot be constrained with the X-ray data available. 
{\bf \it Right:}  
%Results from the broken power-law fits to the X-ray light curve of \srcs, which can be split into three evolution phases characterized by different indices. 
Marginalized posterior distributions for each model parameter are shown on the diagonal, where dashed lines indicate 
the median and 68\% confidence interval.
            }
    \label{fig:MCLC2}
\end{figure*}

Figure \ref{fig:MCLC2} shows the X-ray light curve of \src (filled red circles and downward arrows) and the best-fit model realizations from MCMC fittings (gray curves). 
The light curve is modeled using a smoothed broken power-law function. 
The function takes the following form as presented by \cite{Eftekhari2024}: 
\begin{equation*}
    F_X(t)=F_X\left[\left(\frac{t}{t_{\rm off}}\right)^{-s\alpha_1}+\left(\frac{t}{t_{\rm off}}\right)^{-s\alpha_2}\right]^{-\frac{1}{s}}.
\end{equation*}
%We ignore the EP/WXT data point, as it has a much higher flux. 
%The plateau emission is modeled with a single power-law, followed by 
%a broken power-law to describe the flux decay. 
%the second segment based on Equation (1) from Eftekhari et al.
%The whole light curve can be modeled as follows:

\begin{comment}
    \begin{equation*}
        F_X(t)=
        \left\{
        \begin{aligned}
            & F_c\left(\frac{t}{t_c}\right)^{\alpha_0}, && t<t_c\\
            & F_X\left[\left(\frac{t}{t_{\rm off}}\right)^{-s\alpha_1}+\left(\frac{t}{t_{\rm off}}\right)^{-s\alpha_2}\right]^{-\frac{1}{s}}, && t\geqslant t_c\\
        \end{aligned}
        \right.
    \end{equation*}
\end{comment}
% where $F_c=F_X(t_c)$ and $t_c$ marks the transition from slow evolution to fast power-law decay. Following \citet{Eftekhari2024}, we fix the smoothing parameter $s=10$ and allow the remaining six parameters to vary in the fittings: the powerlaw indices $\alpha_0$, $\alpha_1$ and $\alpha_2$, flux nomalization $F_X$, and two break times are denoted as $t_c$ and $t_{\rm off}$.
\par
In order to incorporate the information provided by the upper limit points into the fitting process, we redefine the likelihood function \citep{Eftekhari2024},
\begin{equation*}
    \mathcal{L}=\prod_i p(e_i)^{\delta_i}F(e_i)^{1-\delta_i},
\end{equation*}
\noindent
where $\delta_i=0$ denotes a data point without detection, and $\delta_i=1$ represents the detection data point. We have
\begin{equation*}
    F(e_i)=\frac{1}{2}\left[1+\mathrm{erf}\left(\frac{e_i}{\sqrt{2}\sigma_i}\right)\right],
\end{equation*}
where $\mathrm{erf}$ is the error function and $\sigma_i$ represents the Poisson single-sided upper limits for data points without detection.

We find that the fit cannot adequately capture the last two upper-limit points.
To address this issue, we introduce a third power-law segment into the fitting formula and omit the smoothing factor. The resulting functional form is

\begin{equation*}
    F_X(t)=
    \left\{
    \begin{aligned}
        & F_{01} \left(\frac{t}{t_{01}}\right)^{\alpha_0}, && t<t_{01} \\
        & F_{01} \left(\frac{t}{t_{01}}\right)^{\alpha_1}, && t_{01}\leqslant t<t_{02} \\
        & F_{01}\left(\frac{t_{12}}{t_{01}}\right)^{\alpha_1} \left(\frac{t}{t_{12}}\right)^{\alpha_2}, && t\geqslant t_{12}\, . \\
    \end{aligned}
    \right.
\end{equation*}
\noindent
Here, $\alpha_0$, $\alpha_1$, and $\alpha_2$ denote the power-law indices of the three segments, $t_{01}$ and $t_{12}$ represent the respective times of the first and second breaks, and $F_{01}$ denotes the flux at the first break time.

\begin{figure*}[htbp!]
    %	\epsscale{0.8}
    \includegraphics[width=0.5\textwidth]{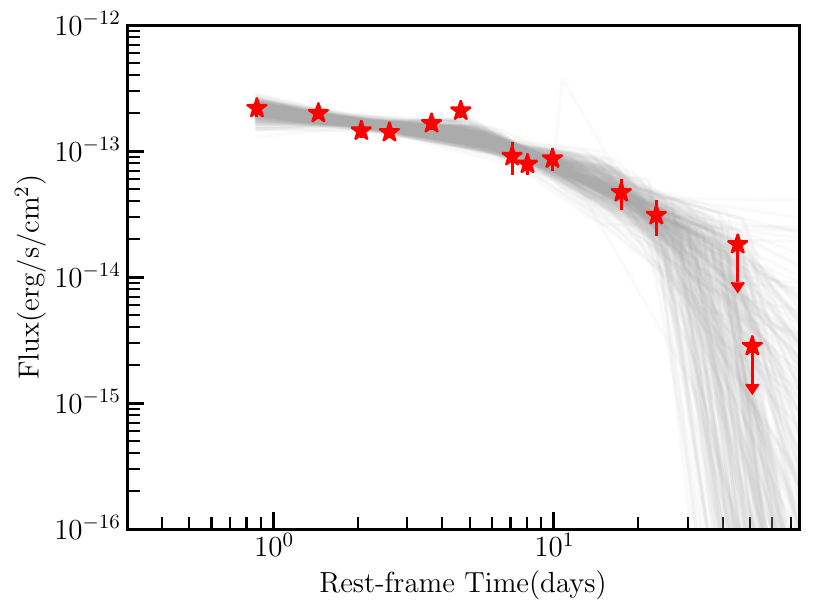}
    \raisebox{0.14cm}{\includegraphics[width=0.5\textwidth]{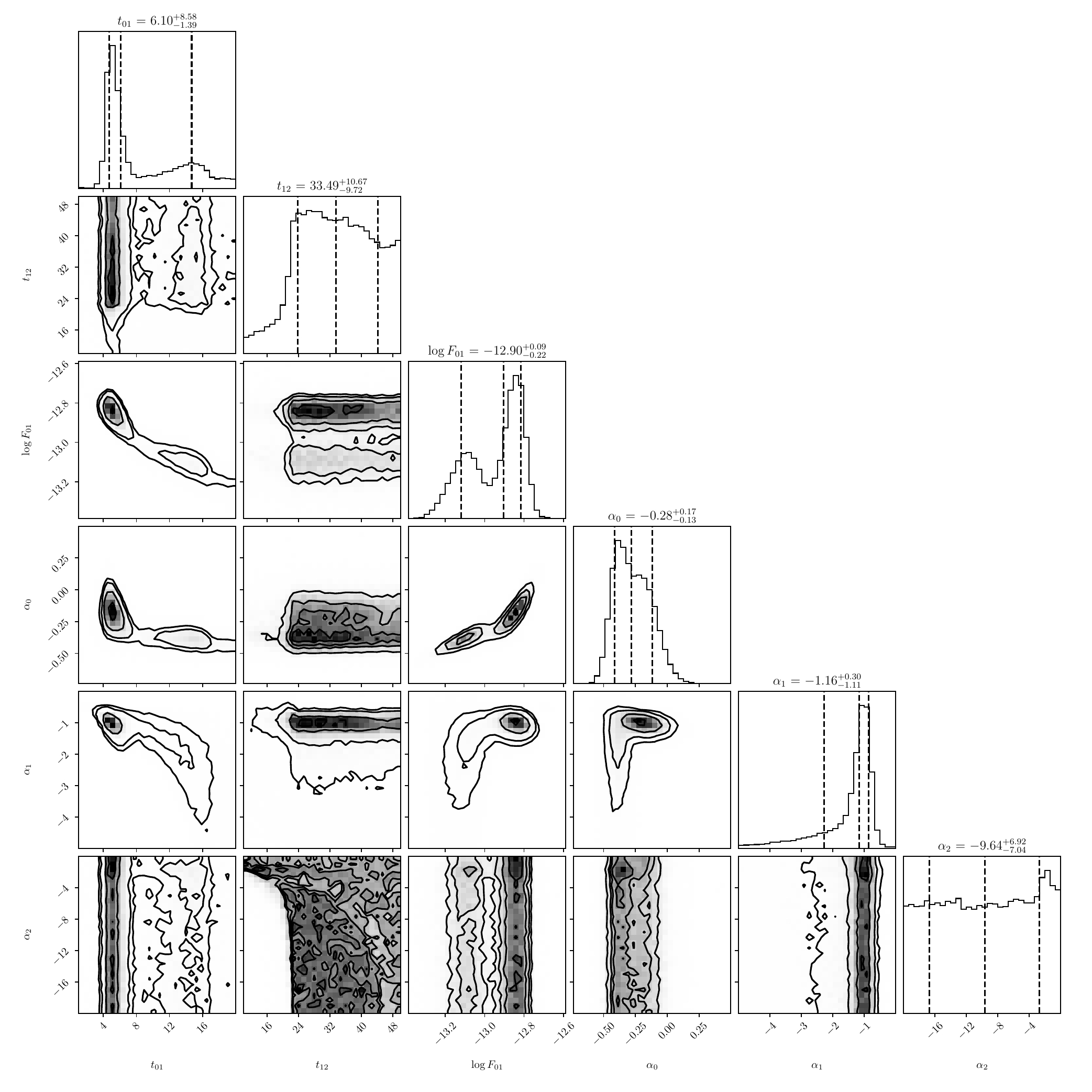}}
    \caption{
    %    {\it \bf Left panel:} 
 The same as Figure \ref{fig:MCLC2}, but for the model with three piecewise power-laws.  
            }
    \label{fig:MCLC3}
\end{figure*}

In Figure \ref{fig:MCLC3}, we also show the best-fit model realizations to the X-ray light curve using the three piecewise power-law model.  
In comparison with the broken power-law function, we found that the Bayes factor $K=4.12$, 
where $K=\exp\left(\log Z_1-\log Z_2\right)$, and $Z_1$ and $Z_2$ are known as the marginal likelihood or Bayesian evidence \citep{Buchner2023}.
This suggests that the fittings with the piecewise power law is statistically better than the broken power-law function. 
Therefore, our analysis of the X-ray light curve was based mainly on the fitting results with the piecewise power law.

\section{Parameter Constraints for the Three-jets Model from the Light-Curve Fitting}
%\appendices
\setcounter{figure}{0}
\renewcommand{\thefigure}{C\arabic{figure}}
\setcounter{table}{0}   %从0开始编号，显示出来表会A1开始编号
\renewcommand{\thetable}{C\arabic{table}}

\begin{figure*}[htbp!]
%	\epsscale{0.8}
    \plotone{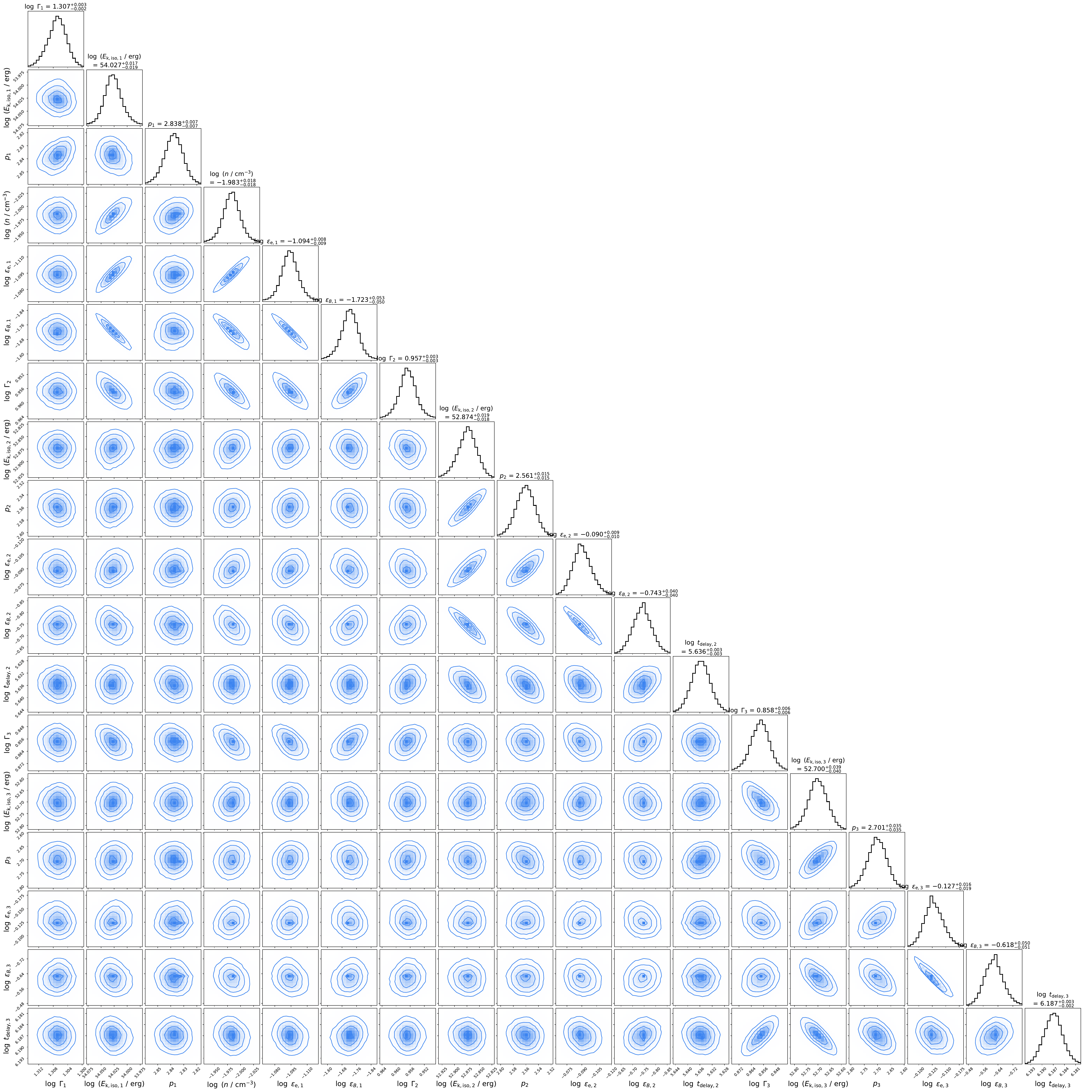}
    \caption{
        Parameters derived for EP241021a by using the three-jet model. The contour curves represent the $1\sigma-2\sigma-3\sigma$ confidence levels. The best-fit parameter values, along with their $1\sigma$ uncertainties, are indicated above the panels of their respective posterior distributions.		
T            }
    \label{fig:corner}
\end{figure*}

In this section, we show the corner plot of the multiwavelength fitting results using the model consisting of three jet ejections (Figure \ref{fig:corner}), and the constraints on the jet parameters (Table \ref{tab:threejet}).

\begin{table}[h!]
	\centering
	\tabcolsep=20pt
	\renewcommand\arraystretch{1.5}
	\caption{Parameter Constraints for the Three-jets Model from the Light-Curve Fitting. }
	\label{tab:fit}
	\begin{tabular}{lccc}
		%           \tablewidth{0pt}
		\hline
		\hline
		Parameters$^{a}$    &   First Jet &   Second jet &   Third jet  \\
		\hline
		$\log \Gamma$ & 1.307 $\pm$ 0.002 & 0.957 $\pm$ 0.003 & 0.858 $\pm$ 0.006 \\
		$\log~(E_{\mathrm{k, iso}} / \mathrm{erg})$ & 54.027 $\pm$ 0.018 & 52.874 $\pm$ 0.018 & 52.7  $\pm$ 0.04 \\
		$p$     & 2.838 $\pm$ 0.007 & 2.561 $\pm$ 0.015 & 2.701 $\pm$ 0.035 \\
		$\log \epsilon_{e}$ & -1.094 $\pm$ 0.008 & -0.09 $\pm$ 0.01  & -0.127 $\pm$ 0.018 \\
		$\log \epsilon_{B}$ & -1.723 $\pm$ 0.052 & -0.743 $\pm$ 0.04  & -0.618 $\pm$ 0.05 \\
		$\log~(n / \mathrm{cm}^{-3})$ & -1.983 $\pm$ 0.018 &  ...  & ... \\
		$\log~(t_{\mathrm{delay}} / \mathrm{s})$ & ... & 5.636 $\pm$ 0.003 & 6.187 $\pm$ 0.002 \\
		\hline
	\end{tabular}
	\begin{flushleft}
		\textbf{Notes.}
		\tablenotetext{a}{The parameters involved in the three-jets model: $\Gamma$ and $E_{\mathrm{k, iso}}$ denote the initial Lorentz factor and the isotropic kinetic energy, respectively; $p$ represents the power-law index of the electron energy distribution; $\epsilon_{e}$ and $\epsilon_{B}$ are the fractions of energy carried by electrons and magnetic fields, respectively; $n$ is the number density of the circumburst medium; $t_{\mathrm{delay}}$ indicates the time delay between the emergence of the first jet and the subsequent two jets. }
	\end{flushleft}
    \label{tab:threejet} 
\end{table}

\end{document}